\documentclass{amsart}

\usepackage{helvet}
\usepackage{amsmath}
\usepackage{amsfonts}
\usepackage{mathtools}

\usepackage{times}
\usepackage{helvet}
\usepackage[OT1]{fontenc}
\usepackage{bm}
\usepackage{bbm}
\usepackage{natbib}
\usepackage{subcaption}
\usepackage[dvips]{graphicx}
\usepackage{color}

\DeclareMathOperator*{\argmax}{arg\,max}

\newcommand{\gt}{\mathbbmtt{g}}
\newcommand{\et}{E^0}
\newcommand{\htr}{H^0}

\newtheorem{lemma}{Lemma}
\newtheorem{theorem}{Theorem}

\title{On a Variational Approximation based Empirical Likelihood ABC Method}

\author{Sanjay Chaudhuri} 
\address{Department of Statistics and Applied Probability, National University of Singapore, \quad Singapore 117546.}
\email{stasc@nus.edu.sg}
\author{Subhroshekhar Ghosh}
\address{Department of Mathematics, National University of Singapore, Singapore 119076.}
\email{subhrowork@gmail.com}
\author{David J. Nott}
\address{Department of Statistics and Applied Probability, National University of Singapore, \quad Singapore 117546.}
\email{standj@nus.edu.sg}
\author{Kim Cuc Pham}
\address{Department of Statistics and Applied Probability, National University of Singapore, \quad Singapore 117546.}
\email{staptkc@u.nus.edu}

\begin{document}

%\jname{Biometrika}
%% The year, volume, and number are determined on publication
%\jyear{2016}
%\jvol{Vol}
%\jnum{1}
%% The \doi{...} and \accessdate commands are used by the production team
%\doi{10.1093/biomet/asm023}
%\accessdate{Advance Access publication on date Month 2016}
%\copyrightinfo{\Copyright\ 2012 Biometrika Trust\goodbreak {\em Printed in Great Britain}}

%% These dates are usually set by the production team
%\received{ Month  2016}
%\revised{ Month  2016}

%% The left and right page headers are defined here:
%\markboth{Sanjay Chaudhuri, Subhroshekhar Ghosh, David J. Nott \and Kim Cuc Pham}{EL based methods in ABC}

%% Here are the title, author names and addresses
%\title{An easy-to-use empirical likelihood ABC method}
%\title{\textcolor{blue}{On a Variational Approximation based Empirical Likelihood ABC Method}}

%\author{Sanjay Chaudhuri} 
%\affil{Department of Statistics and Applied Probability, National University of Singapore, \quad Singapore 117546. \email{stasc@nus.edu.sg}}
%\author{Subhroshekhar Ghosh}
%\affil{Department of Mathematics, National University of Singapore, Singapore 119076.\email{subhrowork@gmail.com}}
%\author{David J. Nott \and Kim Cuc Pham}
% \affil{Department of Statistics and Applied Probability, National University of Singapore, \quad Singapore 117546. \email{standj@nus.edu.sg} \email{staptkc@u.nus.edu}}

\maketitle

\begin{abstract}

Many scientifically well-motivated statistical models in natural, engineering and environmental sciences are specified through a generative process. However, in some cases it may not be possible to
write down the likelihood for these models analytically.  Approximate Bayesian computation (ABC) methods allow Bayesian inference in such situations. The procedures are nonetheless 
typically computationally intensive.  Recently, computationally attractive empirical likelihood based ABC methods have been suggested in the literature.
All of these methods rely on the availability of several suitable analytically tractable estimating equations, and this is sometimes problematic.  
We propose an easy-to-use empirical likelihood ABC method in this article.  First, by using a variational approximation argument as a motivation, we show that the target log-posterior can be approximated as a sum of an expected joint log-likelihood and the differential entropy of the data generating density.  
The expected log-likelihood is then estimated by an empirical likelihood where the only inputs required are a choice of summary statistic, it's observed value, and the ability to simulate the chosen summary
statistics for any parameter value under the model.  The differential entropy is estimated from the simulated summaries using traditional methods.  Posterior consistency is established for the method, and we discuss the bounds for the required number of simulated summaries in detail.  The performance of the proposed method is explored in various examples.

\end{abstract}

%\begin{keywords}
%\textcolor{blue}{
%Approximate Bayesian Computation; Bayesian inference; Variational approximation; Differential entropy; Empirical likelihood; Estimating equation.}
%\end{keywords}

\section{Introduction}

%Suppose that $X$ is a random vector with density function $f_\theta(x)$ where $\theta\in\Theta$ denotes an unknown parameter that we wish to learn about.  
%Bayesian inference with a prior density $\pi(\theta)$ for $\theta$ is considered.  
%The observed data are written as $X_o=\left(X_{o1},\ldots,X_{on}\right)$, and are assumed to be drawn from $f_{\theta_0}(x)$, so that
%$\theta_0$ denotes the true parameter value.  We suppose that 
%the likelihood $f_{\theta}(X_o)$ cannot be written down analytically, but that for any $\theta$ we can draw independent samples according to $f_\theta(x)$.

The concept of likelihood is central to parametric statistical inference. However, for many models encountered in natural, 
engineering and environmental sciences, it is difficult to express the likelihood analytically.  These models are often specified in a generative fashion, 
so that independent samples can be generated from them for any value of the model parameters.
Approximate Bayesian computation (ABC) methods are useful for Bayesian inference in situations
like these
\citep{tavare+bgd97,beaumont+zb02,marin+prr11,fearnhead+p12,blum+nps13}.  
Simple ABC approaches involve first simulating parameter values and data from the prior, and then reducing the data 
to a lower-dimensional summary statistic which is informative for the parameter.
Following this, a comparison is made between
simulated and observed summary statistics.  For 
simulated summary statistics sufficiently close to the observed value, the corresponding parameter value
is accepted as an approximate draw from the posterior.  Other generated values of the parameters are discarded.   
This basic rejection ABC algorithm can be cast as a special case of importance sampling for
a kernel approximation of a summary statistic likelihood, and there is a known curse of dimensionality
associated with use of such methods.  
More sophisticated sampling algorithms somewhat improve efficiency 
\citep{marjoram+mpt03,sisson+ft07,beaumont+rmc09}, but even state-of-the-art ABC methods are computationally demanding 
in high-dimensional cases.  

Partly in response to the above difficulties, various pseudo-likelihood based methods have been considered.  Several such likelihoods have already been used for non-generative models by various authors \citep{monahan+b92,lazar03,chaudhuri+g11}.  Many of these approaches can also be employed in cases where a generative model exists but the associated likelihood is intractable.

Among the pseudo-likelihood methods used for generative models, perhaps the most popular is the synthetic likelihood introduced by \citet{wood10}, which uses a working multivariate normal model for the summary statistics.  Its 
Bayesian implementation is discussed in detail in \citet{price+dln16}.
The synthetic likelihood sometimes performs poorly when the normal approximation of the distribution of the summary statistics is inaccurate.  \citet{wood10} explores marginal transformation of the summaries to make the normality assumption more reasonable. However, such marginal transforms cannot usually achieve multivariate normality when the dependence structure is non-normal, or guarantee validity of the normal approximation over the whole parameter space.  
%The validity of the synthetic likelihood depends on the accuracy of the normal approximation of the distribution of the summary statistics, and even though \citet{wood10} explores marginal transformations of the summaries to improve this accuracy, the assumption of normality in synthetic likelihood has 
%been a concern for many researchers.
Extensions that relax the requirement of normality have been a continuous topic of interest for many researchers in this area.
%Ways of relaxing the normality requirement have been a continuing focus of research
\citet{Fasiolo2016} consider an extended saddlepoint approximation, whereas \citet{Dutta2016} propose
a method based on logistic regression.
\citet{anNottDrovandi2020} and \citet{priddleDrovandi2020} consider semi-parametric extensions of synthetic likelihood making use of transformations.  \citet{drovandi+pl15} describe an encompassing framework for many of the above suggestions, which they call parametric Bayesian indirect inference.  
\citet{frazier2020robust} have recently proposed a robustified version of synthetic likelihood able to detect misspecification and provide some degree of robustness to misspecification.

A fast empirical likelihood based ABC approach was recently suggested by
\citet{mengersen+pr13}, where the intractable likelihood for the generative process
was replaced by an appropriate non-parametric empirical likelihood. %for performing Bayesian inference. 
Empirical likelihood \citep{owen01} is computed from a constrained estimator of the joint empirical distribution function of the data. 
By using this likelihood \citet{mengersen+pr13} could avoid any assumption of normality of the summary statistics.  However, in their proposal constraints based on analytically tractable estimating functions of both the data and the parameters were required.  Since such functions are not readily available, their proposed method is not always easy to apply. 
%is not always easy to apply.  The reason for this is that it requires an appropriate analytically tractable estimating function, which may not be available.  

In this article, we introduce an easy-to-use empirical likelihood based ABC method, where the only required inputs are a choice of summary statistic, it's observed value, 
and the ability to simulate that particular statistic under the model for any parameter value.  
Although we refer to our method as an empirical likelihood ABC approach, it differs from the classical ABC algorithms, in the sense
that no kernel approximation of the summary statistic likelihood is involved.  Furthermore, unlike \citet{mengersen+pr13}, the proposed method does not require analytically tractable estimating functions involving the parameters. %That is, the proposed method is an interpretable likelihood-based, completely data dependent ABC procedure.

  The proposed method is motivated by information projection or variational approximation arguments.  We estimate the true posterior density of the parameter given the observed summary in the following way.  By assuming that the replicated summary is a nuisance parameter we first approximate the true joint conditional density of the replicated summary and the parameter given the observed summary.  
The analytic form of the approximation, which is motivated by results from information projection or variational approximation theory, is derived. A variational approximation of the required posterior can be obtained by marginalising the above information projection over the replicated summary.  This approximation can then be estimated from the observed and replicated summaries.
  The true posterior can be analytically expressed using two tractable terms. The first term is the expectation of an estimate of the log-joint density with respect to the true density of the data generating process.  This is a function of the parameter and the observed summary.    
The second function is the differential entropy of the data generating density which is a function of only the parameter. 
The expectation is estimated from the data using the empirical likelihood based method described above.  We employ a weighted version of the Kozachenko-Leonenko estimator \citep{kozLeo87} due to \citet{berrettSamworthMing2019} to estimate the differential entropy.

The proposed estimate of the posterior is based on an empirical likelihood which differs from what is traditionally used in the literature \citep{owen01}.  The estimated posterior is shown to be consistent for true value of the parameter when both the sample size and the number of replications grow unbounded.  
Furthermore, by invoking the results from \citet{ghosh2019empirical}, we explore the properties of the proposed empirical likelihood when the number of replications increases, but the sample size is held fixed.

In the next section we describe the basic intuition of the approach including the variational approximation of the required posterior, and Section 3 gives the definition of our proposed empirical likelihood approximation and estimate of the differential entropy.  Section 4 discusses the choice of estimating equations, 
and Section 5 describes basic asymptotic properties of the method, proving posterior consistency under reasonable conditions.  We also discuss some choices of the required number of summaries to be generated from the process. 
Section 6 considers five examples and Section 7 gives some concluding discussion.

\section{ABC Empirical Likelihood Posterior}\label{sec:multsamp}
%\section{Construction of ABC empirical likelihood}

In this section we explain the basic idea of the proposed method.  This involves
 finding the functional form of a variational approximation of the required posterior.  This approximate posterior is then estimated from the data using an empirical likelihood based method. 

%considering an artificial experiment incorporating some data replicates.  %similar to
%data cloning methods \citep{doucet+gr02,lele+dl07}.  
%  The likelihood for the replicates can then be related to the original observed data likelihood.  

  We consider a set of $n-$dimensional random vectors $\left\{X_i(\theta), i\in\mathbb{M}_o,\theta\in\Theta\right\}$, where $\mathbb{M}_o=\{o\}\cup\mathbb{N}$, i.e. the set of positive integers appended with symbol $o$.  For every $\theta$, $\left\{X_i(\theta), i\in\mathbb{M}_{o}\right\}$ are i.i.d. with an unknown density $f_0(X_i\mid \theta)$.  The observed data is generated with $\theta=\theta_o$, and would be denoted by $X_o(\theta_o)$ (or $X_o$ for brevity).  
The parameter $\theta$ is assumed to take values in the set $\Theta$. For each $\theta\in \Theta$, $m$ replicates $X_i(\theta)$, $i=1$, $2$, $\ldots$, $m$, are drawn from the data generating process. 
Suppose $\mathcal{Q}_{\Theta}$ is the set of all densities defined on $\Theta$.  We assign a prior distribution $\pi\in\mathcal{Q}_{\Theta}$ on the parameter $\theta$.

Suppose $g(x)=(g_1(x),\dots, g_r(x))^T$ is a vector of deterministic functions of the observations.  For any $\theta\in\Theta$, $g(X_i(\theta))$, $i\in\mathbb{M}_o$ are i.i.d. following an unknown density $f_0\left(g(X_i)\mid \theta\right)$. 
For a pre-specified $g$ and a prior $\pi$, for each $i=1$, $2$, $\ldots$, $m$ the \emph{true} joint distribution of $(\theta,g(X_i),g(X_o))$ is defined as:
\begin{equation}\label{eq:trueJ}
f_0\left(\theta,g(X_i),g(X_o)\right)=f_0\left(g(X_i)\mid\theta\right)f_0\left(g(X_o)\mid\theta\right)\pi(\theta). 
\end{equation}
From this, we define the \emph{true} marginal densities of $(\theta,g(X_o))$ and $g(X_o)$ respectively as:
\[
f_0\left(\theta,g(X_o)\right)=\int f_0\left(\theta,g(X_i),g(X_o)\right)dg(X_i)\text{ and }f_0\left(g(X_o)\right)=\int f_0\left(\theta,g(X_o)\right)d\theta.
\]
Our goal is to estimate the \emph{true} posterior defined as: 
\begin{equation}\label{eq:truePost}
\Pi(\theta\mid g(X_o))=\frac{f_0(\theta,g(X_o))}{f_0(g(X_o))}=\frac{f_0(g(X_o)\mid\theta)\pi(\theta)}{\int f_0(g(X_o)\mid\theta)\pi(\theta)d\theta}=f_0(\theta\mid g(X_o))
\end{equation}
from the observed data $g(X_o)$ and the replicates $g(X_i(\theta)))$, $i=1$, $2$, $\ldots$, $m$ obtained from the data generating process. In what follows, we first find the functional form of a constrained variational approximation of $\Pi(\theta\mid g(X_o))$ which can then be estimated from the available data.

\subsection{Functional form of the Variational Approximation}
In order to specify the motivating variational approximation, let $X(\theta)$ be a generic observation generated at $\theta$.  Furthermore, for notational convenience, suppose we denote $\gt=g(X(\theta))$ and $g_o=g(X_o)$.  At this stage we treat $\gt$ as a nuisance parameter.

Let $\mathcal{Q}$ and $\mathcal{F}$ be the set of all densities defined respectively on $(\theta,\gt)$ and $(\theta,\gt,g_o)$.
For any density $f(\theta,\gt,g_o)\in\mathcal{F}$, let $f(\theta,\gt\mid g_o)$ be the corresponding conditional density of $(\theta,\gt)$ given $g_o$.

Suppose $\mathcal{Q}^{\prime}$ is a subset of $\mathcal{Q}$ defined as:
\begin{equation}\label{def:qp}
  \mathcal{Q}^{\prime}=\left\{q^{\prime}(\theta)f_0(\gt\mid \theta)~:~q^{\prime}(\theta)\in\mathcal{Q}_{\Theta}\right\}.
  \end{equation}
Since $f_0(\gt\mid\theta)$ is the density of the replication generating process, the \emph{true} conditional distribution $f_0(\theta,\gt\mid g_o)=f_0(\gt\mid\theta)\Pi(\theta\mid g_o)\in\mathcal{Q}^{\prime}$.  Our goal is to estimate this true conditional density from the available data.

%We motivate our proposed method as follows. Suppose for some $f\in\mathcal{F}$, $f(\theta,\gt\mid g_o)$ is a candidate estimate of $f_0(\theta,\gt\mid g_o)$,which possibly depends on certain model parameters.  If $f(\theta,\gt\mid g_o)\not\in\mathcal{Q}^{\prime}$, which is the most likely scenario, the density is specified wrong. 
%For handling such wrongly specified densities a common procedure would be to first find the functional form of the projection of $f(\theta,\gt\mid g_o)$ onto $\mathcal{Q}^{\prime}$.  This projection, which is a density in $\mathcal{Q}^{\prime}$, depends on the so called \emph{variational parameters}, which are functions of $\theta$, $\gt$, $g_o$ and the model parameters associated with $f(\theta,\gt\mid g_o)$.  The projection is then \emph{estimated} by evaluating the variational parameters from the available data.

%using some specified criterion (see e.g. \citet{ormerodWand2010})
% ensures the estimated density is as close as possible to the true density. 
%Furthermore, a popular criterion (e.g. \citet{akaike74}) used for such projections is the minimum Kullback-Leibler divergence between the density $f(\theta,\gt\mid g_o)$ and the set $\mathcal{Q}^{\prime}$.

As motivation, suppose for some $f\in\mathcal{F}$, $f(\theta,\gt\mid g_o)$ is a candidate approximation of $f_0(\theta,\gt\mid g_o)$.  If $f(\theta,\gt\mid g_o)\not\in\mathcal{Q}^{\prime}$,  we consider the functional form of the projection of $f(\theta,\gt\mid g_o)$ onto $\mathcal{Q}^{\prime}$.  This functional form will involve some unknown terms, which we will in turn approximate in some way, discussed further below, to obtain an approximation of $f_0(\theta,\gt\mid g_o)$.

%A popular criterion (e.g. \citet{akaike74}) used for such projections is the minimum Kullback-Leibler divergence between the density $f(\theta,\gt\mid g_o)$ and the set $\mathcal{Q}^{\prime}$.

Our projections are computed by minimising Kullback-Leibler divergence between the density $f(\theta,\gt\mid g_o)$ and the set $\mathcal{Q}^{\prime}$ (see e.g. \citet{akaike74}).  Suppose $q(\theta,\gt)\in\mathcal{Q}^{\prime}$.  The Kullback-Leibler divergence between $q(\theta,\gt)$ and $f(\theta,\gt\mid g_o)$ is defined as:
\[
D_{KL}\left(q(\theta,\gt)\mid\mid f(\theta,\gt\mid g_o)\right)=\int q(\theta,\gt)\log\left(\frac{q(\theta,\gt)}{f(\theta,\gt\mid g_o)}\right)d\gt d\theta.
\]

By using the above definition, $q^{\star}(\theta,\gt)$, i.e. the \emph{information projection} \citep{coverThomasBook} or the \emph{variational approximation} of $f(\theta,\gt\mid g_o)$ onto $\mathcal{Q}^{\prime}$ is given by:
\[
  q^{\star}(\theta,\gt)=\min_{q(\theta,\gt)\in\mathcal{Q}^{\prime}}D_{KL}\left(q(\theta,\gt)\mid\mid f(\theta,\gt\mid g_o)\right)
\]

Next we find the analytic expression of $q^{\star}(\theta,\gt)$. %the variational approximation of $f(\theta,\gt\mid g_o)$ onto $\mathcal{Q}^{\prime}$. 
\begin{theorem} \label{thm:postAprx}
For any density $f\in\mathcal{F}$, let $\et_{\gt\mid\theta}\left[\log f(\theta,\gt,g_o)\right]=\int f_0(\gt\mid \theta)\log f(\theta,\gt,g_o) d\gt$ and $\htr_{\gt\mid\theta}(\theta)=-\int f_0(\gt\mid \theta)\log f_0(\gt\mid \theta)d\gt$ be the differential entropy of the density $f_0(\gt\mid \theta)$. Furthermore, let us define:
\[
f^{\prime}(\theta\mid g_o)\coloneqq\frac{e^{\et_{\gt\mid\theta}[\log f(\theta,\gt,g_o)]+\htr_{\gt\mid\theta}(\theta)}}{\int e^{\et_{\gt\mid t}[\log f(t,\gt,g_o)]+\htr_{\gt\mid t}(t)}dt}.
\]
Then $q^{\star}(\theta,\gt)=f^{\prime}(\theta\mid g_o)f_0(\gt\mid\theta)$. %the information projection or the variational approximation of $f(\theta,\gt\mid g_o)$ onto $\mathcal{Q}^{\prime}$ is given by $f^{\prime}(\theta\mid g_o)f_0(\gt\mid\theta)$.
\end{theorem}

The proof of above theorem is presented in the Appendix.  We show that, for any $q(\theta,\gt)=q^{\prime}(\theta)f_0(\gt\mid \theta)\in\mathcal{Q}^{\prime}$, such that $q^{\prime}\in\mathcal{Q}_{\Theta}$, the relationship:
\[
D_{KL}\left(q(\theta,\gt)\mid\mid f(\theta,\gt\mid g_o)\right)=D_{KL}\left(q^{\prime}(\theta)\mid\mid f^{\prime}(\theta\mid g_o)\right)+C
\]
holds, where $C$ is a non-negative function which does not depend on $q$ or $q^{\prime}$.  Now the L.H.S. is minimum when $q^{\prime}(\theta)=f^{\prime}(\theta\mid g_o)$, from which the result follows.

Having specified the variational approximation $q^{\star}(\theta,\gt)$ of $f(\theta,\gt\mid g_o)$, the variational approximation of $f(\theta\mid g_0)=\int f(\theta,\gt\mid g_o)d\gt$ is defined as $\int f^{\prime}(\theta\mid g_o)f_0(\gt\mid\theta)d\gt=f^{\prime}(\theta\mid g_o)$.

The proposed approach of posterior approximation differs from the approach taken by \citet{wood10} in constructing the synthetic likelihood.  The latter assume that $f_0(g_o\mid \theta)$ as well as $f_0(\gt\mid \theta)$ are the same Gaussian density with mean and variance depending on $\theta$.
The posterior is then constructed by plugging in an estimate of the mean and covariance matrix of $\gt$ based on the generated replications at $\theta$.
In the proposed variational approximation based approach, other than the data generative model, no user-specified models for either $f_0(g_o\mid \theta)$ or $f_0(\gt\mid \theta)$ are assumed.  Further, at the the outset, it is recognised that the trial estimate $f(\theta,\gt\mid g_o)$
is specified wrongly and its information projection on a set of densities which contains the true density is used for statistical analysis.  
%When compared to the synthetic likelihood, it is easily demonstrated that the proposed approach would produce different estimate of the posterior.  

Note that, Theorem \ref{thm:postAprx} holds for any $f\in\mathcal{F}$, with no further assumption required.  
In particular, if under $f$, $g_o$ is conditionally independent of $\gt$ given $\theta$, it follows that:

\begin{equation}\label{eq:findep}
  \et_{\gt\mid \theta}\left[\log f\left(\theta,\gt,g_o\right)\right]+\htr_{\gt\mid \theta}(\theta)=\log f\left(\theta,g_o\right)-D_{KL}\left(f_0\left(\gt\mid\theta\right)\mid\mid f\left(\gt\mid\theta\right)\right).
  %& \log f\left(\theta,g_o\right)+\et_{\gt\mid \theta}\left[\log f\left(\gt\mid\theta\right)\right]+\htr_{\gt\mid \theta}(\theta)\nonumber\\
\end{equation}
That is, under the conditional independence the L.H.S. of \eqref{eq:findep} is a variational lower bound of the log-density of $\theta$ and $g_o$, where the equality holds iff $f\left(\gt\mid\theta\right)=f_0\left(\gt\mid\theta\right)$.  The variational approximation of $f(\theta\mid g_o)$ is given by $f^{\prime}(\theta\mid g_o)$ in Theorem \ref{thm:postAprx}. %and the unknown model parameters of $f(\theta\mid g_o)$ are estimated by minimising $D_{KL}\left(f_0\left(\gt\mid\theta\right)\mid\mid f\left(\gt\mid\theta\right)\right)$.  
  
  %\[
  %\exp\left\{\et_{\gt\mid \theta}\left[\log f\left(\theta,\gt,g_o\right)\right]+\htr_{\gt\mid \theta}(\theta)\right\}/\int \exp\left\{\et_{\gt\mid \theta}\left[\log f\left(\theta,\gt,g_o\right)\right]+\htr_{\gt\mid \theta}(\theta)\right\} d\theta,
  %\]

%Study of Kullback-Leibler divergence in the context of model mis-specification has a long and illustrious history in statistics.  In the context of model selection it was first considered by \citet{akaike74}.  In the Bayesian context, the effect of mis-specification
%has been studied by many authors (e.g. \citet{kleijnVanDerWaart2006,kleijnVanderWaart2012,muller2013}).  In recent times \citet{frazier+rr17} and \citet{frazier+mrr18} have considered model mis-specification in the context of conventional ABC.    

If $f(\theta,\gt,g_o)=f_0(\theta,\gt,g_o)$, clearly $f_0(\theta,\gt\mid g_o)\in\mathcal{Q}^{\prime}$, and by definition it is it's own information projection.  That is the variational approximation of $\Pi(\theta\mid g_o)$ is exact.  More importantly we get:
\[
\Pi(\theta\mid g_o)=f^{\prime}_0(\theta\mid g_o)=\frac{e^{\et_{\gt\mid\theta}[\log f_0(\theta,\gt,g_o)]+\htr_{\gt\mid\theta}(\theta)}}{\int e^{\et_{\gt\mid t}[\log f_0(t,\gt,g_o)]+\htr_{\gt\mid t}(t)}dt}.
\]

Furthermore, when $f_0(\gt\mid\theta)$ belongs to a location family $\htr_{\gt\mid\theta}(\theta)$ is not a function of $\theta$.  In that case the expression of $\Pi(\theta\mid g_o)$ simplifies to 
\[
\Pi(\theta\mid g_o)=\frac{e^{\et_{\gt\mid\theta}[\log f_0(\theta,\gt,g_o)]}}{\int_{t\in\Theta}e^{\et_{\gt\mid t}[\log f_0(t,\gt,g_o)]}dt}.
\]

The above equalities can also be established (rather trivially) by noting that:
\[
\et_{\gt\mid\theta}[\log f_0(\theta,\gt,g_o)]+\htr_{\gt\mid\theta}(\theta)=\log f_0(\theta,g_0).
\]
However, Theorem \ref{thm:postAprx} provides a more detailed picture of the proposed procedure, which we now discuss.

The most significant outcome of Theorem \ref{thm:postAprx} is that it motivates an easy two-step procedure for estimating the true posterior $\Pi(\theta\mid g_o)$.  Since $f_0(\theta,\gt,g_0)$ in unknown, at the first step, we find its estimate $\hat{f}_0(\theta,\gt,g_0)$.  %It is often easy to ensure that $\hat{f}_0\in\mathcal{F}$, however, 
Since the analytic form of $f_0(\gt\mid \theta)$ and $\mathcal{Q}^{\prime}$ is unspecified, without further assumptions, it is extremely difficult to ensure that the corresponding conditional density of $(\theta,\gt)$ given $g_o$ is in $\mathcal{Q}^{\prime}$.  
From Theorem \ref{thm:postAprx}, it follows that in the second step, just by estimating $\et_{\gt\mid\theta}[\log \hat{f}_0(\theta,\gt,g_o)]$ and $\htr_{\gt\mid \theta}(\theta)$ a variational approximation of $\Pi(\theta\mid g_0)$ can be obtained.
This implies that in the first step, simple and arguably crude non-parametric or semi-parametric estimators of $f_0(\theta,\gt,g_0)$ can be used.  We are not required to ensure that the corresponding conditional density of $(\theta,\gt)$ given $g_o$ be in $\mathcal{Q}^{\prime}$. However, as we show below, a posterior consistent approximation of the true posterior can still be obtained.  
The proposed estimate of the true posterior requires minimal assumption on the data generating process.  
We have only assumed that for any $\theta\in\Theta$, the replicated and the observed summaries are conditionally independent given $\theta$.  %to each other and to the observed summaries given $\theta$. 

Both $\et_{\gt\mid\theta}[\log f(\theta,\gt,g_o)]$ and $\htr_{\gt\mid \theta}(\theta)$ are tractable terms, and can be estimated from the available data. Of the two, only the first term depends both on $g_o$ and $\gt$.
On the other hand, the differential entropy, which can take both positive or negative values, is not a function of the observed or the replicated summaries.  It is a function of $\theta$ and the density $f_0(\gt\mid\theta)$. 
Under mild assumptions, $|H^0_{\gt\mid\theta}(\theta)|$ remains bounded for all $\theta$, and it has only a minor effect on the asymptotic properties of the posterior, which will be determined by the data dependent term.

%\subsection{An Empirical Likelihood based Estimate of the True Posterior}
\subsection{Posterior Estimation}\label{sec:el}

We now employ empirical likelihood to estimate the true posterior $\Pi(\theta\mid g(X_0))$, using the observed data $g_o$ and the replicates $g(X_i(\theta))$, $i=1$, $2$, $\ldots$, $m$, obtained from the data generating process.  At the outset, we define the estimate  
\[
\hat{\Pi}(\theta\mid g(X_0))\coloneqq\frac{\exp\left(\hat{E}^0_{\gt\mid \theta}\left[\log \hat{f}_0(\theta,\gt,g_o)\right]+\hat{H}^0_{\gt\mid \theta}(\theta)\right)}{\int_{t\in\Theta} \exp\left(\hat{E}^0_{\gt\mid t}\left[\log \hat{f}_0(t,\gt,g_o)\right]+\hat{H}^0_{\gt\mid t}(t)\right) dt}.
\]
which requires estimating three terms.  First, the true log-joint density of the observed summary, the summaries of the i.i.d. replicates and the parameter have to be estimated.  Second, we need to estimate the expectation of the above log-joint density with respect to the distribution of the data generating process.  
Finally, the differential entropy of the data generating density needs to be estimated from the $m$ replicates $g\left(X_i\left(\theta\right)\right)$, $i=1$, $2$, $\ldots$, $m$.   

For simplicity let us assume assume that an estimate of each $f_0\left(\theta,g(X_i),g_0\right)$, $i=1$, $2$, $\ldots$, $m$ (denoted by $\hat{f}_0\left(\theta,g(X_i),g_0\right)$ and discussed below) is available.  Since the i.i.d. replicates $g(X_i)$ for $i=1$, $2$, $\ldots$, $m$ are available, a natural estimate of $\et_{\gt\mid \theta}\left[\log \hat{f}_0(\theta,\gt,g_o)\right]$ is the sample mean of $\log\hat{f}_0\left(\theta,g(X_i),g_0\right)$, $i=1$, $2$, $\ldots$, $m$.  So we can set:
\begin{equation}\label{eq:estmain}
\hat{E}^0_{\gt\mid\theta}\left[\log \hat{f}_0(\theta,\gt,g_o)\right]=\frac{1}{m}\sum^m_{i=1}\log\hat{f}_0(\theta,g(X_i),g_o).
%=\frac{1}{m}\sum^m_{i=1}\log\widehat{f_0(\theta,g(X_i(g(X_o(\theta)))}+\log\pi(\theta),
\end{equation}   

\subsubsection{Empirical Likelihood based Estimator of the Mean}
We now propose an empirical likelihood based estimator for the sample mean on the R.H.S. of \eqref{eq:estmain}.  We first note that:
\begin{equation}\label{eq:estmain2} 
\frac{1}{m}\sum^m_{i=1}\log\hat{f}_0(\theta,g(X_i),g_o)=\frac{1}{m}\sum^m_{i=1}\log\hat{f}_0(g(X_i),g(X_o)\mid \theta)+\log\pi(\theta).
\end{equation}

Furthermore when $\theta=\theta_o$,  $g(X_o)$, $g(X_1)$, $\ldots$, $g(X_m)$ are identically distributed, then for any $i=1,\dots, m$,
\begin{equation}
 % E\left[g_k\left(X_{i}(\theta),\gamma_k\right)-g_k\left(X_{o},\gamma_k\right)\right]=
  \et_{\gt\mid\theta_o}\left[g\left(X_{i}(\theta_o)\right)-g\left(X_{o}(\theta_o)\right)\right]=0. \label{eq:ex}
\end{equation}

The empirical likelihood based posterior is constructed using constraints based on the expectation in \eqref{eq:ex}.  For any $\theta\in\Theta$ and for each $i=1$, $2$, $\ldots$, $m$, define 
\begin{equation}\label{eq:h}
  h_i(\theta)=g\left(X_i(\theta)\right)-g\left(X_o(\theta_o)\right),
  \end{equation}
and the random set:
\begin{align}
~&\mathcal{W}_{\theta}=\left\{w~:~\sum^m_{i=1}w_ih_i(\theta)=0\right\} \cap\Delta_{m-1}\label{eq:w1}\\
=&\bigcap^r_{k=1}\left\{w~:~\sum^m_{i=1}w_i\left[g_k\left(X_{i}(\theta)\right)-g_k\left(X_{o}(\theta_o)\right)\right]=0\right\} \cap\Delta_{m-1},\nonumber 
\end{align}  
where $\Delta_{m-1}$ is the $m-1$ dimensional simplex.  %The random set $\mathcal{W}_{\theta}$ depends only on the observations $X_o$, $X_1$, $\ldots$, $X_m$, whose distribution depend on the parameter $\theta$. %determines all constraints required to find the estimate of our proposed likelihood in \eqref{eq:lmlik}.

%Based on observations $(X_o,X_1)$, $\ldots$, $(X_o,X_m)$, the  distribution $F^{\otimes 2}_{\theta}$ is estimated by the empirical distribution constrained by the set $\mathcal{W}_{\theta}$.  This estimate puts weight $\hat{w}_i$ on points $(X_o,X_i)$ for each $i=1,\ldots,m$, where the vector of weights $\hat{w}$ is constrained to be in $\mathcal{W}_{\theta}$.  

We first set the optimal weights $\hat{w}$ as: 
\begin{equation}
  \hat{w}:=\hat{w}(\theta):=\hat{w}(g(X_1),\ldots,g(X_m),g(X_o))=\argmax_{w\in\mathcal{W}_{\theta}}\left(\prod^m_{i=1}mw_i\right). \label{eq:w2}
\end{equation}
If the problem in \eqref{eq:w2} is infeasible, $\hat{w}$ is defined to be zero.
  
Using the optimal $\hat{w}$ we estimate the first summand in the R.H.S. of \eqref{eq:estmain2} as
% \emph{ABC empirical loglikelihood} as:
\begin{equation*}
\frac{1}{m}\sum^m_{i=1}\log\hat{f}_0(g(X_i),g(X_o)\mid \theta)=\frac{1}{m}\sum^m_{i=1}\log(\hat{w}_i(\theta)).
\end{equation*}
Now, in conjunction with the prior $\pi(\theta)$, we get:

%\begin{align}
%\hat{E}_{\gt\mid\theta}\left[\log \widehat{f_0(\theta,\gt,g_o)}\right]=&\frac{1}{m}\sum^m_{i=1}\log\hat{f}_0(\theta,g(X_i),g(X_o))\nonumber\\
%=&\frac{1}{m}\sum^m_{i=1}\log(\hat{w}_i(\theta))+\log\pi(\theta).\nonumber
%\end{align}

\begin{equation*}
\hat{E}_{\gt\mid\theta}\left[\log \hat{f}_0(\theta,\gt,g_o)\right]=\frac{1}{m}\sum^m_{i=1}\log(\hat{w}_i(\theta))+\log\pi(\theta).
\end{equation*}

The proposed empirical likelihood based estimator can be viewed as a constrained joint-empirical distribution function of the $m$ appended observations $\left(g(X_i),g(X_o)\right)$. Here we assume that in one margin the $m$ generated replicates are observed.  
In the other margin the same observation $g(X_o)$ is repeated $m$ times. This construction is similar to the data-replication methods, discussed in \citet{lele+dl07} and \citet{doucet+gr02} (see also \citet{gourieroux1996}).  
The constraints imposed satisfy those in $\mathcal{W}_{\theta}$ and the fact that for any $\theta\in\Theta$ and for any $i=1$, $2$, $\ldots$, $m$, $g(X_i)$ and $g(X_o)$ are conditionally independent given $\theta$.  The procedure is also well motivated by the discussion in equation \eqref{eq:findep} above.

%We construct a empirical likelihood ratio test for the null hypothesis based on the ammended observations $(g(X_i),g(X_o))$, $i=1$, $2$, $\ldots$, $m$, where one margin consists of the replicate observations $g(X_i)$, and in the other margin the observation $g(X_o)$ is repeated $m$ times.  

When viewed as a data-replication or a data-augmentation method, one obvious advantage of using the mean of log-weights instead of their sum is that the estimate of the corresponding Fisher information matrix would reflect the information in one observation rather than the that in $m$ artificial repeats of $\gt$ at $\theta$.  
That is, for appropriate values of $m$, the shape of final proposed estimate of the true log-posterior would be close to that of the true log-posterior $\Pi(\theta\mid g_o)$. %and its true approximation $\Pi^{\star}$. %We now present some empirical evidence in favour of the above statement.   

The estimate $\hat{f}_0(\theta,\gt,g_o)=\pi(\theta)\sum^m_{i=1}\hat{w}_i(\theta)\delta_{g_i}$, where $\delta_{g_i}$ is the delta function at $g_i$, is a crude estimate of $f_0(\theta,\gt,g)$, and it is not ensured that the corresponding $\hat{f}_0(\theta,\gt\mid g_o)$ is in $\mathcal{Q}^{\prime}$.  
However, as we show below, under weak assumptions, (essentially based on \eqref{eq:ex} above) the corresponding estimated posterior $\hat{\Pi}$, which is motivated by theorem \ref{thm:postAprx}, would be posterior consistent.

\subsubsection{Differential Entropy Estimation}
Several estimators of differential entropy have been studied in the literature.  The oracle estimator is given by $-\sum^m_{i=1}\log f_0(g(X_i(\theta)))/m$.  In this article we implement a weighted k-nearest neighbour based Kozachenko-Leonenko estimator \citep{kozLeo87,tsybakovMeulen1996} described in \citet{berrettSamworthMing2019}.

In order to define the estimator, let $||\cdot||$ denote the Euclidean norm on $\mathbb{R}^r$ and we fix an integer $k$ in $\{1,2,\ldots,m-1\}$.  In the language of \citet{berrettSamworthMing2019}, for each $i=1$, $2$, $\ldots$, $m$, let $g(X_{(1),i})$, $g(X_{(2),i})$, $\ldots$ $g(X_{(m-1),i})$ be a
 permutation of $\{g(X_1),g(X_2),\ldots,g(X_m) \}\setminus\{g(X_i)\}$ such that $||g(X_{(1),i})-g(X_i)||\le||g(X_{(2),i})-g(X_i)||$ $\le\cdots\le ||g(X_{(m-1),i})-g(X_i)||$.  Suppose we denote, $\rho_{(k),i}\coloneqq ||g(X_{(k),i})-g(X_i)||$, that is  $\rho_{(k),i}$ is the $k$th nearest neighbour of $g(X_i)$.  Furthermore, for the fixed $k$, define a set of weights $\nu=(\nu_1,\ldots,\nu_k)^T\in \mathbb{R}^k$ as
\begin{align}
\mathcal{V}^{(k)}\coloneqq&\left\{\nu\in\mathbb{R}^k~:~\sum^k_{j=1}\nu_j\frac{\Gamma(j+2l/r)}{\Gamma(j)}=0\text{~for~$l=1$, $\ldots$, $\lfloor r/4\rfloor$,}\right.\nonumber\\
&\left. \sum^k_{j=1}\nu_j=1\text{ and $\nu_j=0$ if $j\not\in\{\lfloor k/r \rfloor,\lfloor 2k/r\rfloor,\ldots,k\}$}\right\}. 
\end{align}   

For a weight vector $\nu\in\mathcal{V}^{(k)}$, \citet{berrettSamworthMing2019} define the weighted Kozachenko-Leonenko estimator of $\htr_{\gt\mid \theta}(\theta)$ as
\begin{equation}\label{eq:hEst}
\hat{H}^0_{\gt\mid \theta}(\theta)=\frac{1}{m}\sum^m_{i=1}\sum^k_{j=1}\nu_j\log\left(\frac{(m-1)\pi^{r/2}\rho^r_{(j),i}}{e^{-\psi(j)}\Gamma(1+r/2)}\right),
\end{equation}
where $\psi$ is the digamma function.

In order to find one entry in $\mathcal{V}^{(k)}$, we solve: 
\begin{equation}\label{eq:euLik}
\hat{\nu}=\arg\mbox{min}_{\nu\in\mathcal{V}^{(k)}}\sum^k_{j=1}(m\nu-1)^2.
\end{equation}
The objective function in \eqref{eq:euLik} is the so called \emph{Euclidean likelihood} (see \citep{owen01}) which has been previously studied by \citet{brown_chen_1998}.

From \citet{berrettSamworthMing2019} it follows that the normalised risk of the proposed estimator converges in a uniform sense to that of the unbiased oracle estimator. Other histogram or kernel based estimators \citep{hallMorton1993,paninskiYazima2008} can be considered.  Due to curse of dimensionality, they don't perform well in high dimensions.  They are also potentially computationally expensive.

If the summary statistics are approximately normally distributed, it is often sufficient and computationally more efficient to directly use the  expression of differential entropy for a normal random vector, which depends only on the determinant of the covariance matrix.   

%Due to the curse of dimensionality histogram or kernel based estimators \citep{hallMorton1993,paninskiYazima2008} don't easily extend to higher dimensions (i.e. when $r\ge 2$).   The behaviour of the k-nearest neighbour based method proposed by \citet{kozLeo87} is similar \citep{tsybakovMeulen1996}.
%A weighted Kozachenko-Leonenko estimator has recently been shown to be uniformly consistent \citet{berrettSamworthMing2019}, but it requires a large sample size to produce any reliable estimate.

\subsubsection{ABC Empirical Likelihood Posterior}
Finally the corresponding \emph{ABC empirical likelihood} (\emph{abcEl}) estimate of the required posterior, i.e. $\hat{\Pi}(\theta\mid g(X_o))$ is given by,
\begin{align}
%\Pi_o(\theta)\deq
  \hat{\Pi}(\theta\mid g(X_o))&=\frac{\left[e^{\left(\frac{1}{m}\sum^m_{i=1}\log\left(\hat{w}_i(\theta)\right)+\hat{H}^0_{\gt\mid\theta}(\theta)\right)}\right]\pi(\theta)}{\int_{t\in\Theta}\left[e^{\left(\frac{1}{m}\sum^m_{i=1}\log\left(\hat{w}_i(t)\right)+\hat{H}^0_{\gt\mid t}(t)\right)}\right]\pi(t)dt}\nonumber\\
  &\propto\left[e^{\left(\frac{1}{m}\sum^m_{i=1}\log\left(\hat{w}_i(\theta)\right)+\hat{H}^0_{\gt\mid\theta}(\theta)\right)}\right]\pi(\theta).%\nonumber%\\
%&=\frac{\left[\left(\prod^m_{i=1}\hat{w}_i(\theta)\right)^{1/m}\right]\pi(\theta)}{\int_{t\in\Theta}\left[\prod^m_{i=1}\left(\hat{w}_i(t)\right)^{1/m}\right]\pi(t)dt}\propto\left[e^{\frac{1}{m}\sum^m_{i=1}\log(\hat{w}_i)}\right]\pi(\theta).%=\left\{\prod^m_{i=1}\hat{w}_i\right\}^{\frac{1}{m}}\pi(\theta).
\label{eq:mpost}
\end{align}
When $\prod^m_{i=1}\hat{w}_i=0$, we define $\hat{\Pi}(\theta\mid g(X_o))=0$.

%Thus the estimated likelihood is bounded, and the abcEl posterior $\hat{\Pi}(\theta\mid g(X_o))$ is proper for any proper prior $\pi$.

No analytic expression for the proposed abcEl posterior exists in general.  However, by construction, each $\hat{w}_i$ is bounded for all values $\theta$.  All components of $\hat{w}$ in \eqref{eq:w2} are strictly positive iff the origin is in the interior of the convex hull defined by the vectors $h_1$, $h_2$, $\ldots$, $h_m$.  
When the origin is at the boundary of this convex hull, the constrained optimisation in \eqref{eq:w2} is still feasible, but some of the estimated weights are zero, so by definition the posterior is zero as well.  In both these cases, $\mathcal{W}_{\theta}$ in \eqref{eq:w1} is non-empty.  If the origin is outside this closed convex hull, 
this optimisation problem is infeasible and the value of the abcEl posterior is zero. It is well-known (see eg. \citet{chaudhuri+my17}) the that support of the BayesEl posteriors are in general non-convex.  It is expected that the proposed abcEl posterior will suffer from the same deficiency as well.

As we have discussed above the proposed method is more general than the synthetic likelihood.  
The latter assumes normality of the joint distribution of the summary statistics.  
Even though many summary statistics are asymptotically normally distributed, this is not always the case, and  
in some cases involving non-normal summary statistics the synthetic likelihood can perform poorly (see e.g. Section \ref{sec:arch} below).  
\citet{mengersen+pr13} use Bayesian empirical likelihood in an ABC setting. However, the estimating equations they use directly depend on the parameter, 
and these equations must be analytically specified.  Such estimating equations may not be available in many problems.  
In our empirical likelihood approximation, we only require the observed data $X_o$ and simulated data $X_1,\dots,X_m$ under the model for a given $\theta$.  
Furthermore, the proposed empirical likelihood can be computed quite easily and usually at a reasonable computational cost.
The proposed empirical likelihood estimates weights by matching the moments of $g(X_1)$, $\ldots$, $g(X_m)$ with that of $g(X_o)$, without requiring a direct relationship with the parameter.

The proposed posterior in \eqref{eq:mpost} however, is different from the original Bayesian empirical likelihood (BayesEl) posterior used in usual applications \citep{lazar03, chaudhuri+g11}.  The abcEl posterior is defined with the mean of the log-weights (see \eqref{eq:mpost}).  This is different from the usual BayesEl posteriors \citep{chaudhuri+my17}, where the sum of the log-weights are used.  
Because of this difference in construction, the proposed abcEl posterior differ from the usual BayesEL posterior in both asymptotic and finite sample properties, which we will discuss in subsequent sections.

%\subsection{Sampling the Proposed abcEl Posterior}
Since no analytic form is available, any inference about the true value of the parameter has to be drawn by generating samples from the abcEl posterior $\hat{\Pi}(\theta\mid X_o)$. Such a sample can be drawn using Markov Chain Monte Carlo (MCMC) techniques.  
This is sufficient for making posterior inferences. 

%\textcolor{blue}{I think we need some comments about the convergence of the Markov Chain, Does it converge to $\hat{\Pi}(\theta\mid X_o)$? 
%What does it mean to converge to $\hat{\Pi}(\theta\mid X_o)$? If we can make a formal statement, we can make a subsection of this part.}   

\subsubsection{Example} 
\begin{figure}[t]
\begin{center}
\begin{subfigure}{.45\columnwidth}
\resizebox{2.6in}{2.6in}{\includegraphics{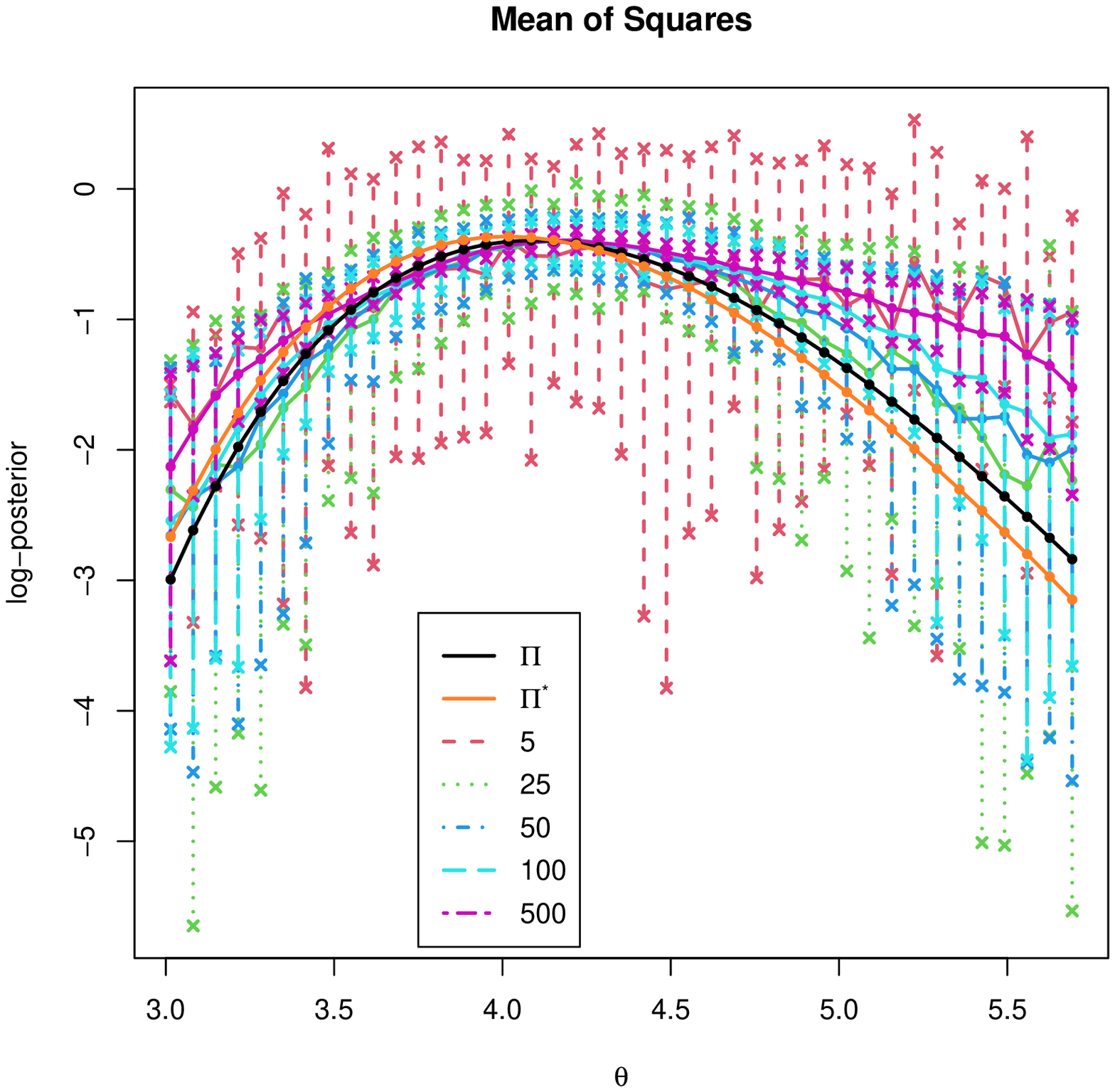}}
\caption{$\hat{\Pi}(\theta\mid g_1(X_o(\theta_o)))$}
\label{fig:hatSigma}
\end{subfigure}\hfill\begin{subfigure}{.45\columnwidth}
\resizebox{2.6in}{2.6in}{\includegraphics{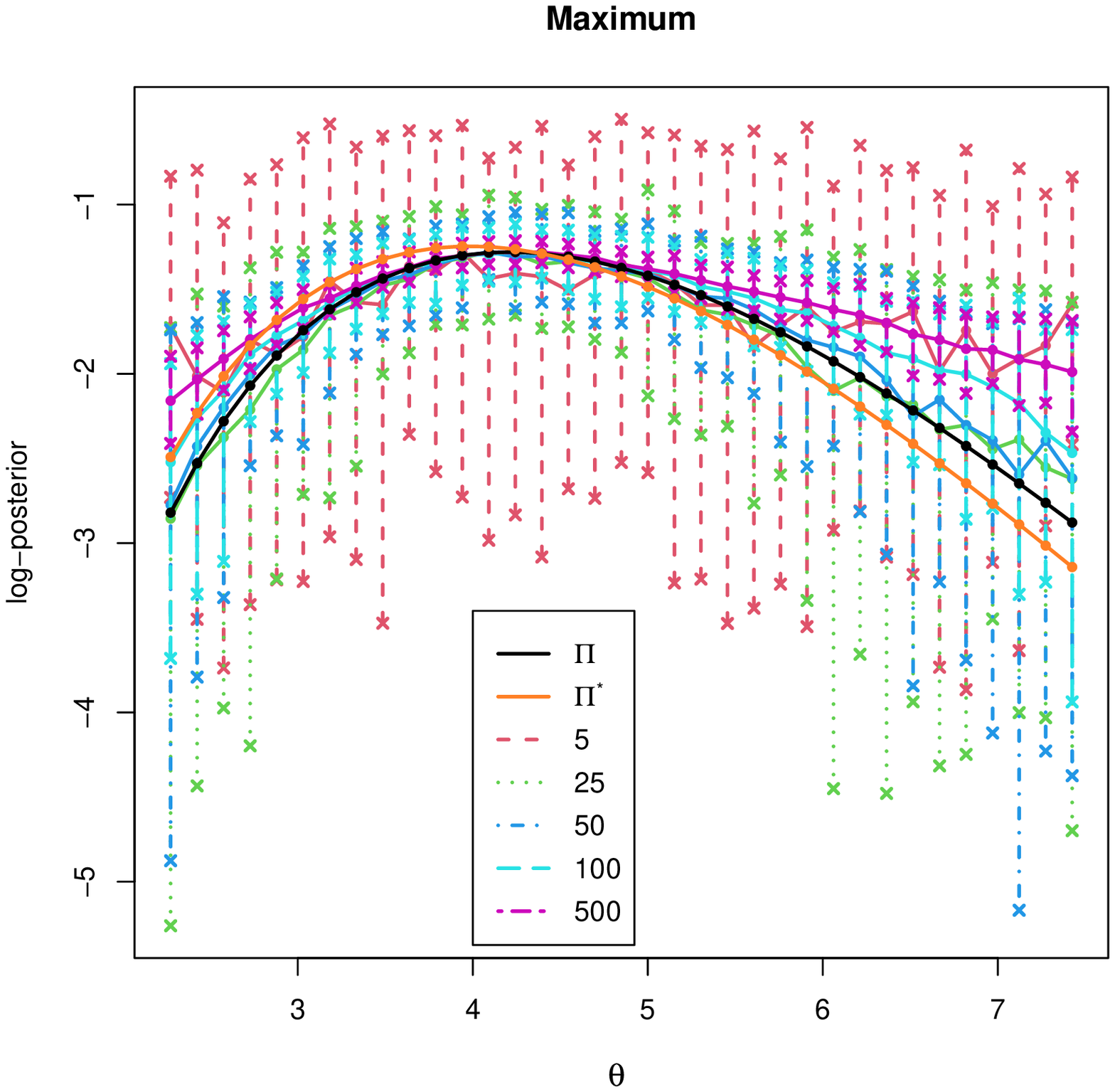}}
\caption{$\hat{\Pi}(\theta\mid g_2(X_o(\theta_o)))$}
\label{fig:hatMax}
\end{subfigure}
\caption{Comparison of the true log-posterior with the logarithm of the proposed estimator for different values of $m$.  The samples of size $n=100$ were drawn from $N(0,\theta)$ distribution.  The true value of $\theta$ ie. $\theta_o$ was $4$.  We chose (a) $g_1(x_i)=\sum_j X^2_{ij}/n$ and (b) $g_2(x_i)=\max_j(X_{ij})$ and a $U(0,10)$ prior on $\theta$.  
The true log-posterior is shown in black.  For each value of $\theta$ and $m$ the means and the limits of the $95\%$ confidence intervals of the estimated log-posterior based on $100$ repetitions are shown.}
\label{fig:post}
\end{center}
\end{figure}

In Figure \ref{fig:post} we compare the shape of the abcEl log-posteriors with the true log-posteriors $\Pi$ for the variance of a Normal distribution with zero mean conditional on (a) $g_1(X_i)=\sum_jX^2_{ij}/n$ (Figure \ref{fig:hatSigma}) and (b) $g_2(X_i)=\max_j(X_{ij})$ (Figure \ref{fig:hatMax}).  
Here, for each $i=1$, $2$, $\ldots$, $m$, and $j=1$, $2$, $\ldots$, $100$, the observation $X_{ij}$ is drawn from a $N\left(0,\theta\right)$.  
The true value of the parameter i.e. $\text{Var}(X_o)$ was fixed at $4$. We assume that the parameter $\theta$ follows a $U(0,10)$ prior.  In order to compare the contribution of the differential entropy term, we also display the function $\log \Pi^{\star}_0(\theta\mid g_o)=\et_{\gt\mid\theta}\left[\log f_0(\theta,\gt,g_o)\right]-\log\int_{t\in\theta}\exp(\et_{\gt\mid t}\left[\log f_0(t,\gt,g_o)\right])dt$.

%In Figures \ref{fig:hatSig} and \ref{fig:hatMax} the proposed abcEl posteriors are compared.  %In Figures \ref{fig:tildeSig} and \ref{fig:tildeMax}, we compare another \emph{partial} estimate $\tilde{\Pi}^{\star}$ of $\Pi^{\star}$.
%Here, the expectaion $E_{\gt\mid\theta}\left[\log f_0(\theta,\gt,g_o)\right]$ is estimated by $\sum^m_{i=1}\log f_0(\theta,g(X_i),g(X_o))/m$, but unlike the abcEl posterior the true joint density was used instead of its empirical likleihood based estimate. 

The log-posteriors were compared on a grid of parameters whose true posterior value were larger than the $.05$.  % of the maximum of the true posterior. 
Based on $100$ repetitions, At each value of $\theta$ and $m$, the mean and the endpoints of the symmetric $95\%$ confidence intervals are shown in the figure.
To make the comparison of the shapes easier, for each $m$, maximum of the mean of abcEl log-posterior was matched with the maximum value of the true log-posterior.  %For the same reason the maximum of $\Pi^{\star}$ was matched with the maximum of maximum of $\Pi$.  

From Figure \ref{fig:post} it follows that for $m=25$ and $m=50$, for each value of $\theta$ the means of the estimated log-posteriors (solid coloured lines) are very close to the true log-posterior (solid black line) for both $g_1(X_o)$ and $g_2(X_o)$.  
Furthermore the $95\%$ confidence bands always cover the corresponding true value of the log-posterior.  
It is evident that the proposed abcEl posterior is a good approximation of the true posterior up to a scaling constant.   This is even true for the summary function $g_2(X_o)$, which unlike $g_1(X_o)$, asymptotically does not converge to a normal random variable under any centring or scaling.

From figures \ref{fig:hatSigma} and \ref{fig:hatMax} it is evident that $\log f^{\prime}_0(\theta\mid g_o)$ closely approximates the true log-posterior $\Pi$.  That is, the differential entropy term has arguably minimal contribution to the true posterior.

As the number of replicates i.e. $m$ increases (see $m=500$), in Figure \ref{fig:post} the log-posterior, tends to get more flat in shape.  However, the confidence bands get narrower.  
This is a known property of a mis-specified empirical likelihood \citep{ghosh2019empirical}, which naturally occurs with high probability in the computation of abcEl posterior.  We discuss this phenomenon and use it to select an appropriate $m$ in Section \ref{sec:m} below.  

%also show that the choice of an appropriate $m$ involves a bias-variance trade-off. On the one hand, for smaller values of $m$ (say $m=5$), for each $\theta$ the mean of the proposed estimate seems to be close to the true value of the log-posterior. 
%However, the corresponding confidence intervals are long, which hints at high variance.  As $m$ increases (see $m=500$) the lengths of the confidence bands shorten, however, the means of the estimated log-posteriors are generally larger than the corresponding true log-posteriors.   
 
%%%%%%%%%%%%%%%%%%%%%%%%%%%%%%%%%%%%%%%%%%%%%%%%%%%%%%

%\begin{figure}[ht]
%\begin{center}
%\begin{subfigure}{.45\columnwidth}
%\resizebox{2.75in}{2.75in}{\includegraphics{meanOfSqPost2.ps}}
%\caption{$\tilde{\Pi}^{\star}(\theta\mid g_1(X_o(\theta_o)))$}
%\label{fig:tildeSig}
%\end{subfigure}
%\hfill
%\begin{subfigure}{.45\columnwidth}
%\resizebox{2.75in}{2.75in}{\includegraphics{maxPost2.ps}}
%\caption{$\tilde{\Pi}^{\star}(\theta\mid g_2(X_o(\theta_o)))$}
%\label{fig:tildeMax}
%\end{subfigure}
%\begin{subfigure}{.45\columnwidth}
%\resizebox{2.75in}{2.75in}{\includegraphics{meanOfSqPost.ps}}
%\caption{$\hat{\Pi}(\theta\mid g_1(X_o(\theta_o)))$}
%\label{fig:hatSig}
%\end{subfigure}
%\hfill
%\begin{subfigure}{.45\columnwidth}
%\resizebox{2.75in}{2.75in}{\includegraphics{maxPost.ps}}
%\caption{$\hat{\Pi}(\theta\mid g_2(X_o(\theta_o)))$}
%\label{fig:hatMax}
%\end{subfigure}

%That is, unlike the original empirical likelihood the Wilk's statistics corresponding to the proposed likelihood, would not have an asymptotic chi-squared distribution.% under mild conditions. The same would not be true for the Wilk's statistic corresponding to the proposed likelihood. 

\subsection{Choice of Estimating Equations}
It is clear that much depends on the correct specification of the constraints imposed on the weights which determine the empirical likelihood. 
In most applications of Bayesian empirical likelihood, these constraints directly depend on the parameter $\theta$ through an analytically specified estimating equation.  
However, the structure of our proposed empirical likelihood allows us to specify constraints without involving the parameter except through the simulation of the observations $X_i=X_i(\theta)$.
Many choices for the  constraint functions are possible.  We outline some simple choices below. %although often these will not be adequate in examples involving complicated dependent data.

From now on, we assume that for $i\in\{o,1,\dots,m\}$, $X_i\in\mathbb{R}^n$.  For some $k$ and some positive deterministic $\gamma_k$, for each $i$ we may define,
\begin{equation}
g_k\left(X_{i}\right)=\frac{1}{n}\sum^n_{j=1}X^{\gamma_k}_{ij}, \label{eq:mnt}
\end{equation}
so that $g_k$ is the $\gamma_k$th raw sample moment.  Provided $E[X_{ij}^{\gamma_k}]$ exists, such a choice of $g_k$ would constrain the underlying distribution through its moments.  
Similarly the $\gamma_k$ sample quantile of $X_i$ may be used for any $\gamma_k \in[0,1]$, which would directly put a constraint on the distribution through its quantiles.
Another possibility is the proportion of times $X_i$ is larger than $\gamma_k$, 
\begin{equation}
g_k\left(X_{i}\right)=\frac{1}{n}\sum^n_{j=1}\mathbf{1}_{\{X_{ij}\ge \gamma_k\}}. \label{eq:upt}
\end{equation}
Other than these generic choices, one can base the constraints on functionals of transformed variables. For example, in certain situations constraints based on the spectral distribution of the data could be used. 

With these choices of $g$, the likelihood is estimated by matching the marginal moments, quantiles and up-crossings of the generated vectors with those of the observed values.  %Even though  
%While these are some useful generic functions of the data, choices of summary statistic that use insights about the model may be better
%in particular cases.
In complex data models, where the $X_i$ and $X_o$ have non-identically distributed and dependent components,
looking at simple marginal properties of the components of $X_i$ and $X_o$ may not be adequate and some insight about the model could be used to choose the constraints.  In such cases, constraints 
can be based on joint moments, joint quantiles or joint up-crossings of subsets of $\{X_{i1},\ldots,X_{in}\}$, as we illustrate later.  
Any summary statistics used in traditional ABC analyses can also be used in the proposed empirical likelihood approach.

%Finally, it can be easily seen that, replacing the constraints based on averages with those based on totals do not change the likelihood or the posterior, so the constraints based on the totals can be used as well (see Section \ref{sec:graph}).    

\section{Properties of the ABC Empirical Likelihood Posterior}
The asymptotic properties of conventional ABC methods have been a topic of much recent research
\citep{frazier+mrr18,li+f18a,li+f18b}.  Here 
we investigate some basic asymptotic properties of our proposed empirical likelihood method. The proofs of the results are deferred to the supplement. 

Following \citet{owen01} the weights in \eqref{eq:w2} can be obtained by maximising the objective function:

\[
L(w)=\sum^m_{i=1}\log(mw_i)-\alpha\left(\sum^m_{i=1}w_i-1\right)-n\lambda^T\sum^m_{i=1}w_ih_i,
\]
where $\alpha$ and $\lambda$ are the Lagrange multipliers associated with the constraints.  It is easily shown that $\alpha=1$ and the optimum weights are given by
\[
\hat{w}_i=\frac{1}{m}\frac{1}{1+\hat{\lambda}^Th_i},
\]
where $\hat{\lambda}$ is obtained by solving the equation

\begin{equation}\label{eq:lambda}
\sum^m_{i=1}\frac{h_i}{1+\hat{\lambda}^Th_i}=0.
\end{equation}

\subsection{Posterior Consistency}\label{sec:postcons}
In what follows below, we consider limits as $n$ and $m=m(n)$ grow unbounded.  Furthermore, for convenience, we make the dependencies of $X_o$ and $X_1$, $X_2$, $\ldots$, $X_m\in\mathbb{R}^n$ on sample size $n$ and parameter $\theta$ explicit. %, and suppress their dependence on $\gamma$.  
In what follows, a sequence of events $\{E_n, n\ge 1\}$ is said to occur with high probability, if $P(E_n)\rightarrow 1$ as $n\rightarrow\infty$.

Suppose that we define
\[
h^{(n)}_i\left(\theta\right)=\left\{g\left(X^{(n)}_i(\theta)\right)-g\left(X^{(n)}_o(\theta_o)\right)\right\}, 
\]  
and assume $\et_{\gt\mid\theta}[g(X^{(n)}_i(\theta))]$ is finite so that we can write 
\[
g\left(X^{(n)}_i(\theta)\right)=\et_{\gt\mid\theta}\left[g\left(X^{(n)}_i(\theta)\right)\right]+\xi^{(n)}_i(\theta)=\mathfrak{g}^{(n)}(\theta)+\xi^{(n)}_i(\theta),
\] 
where $\et_{\gt\mid\theta}[\xi^{(n)}_i(\theta)]=0$ for all $i$, $n$ and $\theta$.

We make the following assumptions.
\begin{itemize}
%\item Assume that $\theta_o$ is in the interior of the support of the likelihood.
\item[(A1)] (Identifiability and convergence) There is a sequence of positive increasing real numbers $b_n\rightarrow\infty$, such that:
\[
\mathfrak{g}^{(n)}(\theta)=b_n\left\{\mathfrak{g}(\theta)+o(1)\right\},
\]   
where $\mathfrak{g}(\theta)$ is a one-to-one function of $\theta$ that does not depend on $n$.  Furthermore, $\mathfrak{g}(\theta)$ is continuous at $\theta_o$ and for each $\epsilon>0$,  and for all $\theta\in\Theta$, there exists $\delta>0$, such that whenever $\mid\mid\theta-\theta_o\mid\mid>\epsilon$, $\mid\mid \mathfrak{g}(\theta)-\mathfrak{g}(\theta_o)\mid\mid>\delta$.

\item[(A2)] (Feasibility) For each $\theta$, $n$ and $i=o,1$, $\ldots$, $m(n)$, the vectors $\xi^{(n)}_i(\theta)$ are identically distributed, supported over the whole space, and their distribution puts positive mass on every orthant, $\mathcal{O}_s$ of $\mathbb{R}^r$, $s=1$, $2$, $\ldots$, $2^r$.  Furthermore, for every orthant $\mathcal{O}_s$, as $n\rightarrow\infty$, 
\[
\sup_{\{i~:~\xi^{(n)}_i(\theta)\in\mathcal{O}_s\}}\mid\mid \xi^{(n)}_i(\theta)\mid\mid\longrightarrow\infty
\]
in probability, uniformly in $\theta$.   
\item[(A3)] (Growth of extrema of Errors) As $n\rightarrow\infty$, 
\[
\sup_{i\in\{o,1,2,\ldots, m(n)\}}\frac{\mid\mid \xi^{(n)}_i(\theta)\mid\mid}{b_n}\rightarrow 0
\]
in probability, uniformly in $\theta\in\Theta$.
%\item For each $n$, there exists $\delta_n$ such that whenever $\mid\mid\theta-\theta_o\mid\mid\le\delta_n$, $\mid\mid\mathfrak{g}(\theta)-\mathfrak{g}(\theta_o)\mid\mid\le b^{-1}_n$.  
\end{itemize}

  %Suppose $\hat{H}^{0(n)}_{\gt\mid\theta}(\theta)$ is the weighted Kozachenko-Leonenko estimator of the diferential entropy of $f_0(\gt\mid\theta)$ at $\theta$, based on $m(n)$ observations.  For $\theta\in \Theta$ and $\epsilon>0$, by $B(\theta,\epsilon)$ we denote the ball of radius $\epsilon$ around $\theta$.  We further assume that:}
%\begin{itemize}
%\item[(A4)]\textcolor{blue}{For a fixed $\epsilon>0$, and for each $\theta\in B(\theta_o,\epsilon)$, $\mid \hat{H}^{0(n)}_{\gt\mid\theta}(\theta)\mid$ is bounded with a high probability.}
%\end{itemize}
%Although (A1) seems quite strong , this assumption is 
%easily weakened to convergence of $\mathfrak{g}^{(n)}(\theta)/b_n-\mathfrak{g}(\theta)$ to zero uniformly in
%$\theta$.  

Assumption (A1) ensures identifiability and additionally implies that $\mathfrak{g}^{(n)}(\theta)/b_n-\mathfrak{g}(\theta)$ converges to zero uniformly in $\theta$.
Assumption (A2) is important for ensuring that with high probability the empirical likelihood ABC posterior is a valid
probability measure for $n$ large enough.  Assumptions (A2) and (A3) also link the number of simulations $m$ to $n$ 
and ensure concentration of the posterior with increasing $n$. %\textcolor{blue}{Assumption (A4) ensures that, with high probability $exp(\hat{H}^{0(n)}_{\gt\mid\theta}(\theta))\ne 0$ when $l_n(\theta)>0$.} 
The proofs of the results below are given in the Appendix.  The main result, Theorem 1, shows posterior consistency for the proposed empirical likelihood method.  

Let $l_n(\theta):=\exp(\sum^{m(n)}_{i=1}\log\left(\hat{w}_i(\theta)\right)/m(n))$ and for each $n$, we define:
\[
\Theta_n=\left\{\theta~:~\mid\mid\mathfrak{g}(\theta)-\mathfrak{g}(\theta_o)\mid\mid\le b^{-1}_n\right\}.
\]
By continuity of $\mathfrak{g}$ at $\theta_0$, $\Theta_n$ is nonempty for each $n$.  Furthermore, since $b_n$ is increasing in $n$, $\Theta_n$ is a decreasing sequence of sets in $n$. 

\begin{lemma}\label{lem:1}
  Under assumptions (A1) to (A3), with high probability, the likelihood $l_n(\theta)>0$ for all $\theta\in\Theta_n$.
  %Under the assumptions, there exists integer $n_0$, such that for each $n\ge n_0$, there exists a neighbourhood of $\theta_o$, on which the likelihood is strictly positive with high probability.
\end{lemma}

Lemma \ref{lem:1} shows that for large $n$ the estimated likelihood is strictly positive in a neighbourhood of $\theta_0$.  Next, we show that the empirical likelihood is zero outside certain neighbourhood of $\theta_0$.

\begin{lemma}\label{lem:2}
  Under assumptions (A1) - (A3), for every $\epsilon>0$, the empirical likelihood is zero outside $B(\theta_0,\epsilon)$, with high probability.
%Under the assumptions, for any $\epsilon>0$, there exists $n(\epsilon)$ such that for any $n>n(\epsilon)$, and for any $\theta$, such that $\mid\mid\theta-\theta_o\mid\mid >\epsilon$, the empirical likelihood is equal to zero.%and the resulting posterior are equal to zero. 
\end{lemma}

Now suppose we choose $\epsilon=b^{-1}_1$ and $n>n(b^{-1}_1)$ % by Lemma \ref{lem:1}:
%\[
%\int_{\left\{\theta~:~\mid\mid\mathfrak{g}(\theta)-\mathfrak{g}(\theta_o)\mid\mid>b^{-1}_1\right\}}l_n(\theta)\pi(\theta)d\theta=0.
%\]
%Since $\Theta_n\subseteq\Theta_1$, and by lemma \ref{lem:2}, 
such that $l_n(\theta)$ is positive on $\Theta_n$ with high probability.  Furthermore, for all $n$ and for all $\theta$, $\min_{i\ne j}\mid\mid g(X_j(\theta))-g(x_i(\theta))\mid\mid>0$ with probability $1$, which implies $\mid\hat{H}^{0(n)}_{\gt\mid \theta}(\theta)\mid>-\infty$ with probability $1$ as well.  This proves that for large values of $n$, with high probability:
%\[
%\int_{\theta\in\Theta}l_n(\theta)\pi(\theta)d\theta\ge\int_{\theta\in\Theta_n}l_n(\theta)\pi(\theta)d\theta>0
%\]
%with high probability. This proves that for large values of $n$, with high probability:
\[
\int_{\theta\in\Theta}l_n(\theta)e^{\hat{H}^{0(n)}_{\gt\mid \theta}(\theta)}\pi(\theta)d\theta\ge\int_{\theta\in\Theta_n}l_n(\theta)e^{\hat{H}^{0(n)}_{\gt\mid \theta}(\theta)}\pi(\theta)d\theta>0,
\]
and 
\[
\hat{\Pi}_n\left(\theta\mid g(X_o(\theta_o))\right)=\frac{l_n(\theta)e^{\hat{H}^{0(n)}_{\gt\mid \theta}(\theta)}\pi(\theta)}{\int_{t\in\Theta}l_n(t)e^{\hat{H}^{0(n)}_{\gt\mid t}(t)}\pi(t)dt}
\]
is a valid probability measure (with high probability).  The main result, Theorem 1 below, establishes posterior consistency.

\begin{theorem}\label{thm:1}
As $n\rightarrow\infty$, $\hat{\Pi}_n\left(\theta\mid g(X_o(\theta_o))\right)$ converges in probability to $\delta_{\theta_o}$, where $\delta_{\theta_0}$ is the degenerate probability measure supported at $\theta_0$. 
\end{theorem}

\subsection{Behaviour of the Proposed Posterior with Growing Number of Replications}\label{sec:m}

We now consider how the proposed abcEl posterior behaves when the sample size $n$ is kept fixed and the number of replications obtained from the data generating process i.e. $m$ is allowed to grow.  

First of all, from Figure \ref{fig:post} it is evident that the shape of the proposed approximate posterior depends on the choice of $m$.  Large values of $m$ produce flatter but more pointwise concentrated (smaller variance) estimates for each value of the parameter.  
Such behaviour of the proposed estimator of the log-posterior is not unexpected and originates from our construction. 

Recall that the constraints used in the construction of the empirical likelihood are based on the identity in \eqref{eq:ex}, which can only be satisfied when $\theta=\theta_o$.  The properties of empirical likelihood under mis-specified but feasible constraint has been studied by \citet{ghosh2019empirical}.
For fixed $n$, since $g(X_o(\theta_o))$ remains fixed throughout, it is only meaningful to consider expectation of $h^{(n)}_i(\theta)$ conditional on $(\theta,g(X_o(\theta_o)))$ .  Since each $X_i(\theta)$ is conditionally independent of $X_o(\theta_o)$ given $\theta$, for each $i=1$, $2$, $\ldots$, $m$, and $\theta\in\Theta$ we get:
\[
\et_{\gt\mid(\theta,g(X_o(\theta_o)))}\left[h^{(n)}_i(\theta)\right]=\et_{\gt\mid\theta}\left[g(X_i(\theta))\right]-g(X_o(\theta_o))\ne 0 \text{ a.e.},
\]
thus allowing us to invoke the ideas of \citet{ghosh2019empirical} for the mis-specified setting. 

Again by construction, for each $\theta\in\Theta$ and $i\ne j$,  $h^{(n)}_i(\theta)$ is conditionally independent of $h^{(n)}_j(\theta)$ given $\theta$.
It also follows that for all $\theta\in\Theta$:        
\begin{align}
~&\lim_{m\rightarrow\infty}\frac{1}{m}\sum^m_{i=1}h^{(n)}_i(\theta)=\lim_{m\rightarrow\infty}\frac{1}{m}\sum^m_{i=1}g(X^{(n)}_i(\theta))-g(X^{(n)}_o(\theta_o))\nonumber\\
=&\et_{\gt\mid\theta}\left[g(X^{(n)}_1(\theta))\right]-g(X^{(n)}_o(\theta_o))=\et_{\gt\mid(\theta,g(X_o(\theta_o)))}\left[h^{(n)}_1(\theta)\right]\ne 0~~a.e.\nonumber
\end{align}

For fixed $n$, after conditioning on $g(X_o(\theta_o))$, the constraints $h^{(n)}_i(\theta)$, $i=1$, $2$, $\ldots$, $m$ satisfy the assumptions of \citet{ghosh2019empirical} for all $\theta\in \Theta$.  In particular, with $g(X_o(\theta_o))$ fixed,  the constraints in the problem \eqref{eq:w2} are mis-specified for all $\theta\in \Theta$ almost everywhere (even when $\theta\ne\theta_o$).  The constrained optimisation problem in \eqref{eq:w2} however could still be feasible and the resulting estimated posterior could be positive. 

%This does not mean however that constrained optimisation problem in \eqref{eq:w2} is infeasible. 

%.  Since $X_o$ is fixed, the conditional expectation of $h_i(\theta)$ given $g(X_o)$ in \eqref{eq:h} is not equal to zero.  

  Using the notations introduced above, when $r=1$, i.e. there is only one constraint present, 
under conditions similar to those described above, it can be shown that, \citep[Theorem $3.4$]{ghosh2019empirical} for any $\theta\in\Theta$:
\begin{align}\label{eq:m}
l_m(\theta)&\coloneqq\frac{1}{m}\sum^m_{i=1}\log(\hat{w}(\theta))=-\frac{1}{\mathcal{M}_m(\theta)}\left|E[g(X^{(n)}_i(\theta))]-g(X^{(n)}_o(\theta_o))]\right|(1+o_p(1)),\nonumber\\
& =-\frac{b_n}{\mathcal{M}_m(\theta)}\left|(\mathfrak{g}(\theta)-\mathfrak{g}(\theta_o)+o(1))-\frac{\xi^{(n)}_o(\theta_o)}{b_n} \right|(1+o_p(1)),  
\end{align}
where $\mathcal{M}_m(\theta)$ is a non-random sequence such that, as $m\rightarrow\infty$, $\mathcal{M}_m \rightarrow\infty$ and both
\begin{align}
(a)~&~\frac{1}{\mathcal{M}_m(\theta)}\max_{1\le i\le m}\left|\xi^{(n)}_i(\theta)\right|1_{\left\{\xi^{(n)}_i(\theta)>0\right\}}=1+o_p(1),\nonumber\\ %as $m\rightarrow\infty$,
(b)~&~\frac{1}{\mathcal{M}_m(\theta)}\max_{1\le i\le m}\left|\xi^{(n)}_i(\theta)\right|1_{\left\{\xi^{(n)}_i(\theta)<0\right\}}=1+o_p(1).\nonumber %as $m\rightarrow\infty$.
\end{align}   
hold.  We further assume that, $\mathcal{M}_m=o(m)$.

The sequence $\mathcal{M}_m(\theta)$ is the rate at which the maximum of the $g(X_i(\theta))$ grows away from its mean.  As for example, when $\xi^{(n)}_o(\theta_0)$ is a $N(0,\sigma^2_0)$ random variable, $\mathcal{M}_m\sim\sigma_0\sqrt{2\log m}$. In the examples used in Figure \ref{fig:post} for both the functions $g_1$ and $g_2$, there are such non-random sequences $\mathcal{M}_m(\theta)$ satisfying these conditions.  

In the rest of this section we assume that $r=1$.  From \eqref{eq:m}, it is clear that the variance of the expected log-likelihood gets reduced as $m$ increases.
On the other hand an increasing $m$ implies that the $l_m(\theta)$ will be flatter in shape.  This is evident from the Figure \ref{fig:post} where the curve joining the means of the proposed estimated log-posterior progressively flattens with the number of replications.  We provide more justifications of this phenomenon below.

%Furthermore, from the same result it follows that:
%\[
%\lim_{m\rightarrow\infty}\hat{\Pi}(\theta\mid g(X_o))=\pi(\theta).
%\]  
%That is for a flat prior (e.g. in Figure \ref{fig:post}), for larger values of $m$, the estimate of the true posterior will be flatter in shape.  This is evident from the figure where curve joining the means of the proposed estimated log-posterior progressively flattens with the number of replications.  

Using the results from \citet{ghosh2019empirical} it is possible to specify bounds on the rate of growth of the number of replicates with the sample size. %We first deduce a lower bound of $m$ in terms of $n$. 
Since the differential entropy plays a relatively minor role in determining the posterior, in what follows we concentrate on $l_m(\theta)$.

% Suppose $\xi^{(n)}_o(\theta_o)$ is a $N\left(0,\sigma^2_o\right)$ random variable.  Consider a test of the null hypothesis $H_0:~\theta=\theta_o$ vs an unrestricted alternative based on $l_m(\theta)$, which rejects the null when the likelihood ratio is smaller than a fixed constant $C_0\in(0,1)$.   

%When $\theta=\theta_o$, the probability of a Type $1$ error in a test of hypothesis, with the null, $H_0:\theta=\theta_o$, and based on $l_m(\theta)$ should rapidly converge to zero with $n$. Suppose the type $1$ error decrease at the rate of  $n^{-\alpha}$ for some $\alpha>0$.  

  \subsubsection{Testing Under Unrestricted Alternative}
  By construction $l_m(\theta)$ is a random function. However, at least heuristically, $l_m(\theta_o)$ should be larger than $l_m(\theta)$ for any $\theta\ne\theta_o$ with a high probability.  More formally, this implies, we should fail to reject the null in the likelihood ratio test for the hypothesis $\theta=\theta_o$ against the unrestricted alternative.

Since $l_m(\theta)$ is different from the traditional empirical likelihood, it's asymptotic and finite sample properties are of interest by themselves.  The likelihood ratio statistic is given by:
\[
LR(\theta_o)=\frac{\exp(l_m(\theta_o))}{\max_{w\in\Delta_{m-1}}\exp(\sum^m_{i=1}\log(w_i)/m)}.
\]
Clearly, the maximum value the denominator attains is, $1/m$. So the log-likelihood ratio $\log LR(\theta_o)$ turns out to be $l_m(\theta_o)+\log m$.

  The test rejects $H_0$ if $\log LR(\theta_o)$ is smaller than $\log C_0$, for some pre-specified $C_0\in(0,1)$.  Ideally, $C_0$ should be a function of $m$.  However, at this point we assume $C_0$ to be fixed.

  Using \eqref{eq:m}, the probability of rejecting the null hypothesis is given by:
\begin{align}
~&Pr[\log m+l_m(\theta_o)\le \log C_0]=Pr[l_m(\theta_o)\le \log C_0-\log m]\nonumber\\
  =&Pr\left[-\frac{1}{\mathcal{M}_m(\theta_o)}\left|\xi^{(n)}_o(\theta_o)+o(1)\right|(1+o(1))\le \log\left(\frac{C_0}{m}\right)\right]\nonumber\\
  =&Pr\left[\left|\xi^{(n)}_o(\theta_o)+o(1)\right|(1+o(1))\ge -\mathcal{M}_m(\theta_o)\log\left(\frac{C_0}{m}\right)\right]\nonumber
\end{align}

Now Suppose that $\xi^{(n)}(\theta)$ is a $N(0,\sigma^2_0)$ random variable.  Using the tail bounds for a normal distribution, we get:

\begin{align}
  Pr&\left[\left|\xi^{(n)}_o(\theta_o)+o(1)\right|(1+o(1))\ge -\mathcal{M}_m(\theta_o)\log\left(\frac{C_0}{m}\right)\right]\nonumber\\
  &\le exp\left(-\frac{1}{2\sigma^2_o}\left\{\mathcal{M}_m(\theta_o)\log\left(\frac{C_0}{m}\right)\right\}^2\right)
\end{align}
By substituting $\mathcal{M}_m(\theta_o)=\sigma_o\sqrt{2\log m}$ in the exponent of the above expression we get:
\begin{align}
&\frac{1}{2\sigma^2_o}\left\{\mathcal{M}_m(\theta_o)\log\left(\frac{C_0}{m}\right)\right\}^2=(\log m)\left(\log C_0-\log m\right)^2\nonumber\\
=&(\log m)^3-2(\log m)^2\log C_0+(\log m)(\log C_0)^2.\nonumber 
\end{align}

  Clearly, the $(\log m)^3$ term dominates and the probability of rejecting the null hypothesis decreases at the rate of $\exp(-(\log m)^3)$.  This is true even if $C_0$ increases to one with increasing $m$ at a suitable rate.  This is the natural scenario, since with increasing number of replications, the rejection criterion should become more and more stringent.

  Finally, in order to describe some relationship between $m$ and $n$, suppose we would like to ensure, that the probability of rejecting the null hypothesis reduces at the rate of $p_n$.
Then it follows that the number of replications required to ensure such a rate is of the order $m=\exp((-\log p_n)^{1/3})$.

%\[
%l_m(\theta_o)=-\frac{1}{\mathcal{M}_m(\theta_o)}\left|\xi^{(n)}_o(\theta_o)+o(1)\right|(1+o(1)),
%\]
%For large values $m$, the probability of a type $1$ error is given by:
%\[
%p_n= Pr\left[\left|\xi^{(n)}_o(\theta_o)\right|>\mathcal{M}_m\right].
%\]
%Now using the fact that $\xi^{(n)}_o(\theta_o)$ normally distributed, $\mathcal{M}_m=\sigma_o\sqrt{2\log(m)}$. Furthermore, by using standard approximation of the tail probabilities for normal random variables we get: 
%\begin{align}
%p_n=&Pr\left[\left|\xi^{(n)}_o(\theta_o)\right|>\mathcal{M}_m\right]\nonumber\\
%=&Pr\left[\left|\xi^{(n)}_o(\theta_o)\right|>\sigma_o\sqrt{2\log(m)}\right]\nonumber\\
%=&Constant*exp(-\log(m))=O(1/m).
%\end{align}
%This yields a lower bound $m=O(p^{-1}_n)$.
%\end{proof}
%\end{theorem} 

  \subsubsection{Bounds on the growth of the number of replications in terms of sample size}\label{sec:bounds}
Other bounds on the growth rate of $m$ in terms of $n$ can be obtained using \eqref{eq:m}.  
Since the posterior is itself a random probability distribution, in order to ensure the posterior consistency, we need to choose $m$ as a function of $n$ in a way that with high probability two things happen:
  first, $\exp(l_m(\theta))$ converges to zero for all $\theta\ne\theta_o$ and second, for $\theta=\theta_o$, $\exp(l_m(\theta_o))$ does not collapse to zero.   

In order to ensure the first condition, suppose $\theta\ne\theta_o$, and as  $m,n\rightarrow\infty$, and in \eqref{eq:m}, $b_n/\mathcal{M}_m(\theta)$ diverges.  Since by assumption (A3), as $m,n\rightarrow\infty$, $\sup_{i\in\{o,1,2,\ldots,m\}}$ $|\xi^{(n)}_o(\theta_o)|/b_n$ $\rightarrow 0$, 
in probability, uniformly over $\theta$, and by assumption (A1), $|\mathfrak{g}(\theta)-\mathfrak{g}(\theta_o)|>0$, for each $\theta=\theta_o$, the R.H.S. of \eqref{eq:m} diverges to $-\infty$.  So $\exp(l_m(\theta))$ converges to zero.  
That is, an upper bound of the rate of growth of $m$ can thus be obtained by inverting the relation $b_n>\mathcal{M}_m(\theta)$.

Depending on the distribution of $\xi^{(n)}_o$, $m$ can be much larger than $n$.
For example, if $\xi^{(n)}_o$ follows a normal distribution with mean zero and variance $\sigma^2_o$, $b_n=\sqrt{n}$ and $\mathcal{M}_m(\theta)$ is of the order $\sigma_o\sqrt{2\log(m)}$, which allows an upper bound of $m$ as large as $\exp(n/(2\sigma^2_o))$.

Similar to the argument for the upper bound, for posterior consistency $l_m(\theta_o)$ cannot diverge to $-\infty$.  There exists a constant $C_1>0$ such that, $l_m(\theta)>-C_1$ with a high probability.

For \eqref{eq:m}, it follows that when $\theta=\theta_o$:
\begin{equation}\label{eq:thetao}
l_m(\theta_o)=-\frac{\mid\xi^{(n)}_o(\theta_o)\mid}{\mathcal{M}_m(\theta_o)}(1+o_p(1)).
\end{equation}
For simplicity of presentation, we also suppose $\xi^{(n)}_o(\theta_o)$ is a $N(0,\sigma_o^2)$ variable.    

For a fixed $C_1>0$, we first compute $Pr[l_m(\theta_o)\le -C_1]$.  Using the tail bound for a $N(0,\sigma^2_o)$ random variables we get,
\begin{align}
~&Pr[l_m(\theta_o)\le -C_1]= Pr\left[-\frac{\left|\xi^{(n)}_o(\theta_o)\right|}{\mathcal{M}_m(\theta_o)}(1+o_p(1))\le -C_1\right]\nonumber\\
=&Pr\left[\left|\xi^{(n)}_o(\theta_o)\right|\ge C_1\frac{\mathcal{M}_m(\theta_o)}{1+o_p(1)}\right]\le \exp\left(-\frac{1}{2}\left(\frac{C_1\mathcal{M}_m(\theta_o)}{\sigma_o}\right)^2\right).\label{eq:lbound}
\end{align}

Since $\xi^{(n)}_o(\theta_o)$ is normally distributed, $\mathcal{M}_m(\theta_o)=\sigma_o\sqrt{2\log m}$, diverges as $m\rightarrow\infty$.  So the R.H.S. of \eqref{eq:lbound} converges to zero.  That is, for any $C_1>0$, $Pr[l_m(\theta_o)\le -C_1]$ converges to zero.  Furthermore, by substituting the expression for $\mathcal{M}_m(\theta_o)$ in \eqref{eq:lbound} we get:
\begin{equation}\label{eq:lbound2}
Pr[l_m(\theta_o)\le -C_1]\le \exp(-C_1^2\log m)=\frac{1}{m^{C^2_1}}.
\end{equation}

Now as before by setting $p_n=m^{-C^2_1}$, we get $m=p_n^{-1/C^2_1}$.  In particular, if $p_n=n^{-\alpha}$, $m=n^{\alpha/C^2_1}$.  

The bounds for $m$ in terms of $n$ described above strikes a balance between the probability of two events, namely, $\exp(l_m(\theta))$ collapses to zero for fixed $\theta\ne\theta_o$, and $\exp(l_m(\theta_o))$ does not collapse to zero.  From our discussion above, 
the number of replications $m$ growing to infinity by itself ensures that the probability of the latter event increases to one. On the other hand, the condition which ensures that the first event occurs with a high probability involves both $m$ and the sample size $n$.
 
\subsubsection{Behaviour of the log-likelihood when $\mathcal{M}_m(\theta)/b_n$ diverges}
Let us fix $\theta\ne\theta_o$ and suppose $\xi^{(n)}_o(\theta_o)$ follows a $N(0,\sigma^2_o)$ distribution.  Then for a fixed $C_2>0$, it can be shown that:

\begin{align}
Pr[l_m(\theta)\le -C_2]\le& Pr\left[\left|\xi^{(n)}_o(\theta_o)\right|\ge\mathcal{M}_m(\theta)\left\{C_2-\frac{b_n}{\mathcal{M}_m(\theta)}\left|\mathfrak{g}(\theta)-\mathfrak{g}(\theta_o)\right|\right\}\right]\nonumber\\
\le&\exp\left[-\frac{(\mathcal{M}_m(\theta))^2}{2\sigma^2_o}\left\{C_2-\frac{b_n}{\mathcal{M}_m(\theta)}\left|\mathfrak{g}(\theta)-\mathfrak{g}(\theta_o)\right|\right\}^2\right]\nonumber\\
\end{align}
Now by substituting $\mathcal{M}_m(\theta))=\sigma_o\sqrt{2\log m}$ we get:
\begin{equation}\label{eq:lbf}
Pr[l_m(\theta)\le -C_2]\le \left(\frac{1}{m}\right)^{\left\{C_2-\frac{b_n}{\sigma_o\sqrt{2\log m}}\left|\mathfrak{g}(\theta)-\mathfrak{g}(\theta_o)\right|\right\}^2}.
\end{equation}
Now, if $\mathcal{M}_m(\theta)/b_n=\sigma_o\sqrt{2\log m}/b_n$ diverges with $m$ and $n$, clearly, for large values of $m$ and $n$, $Pr[l_m(\theta)\le -C_2]\approx m^{-C^2_2}$.  That is, for any fixed $C_2>0$ and $\theta\ne\theta_o$,  $l_m(\theta)\ge -C_2$ with a high probability, 
and $\exp(l_m(\theta))$ does not collapse to zero with a high probability. % Together with the discussion in section \ref{sec:bounds} above, the above result provides alternative conditions for posterior consistency to hold when $r=1$.  However, the discussion in section \ref{sec:postcons} provides a more general result.  

Furthermore, for a fixed $n$, R.H.S. of \eqref{eq:lbf} is a decreasing function in $m$.  That is if the sample size is kept fixed, increasing the number of replications will increase the probability of $l_m(\theta)\ge -C_2$.  As a result, the log-likelihood will be flatter in shape. This complies with our observations in Figure \ref{fig:post}, and formally explains it.

\section{Illustrative Examples and Applications}

In this section we consider five illustrative examples.  First, however, we comment on computational issues arising in their implementation.
The estimated weights in \eqref{eq:w2}, which define the empirical likelihood, can only be computed numerically in almost all cases. This makes it necessary to use methods such as MCMC to sample from the posterior.  
The support of the posterior may be non-convex \citep{chaudhuri+my17}.  
In the examples below, we use Metropolis-Hastings random walk methods with normal proposal 
for the MCMC sampling, but more sophisticated methods could also be used in the case of a high-dimensional parameter.   

The MCMC sampling procedure from a posterior distribution derived from a likelihood in effect samples from a likelihood estimated using Monte Carlo methods.
Similar to the Bayesian synthetic likelihood \citep{price+dln16}, this approach is related to pseudo-marginal Metropolis-Hastings methods \citep{beaumont03,andrieu+r09,doucet+pdk15} in the sense that the use of a noisy estimate of a likelihood or pseudo-likelihood is involved.
    In pseudo-marginal Metropolis-Hastings algorithms, it is observed that when the variance of the likelihood estimate is large, the MCMC chain mixes poorly. We observe a similar phenomenon with empirical likelihood as well. 
  Hence the number of replicates generated, i.e. $m$ should be chosen judiciously.  This is also true for the synthetic likelihood approach \citep{price+dln16}.  
The choices for $m$ used in the examples below are sufficient to ensure adequate mixing, but they
depend on the dimensionality and distributional properties of the summary statistics, and need to be considered on a case by case basis.

%one can think of the use of
%a noisy likelihood estimate within the MCMC sampling as implementing a pseudo-marginal Metropolis-Hastings method \citep{}.  

%In the case of synthetic likelihood, 
%the results are not sensitive statistically to the number of samples $m$ used for likelihood approximations, and this seems to be true for the empirical likelihood also.  However, the choice of $m$ has computational as well as statistical implications.  As noted in \citet{price+dln16}, and also in the pseudo-marginal Metropolis-Hastings literature \citep{doucet+pdk15}, if the variance of the noisy likelihood estimate is too large, the resulting
%MCMC chain will mix poorly.  The choices for $m$ used in the examples below are sufficient to ensure adequate mixing, but this
%depends on the dimensionality and distributional properties of the summary statistics, and needs to be considered on a case by case basis.

Computation of the empirical likelihood is generally very fast.  Several efficient optimisation methods are available.  We have used the R package {\tt emplik} \citep{zhou+y16} 
in the experiments below.  The computational
effort involved in implementing the proposed approach is similar to the synthetic likelihood in our examples.

Five examples are considered.  The first is a simple normal location example, and we use this to illustrate the effects of different summary statistic choices in the method.
In the second example the proposed method is employed to estimate the underlying edge probability of an Erd\"{o}s-Renyi random graph. 
The third example concerns a $g$-and-$k$ model, which is a standard
benchmark model for ABC inference algorithms.  The fourth one involves dependent data simulated from an ARCH(1) model (also considered in \cite{mengersen+pr13}).  The summary statistics used in this example are non-Gaussian, and
we show that compared to the synthetic likelihood, empirical likelihood is more robust to this non-normality.  
The fifth example is a real example for stereological extremes.  %We use this example for two purposes. First of all,
For this example, we first find summaries for which the proposed method performs comparably to the synthetic likelihood and rejection ABC methods.  Furthermore, in order to illustrate
the importance of the choice of the summary statistics, we consider a set of hard to match summaries, which fit poorly to the assumed model. It is seen that the proposed empirical likelihood does not work well in this situation. However, it is no worse than the synthetic likelihood if implemented with the same summaries.
%We present shortened descriptions below.  More details can be found in the supplement, where an additional example involving a simple normal location family is also presented.    

%  a situation where the empirical likelihood method fails, we choose hard to match summary statistics, which fit poorly to the assumed model.  However, we observe that, in this example, it is difficult to implement the synthetic likelihood with the same
%  summary statistics, as well.

\begin{figure}[t]
  \begin{center}
\begin{subfigure}{.45\columnwidth}
\resizebox{2.75in}{2.75in}{\includegraphics{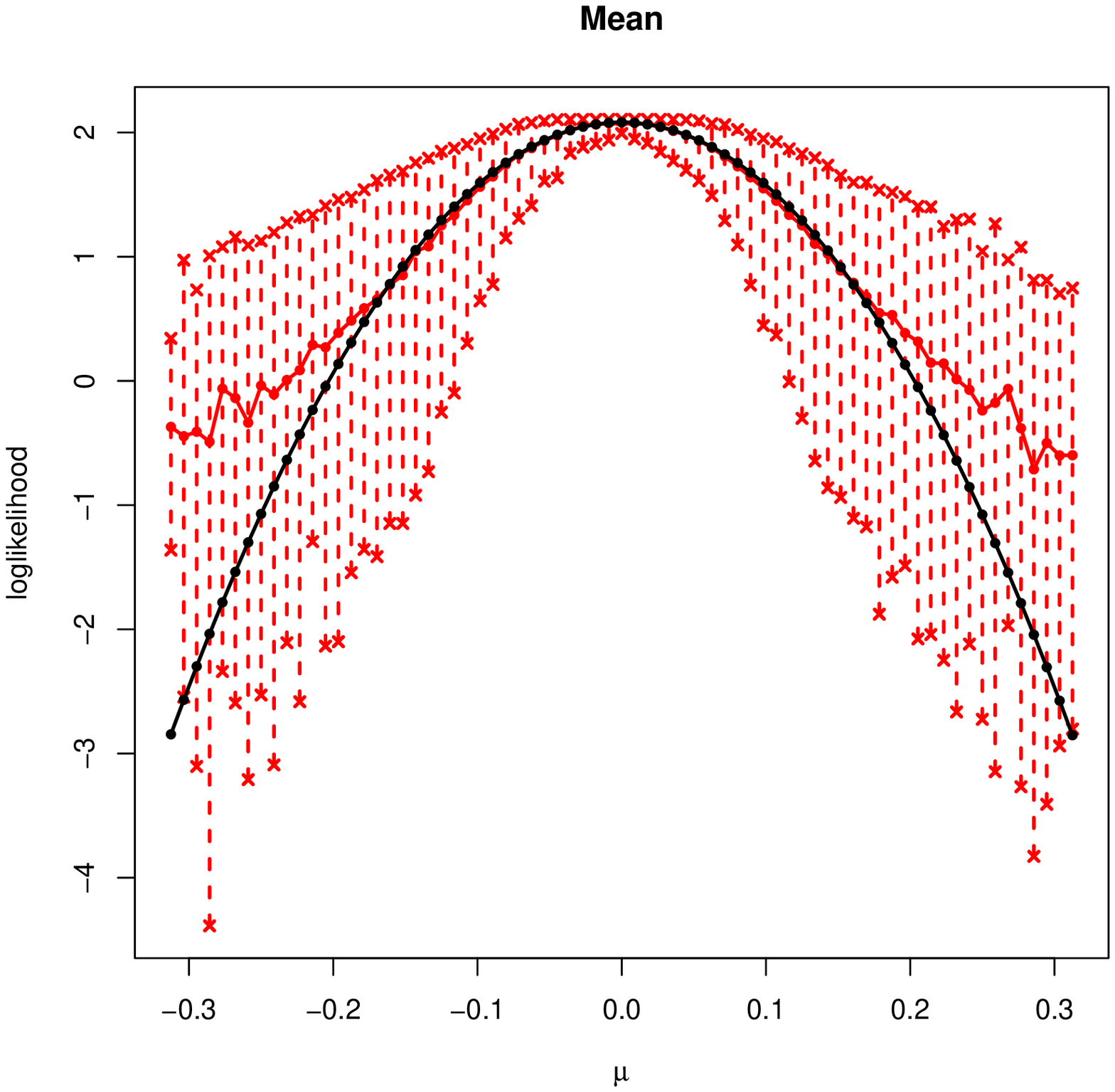}}
\caption{\label{fig:F1a}}
\end{subfigure}
\begin{subfigure}{.45\columnwidth}
\resizebox{2.75in}{2.75in}{\includegraphics{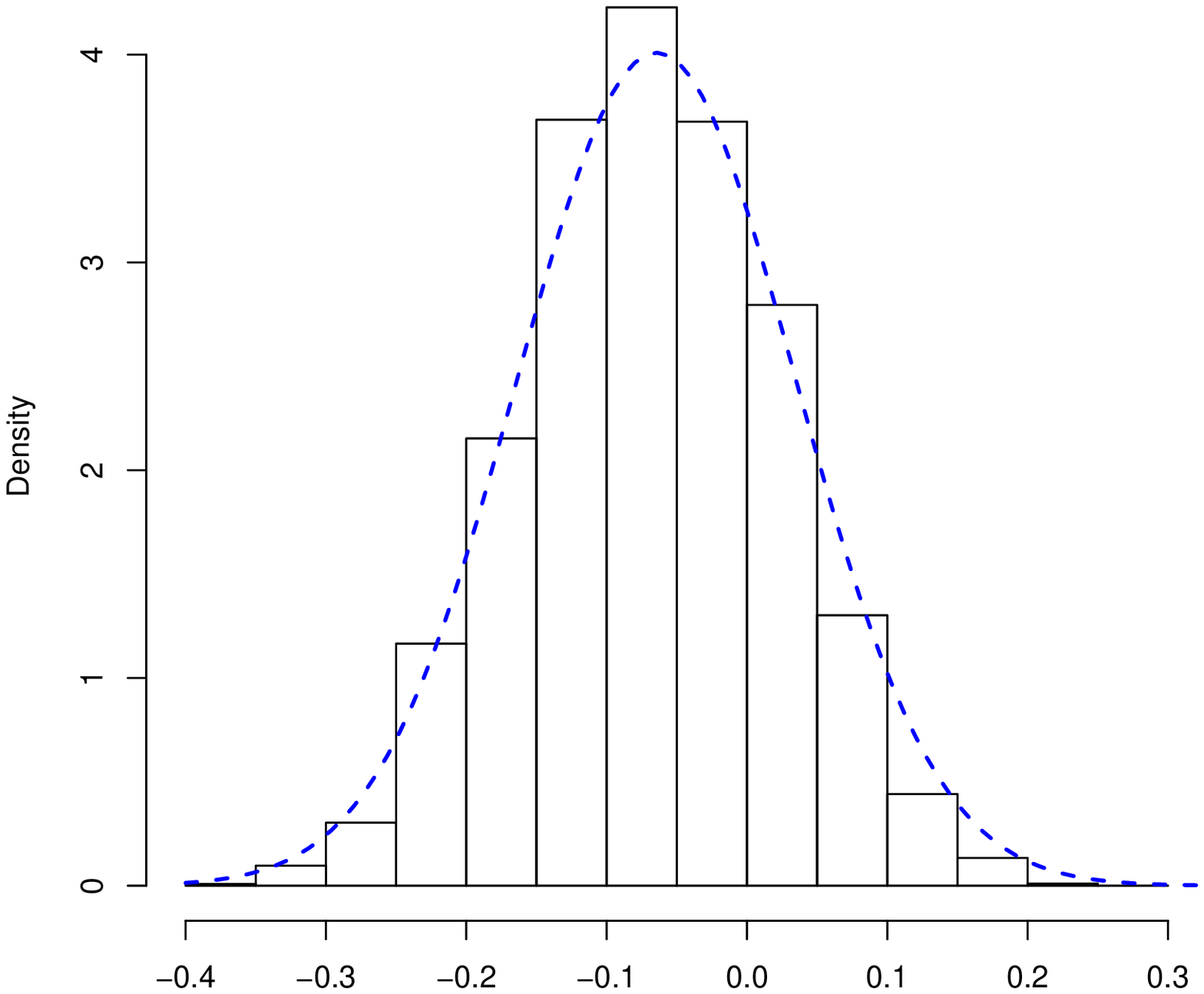}}
\caption{\label{fig:F1b}}
\end{subfigure}
\caption{Comparison of the true posterior of the mean of a Normal distribution with unit variance conditional on the sample mean with our proposed empirical likelihood based ABC posterior. The true posterior is based on samples of size $n=100$.  The proposed posterior is computed with sample mean as the summary and $m=25$.  
Figure \ref{fig:F1a} directly compares the true log-posterior (black curve) with the means and 95\% credible intervals of the proposed approximate posterior based on $1000$ replications for each parameter value (in red).  
Figure \ref{fig:F1b} compares the true posterior (dashed line) with the histogram of the samples drawn from the proposed empirical likelihood based ABC posterior (underlying histogram). 
% for inference about the normal mean based on a sample of size $n=100$ with a known variance $\sigma^2=1$.  Sample mean was used as summary for the latter.
}
\label{F1}
\end{center}
\end{figure}

\subsection{Normal distribution}

Our first example considers inference about a mean $\mu$ for a random sample of size $n=100$ from a normal density, $N(\mu,1)$.  
The prior for $\mu$ is $N(0,1)$.  The observed data $X_o$ is generated with $\mu=0$. The exact posterior for $\mu$ is  
normal, $N(\sum^n_{j=1}X_{oj}/(n+1),(n+1)^{-1})$.  The proposed empirical likelihood based method was implemented with $m=25$.  We considered several choices of constraint functions  $g_1$, $\ldots$, $g_r$.  
Specifically, for $i=o,1,\ldots,m$, we take (a) $g_1(X_i)=\sum^n_{j=1}X_{ij}/n$,  (b) $g_2(X_i)=\sum^n_{j=1}X^2_{ij}/n$, (c) $g_3(X_i)=\sum^n_{j=1}X^3_{ij}/n$, (d) $g_4(X_i)=\sum^n_{j=1}X^4_{ij}/n$, (e) $g_5(X_i)=\mbox{median of }X_i$, (f) $g_6(X_i)=\mbox{first quartile of }X_i$, (g) $g_7(X_i)=\mbox{third quartile of }X_i$.  Here the constrains considered use the
first four raw moments ((a)-(d)) and the three quartiles ((e)-(g)).
Combinations of these constraints are considered within the empirical likelihood procedure.  

The posteriors obtained from our proposed empirical likelihood based ABC method with the above summaries are close to the true posterior. An illustrative example, with sample mean as summary, is presented in Figure \ref{F1}.
Here, the true posterior density, i.e. the dashed line, is quite close to the histogram of the samples drawn from the posterior obtained from the proposed method.

Different constraints are compared based on the coverage and the average length of the $95\%$ credible intervals for $\mu$ obtained from $100$ replicates.  These values give some indication of frequentist coverage of the credible
intervals when $\mu=0$, but the results can also be used to compare with corresponding quantities for the true posterior as one way of checking if the empirical likelihood approach approximates the true posterior well in relevant ways for inference.
For each replicate, MCMC approximations to the posterior are
based on $50,000$ sampling iterations with $50,000$ iterations burn in.  
The results are presented in Table~\ref{Tab2}.

%The coverage and average length of 95\% credible intervals for the true posterior are $0.95$ and $0.39$ (2 d.p.). 
From Table \ref{Tab2}, we see that the proposed method performs quite well when either the mean or median is used as constraint function.  Note that the sample mean is minimal sufficient for $\mu$, and would be an ideal choice of summary statistic
in conventional likelihood-free procedures such as ABC.  Table \ref{Tab2} also shows that when 
many summary statistics are used, the performance of empirical likelihood ABC deteriorates.    
Inclusion of raw moments of higher orders and more quantiles makes both frequentist performance (in terms of coverage) and any correspondence with the true posterior worse.  Simultaneous constraints with the mean and median gives a coverage and average credible interval length quite different to those for the true posterior.  This is consistent with the experiences of \citet{mengersen+pr13}, who implement a Bayesian empirical likelihood based on parametric constraints.  

  Unlike the synthetic likelihood, which can automatically down-weight relatively uninformative summaries through the estimation of their means and covariances, the empirical likelihood based method, as proposed, cannot choose constraints and therefore is more vulnerable to uninformative components.
On the other hand, the empirical likelihood does not assume normality for summary statistics, and performs better in models where normality should not be assumed, (see example in Section \ref{sec:arch} below).
For the proposed empirical
likelihood method, similar to conventional ABC methods, we recommend to use summary statistics that are informative and
of minimal dimension.
Finally, we note that increasing the value of $m$ beyond $25$ seemed to cause no appreciable difference in the results.

\begin{table}[t]
 \caption{\label{Tab2} The coverage and the average length of $95\%$ credible intervals for $\mu$ for various choices of constraint functions when $\mu=0$ and $n=100$.  The coverage for the true posterior is $0.95$ and average length is $0.39$ (2 d.p.).}
  \begin{center}
  %\centering
%\fbox%
\begin{tabular}{lcc}
Constraint Functions & Coverage & Average Length\\
%\\[5pt]
Mean, (a).&$0.93$&$0.34$\\
Median, (e).&$0.93$&$0.43$\\
%$3$ quartiles&$0.76$&$0.2807$\\
%$1st$ moment (mean) &$0.93$&$0.3406$\\
%Mean and Median&$0.76$&$0.2392$\\
First two raw moments, (a), (b).&$0.88$&$0.30$\\
First three raw moments, (a), (b), (c).&$0.85$&$0.27$\\
%Median&$0.93$&$0.4259$\\
Three quartiles, (e), (f), (g).&$0.76$&$0.28$\\
Mean and Median, (a), (e).&$0.76$&$0.24$\\
First four raw moments, (a), (b), (c), (d).&$0.72$&$0.22$\\
  \end{tabular}
  %}
  \end{center}
%\label{Tab2}
\end{table}

%%%%Figure For Example on Graphs%%%%%% 

\subsection{Estimation of Edge Probability of an Erd\"{o}s-Renyi Random Graph}\label{sec:graph}

\begin{figure}[b]
\begin{center}
\begin{subfigure}{.45\columnwidth}
\resizebox{2.75in}{2.75in}{\includegraphics{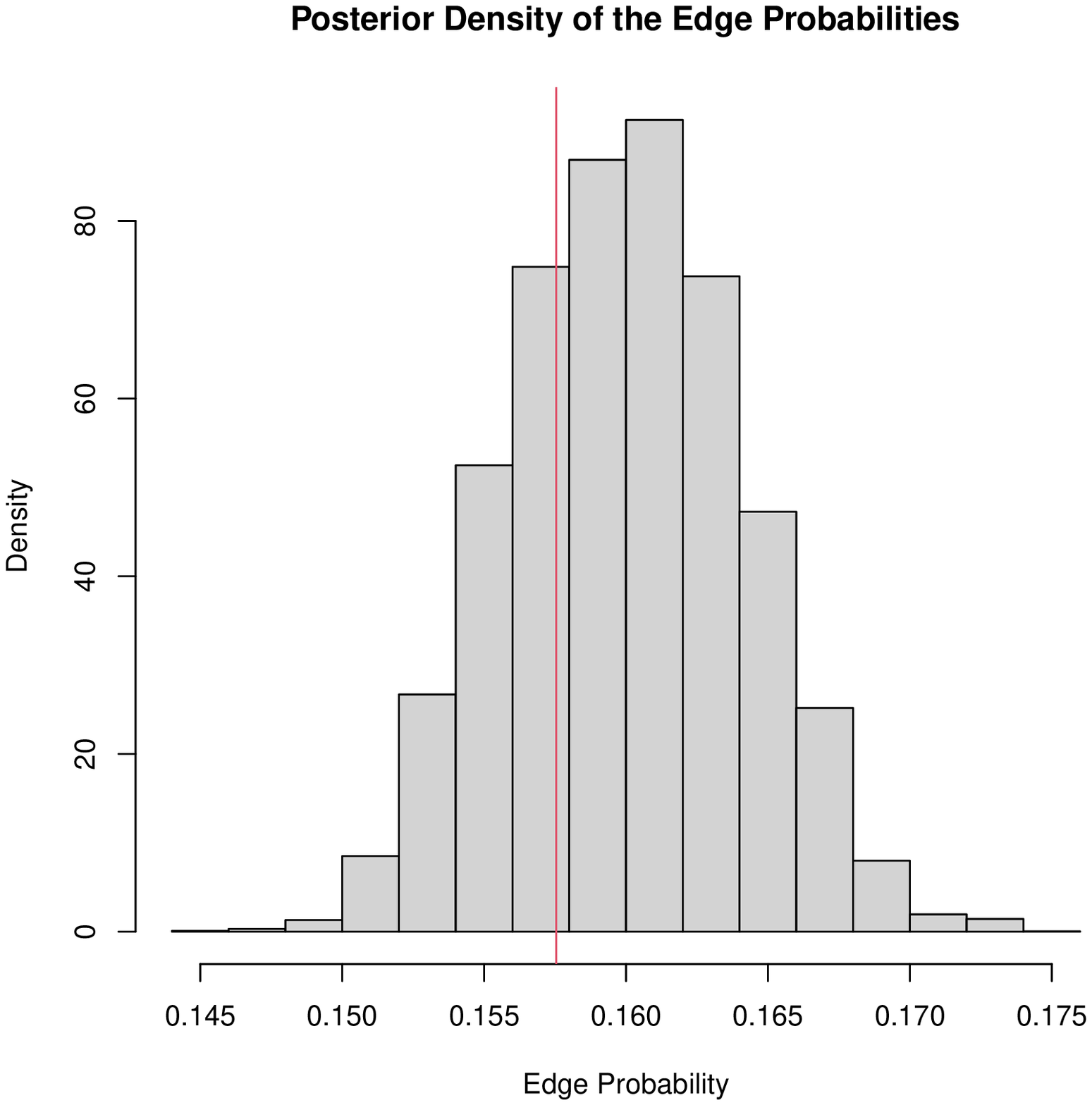}}
\caption{\label{fig:graphA}}
\end{subfigure}
\begin{subfigure}{.45\columnwidth}
\quad\resizebox{2.75in}{2.75in}{\includegraphics{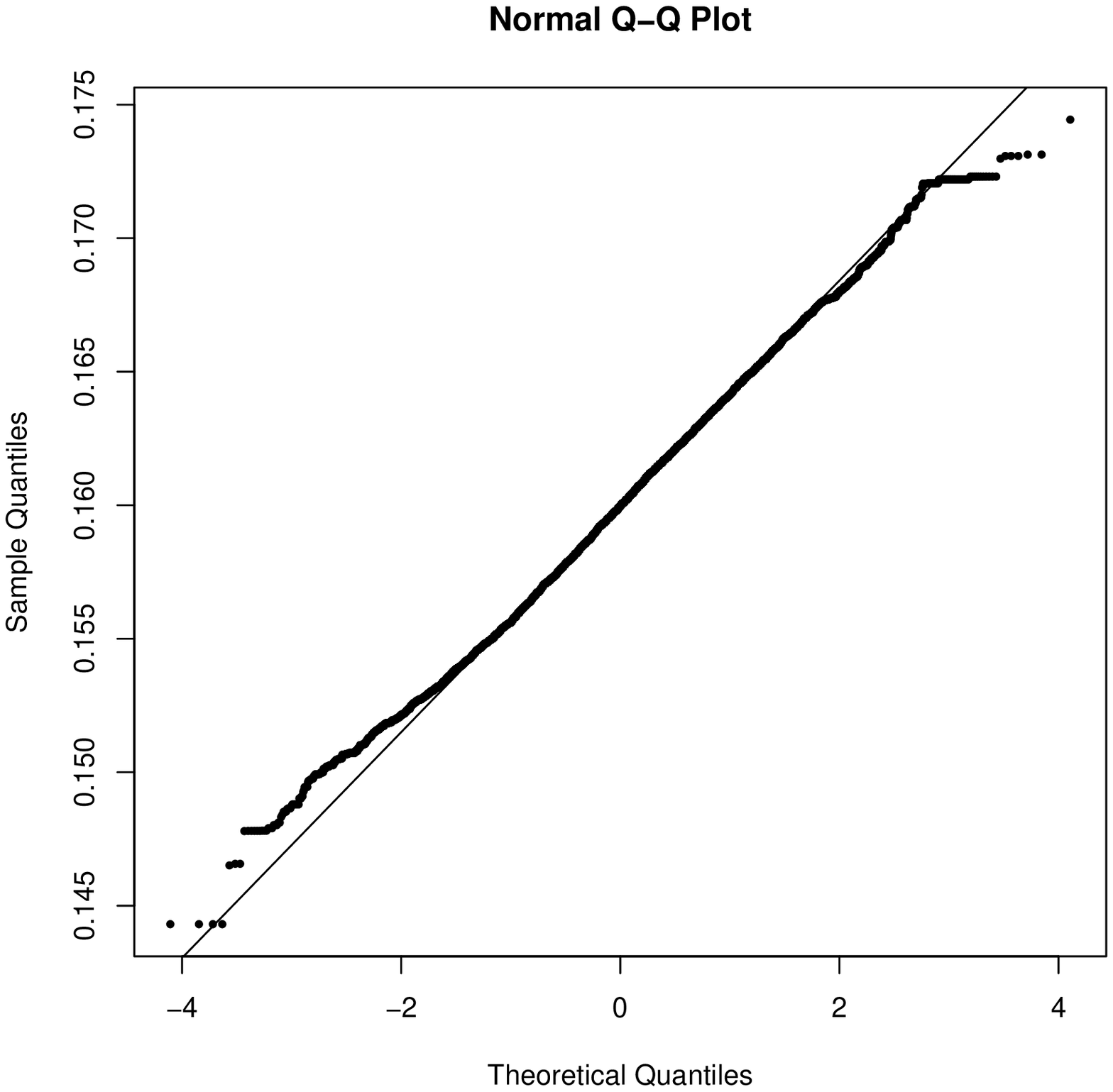}}
\caption{\label{fig:graphB}}
\end{subfigure}
\caption{The histogram (\ref{fig:graphA}) of the samples drawn from the proposed posterior of the edge probability of an Erd\"{o}s-Renyi graph. The graph had $n=100$ nodes.  The abcEl posterior was computed with the number of edges and the number of triangles as estimating equations, and with $m=25$.  The Q-Q plot of the sample against normal distribution is presented in Figure \ref{fig:graphB}.}
\label{fig:graph}
\end{center}
\end{figure}

In our second illustration we estimate the edge probability of an observed Erd\"{o}s-Renyi random graph with $n$ vertices.  Suppose $p$ is the probability of an edge between any two vertices.  We assume that $p$ has a $\text{Beta}(1.5,1.5)$  distribution.
The observed graph had $n=100$ nodes, and the number of edges and the number of triangles were used as two estimating equations.  The posterior was computed using $m=25$ replications.  
Samples from the proposed abcEl posterior were drawn using random walk Markov Chain Monte Carlo with log-odds of the edge probabilities proposed from a normal distribution. 
%\textcolor{blue}{I have kept the description a bit short avoiding the details of ERGM model and exponential tilting based methods. It can be added if necessary.}

The above experiment was repeated $100$ times and the observed coverage of the $95$\% confidence intervals was about $89$\%.  
A typical example of the sampled posterior distribution is presented in Figure \ref{fig:graph}.
In Figure \ref{fig:graph} the histogram of the observation sampled from the posterior is presented. 
The true value of the edge probability ie. $p_o$ is presented by the vertical red straight line.  In Figure \ref{fig:graphB} a Q-Q plot of the sample with normal distribution is presented.  The posterior seems to be slightly lighter-tailed than a normal distribution.

The proposed methodology described here easily generalises to more general exponential random graph models (ERGM) \citep{snijders_pattison_robins_handcock_2006,robins_pattison_kalish_lusher_2007}. For instance, node specific edge probabilities, which depend on covariates can be easily accommodated.  
It allows an alternative way to estimate the model parameters in an ERGM model by avoiding pitfalls of model degeneracies (see. e.g. \citet{fellows_handcock_2017}).

%%%%%%%%%%%%FIGURE FOR g&k MODEL \label{F2}

\begin{figure}[t]
  \begin{center}
    \begin{subfigure}{.45\columnwidth}
      \resizebox{2.5in}{2.5in}{\includegraphics{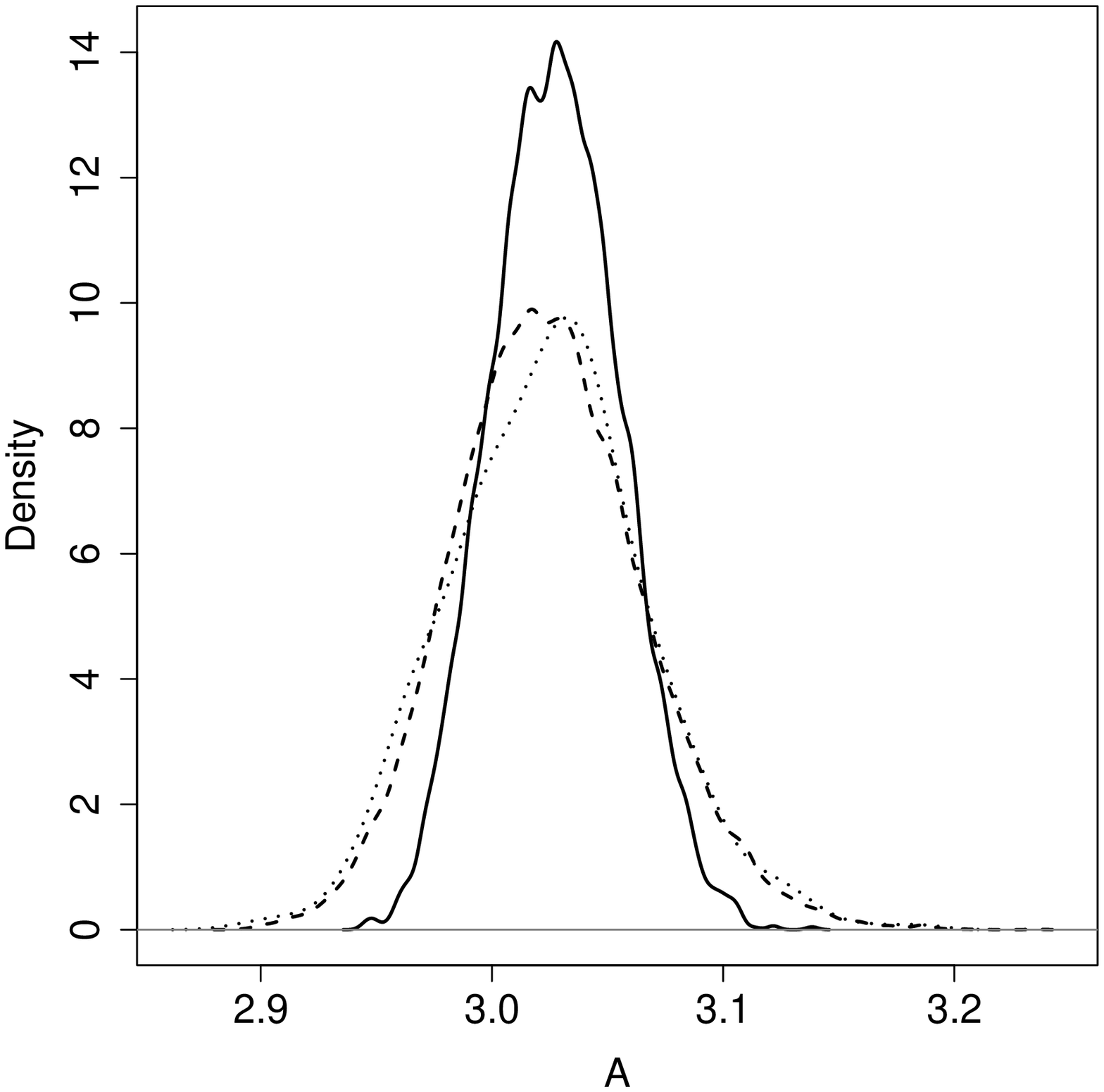}}
      \end{subfigure}\hfill\begin{subfigure}{.45\columnwidth}
    \resizebox{2.5in}{2.5in}{\includegraphics{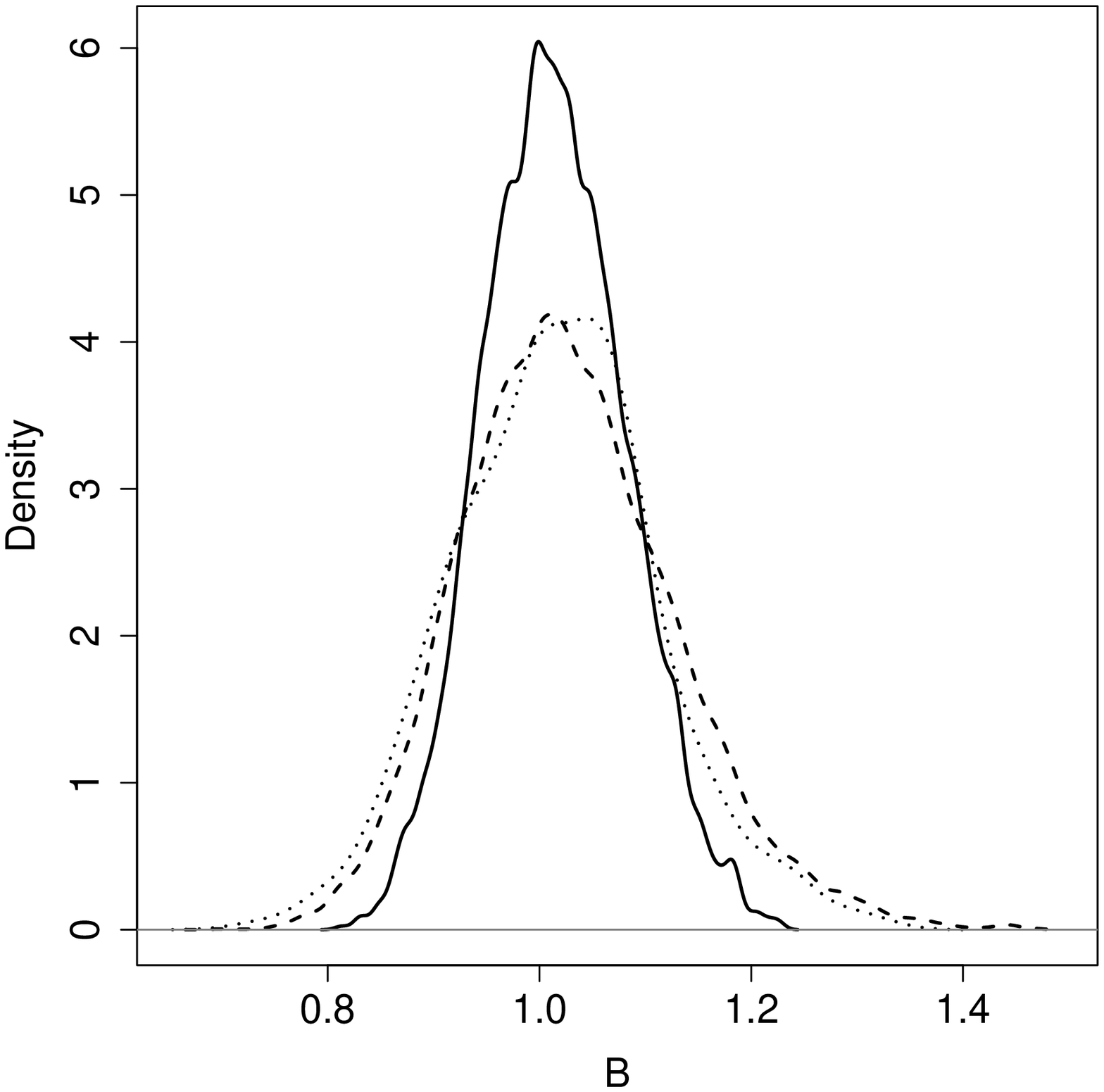}}
    \end{subfigure}\\
    \begin{subfigure}{.45\columnwidth}
      \resizebox{2.5in}{2.5in}{\includegraphics{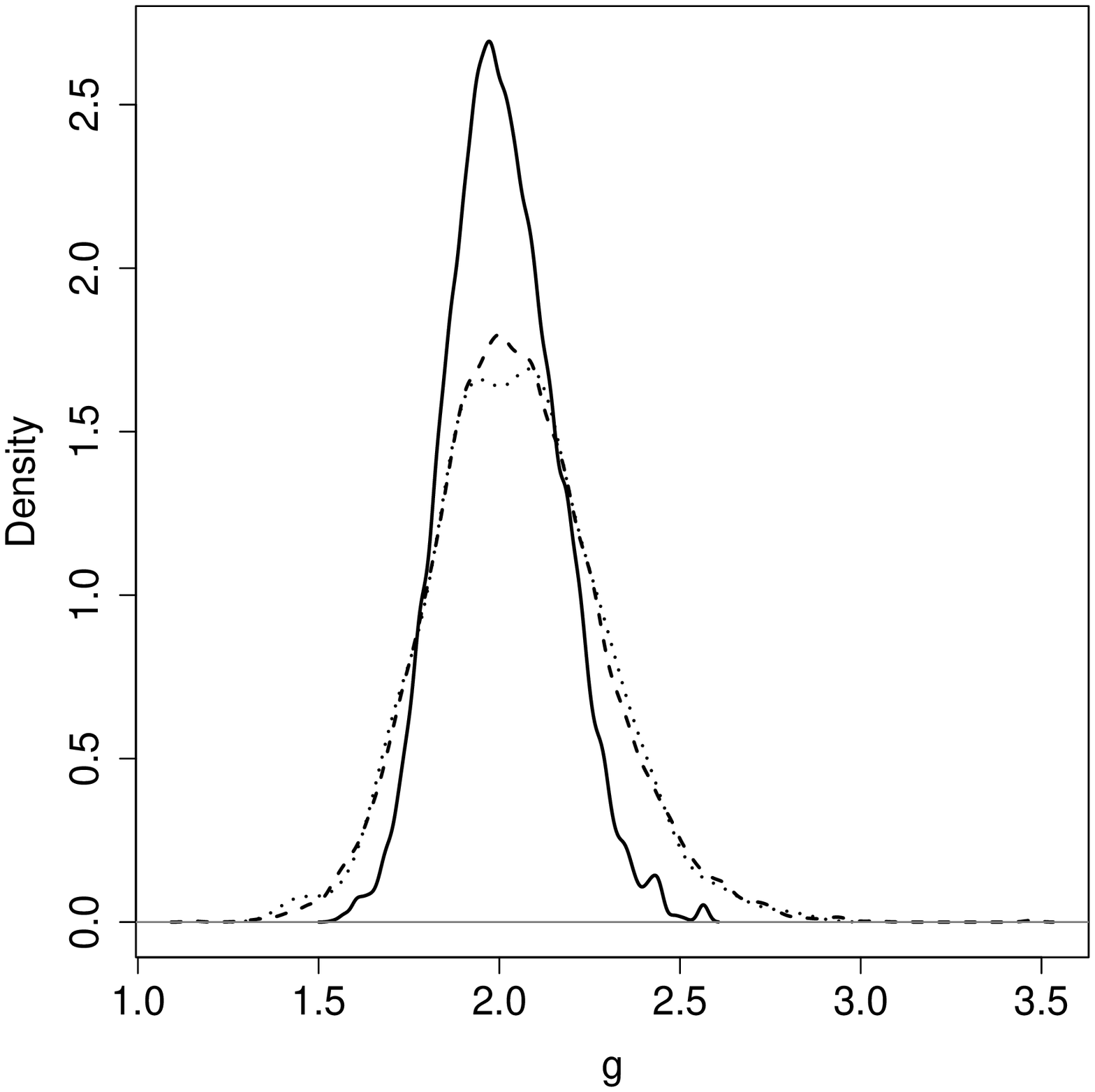}}
      \end{subfigure}\hfill\begin{subfigure}{.45\columnwidth}
    \resizebox{2.5in}{2.5in}{\includegraphics{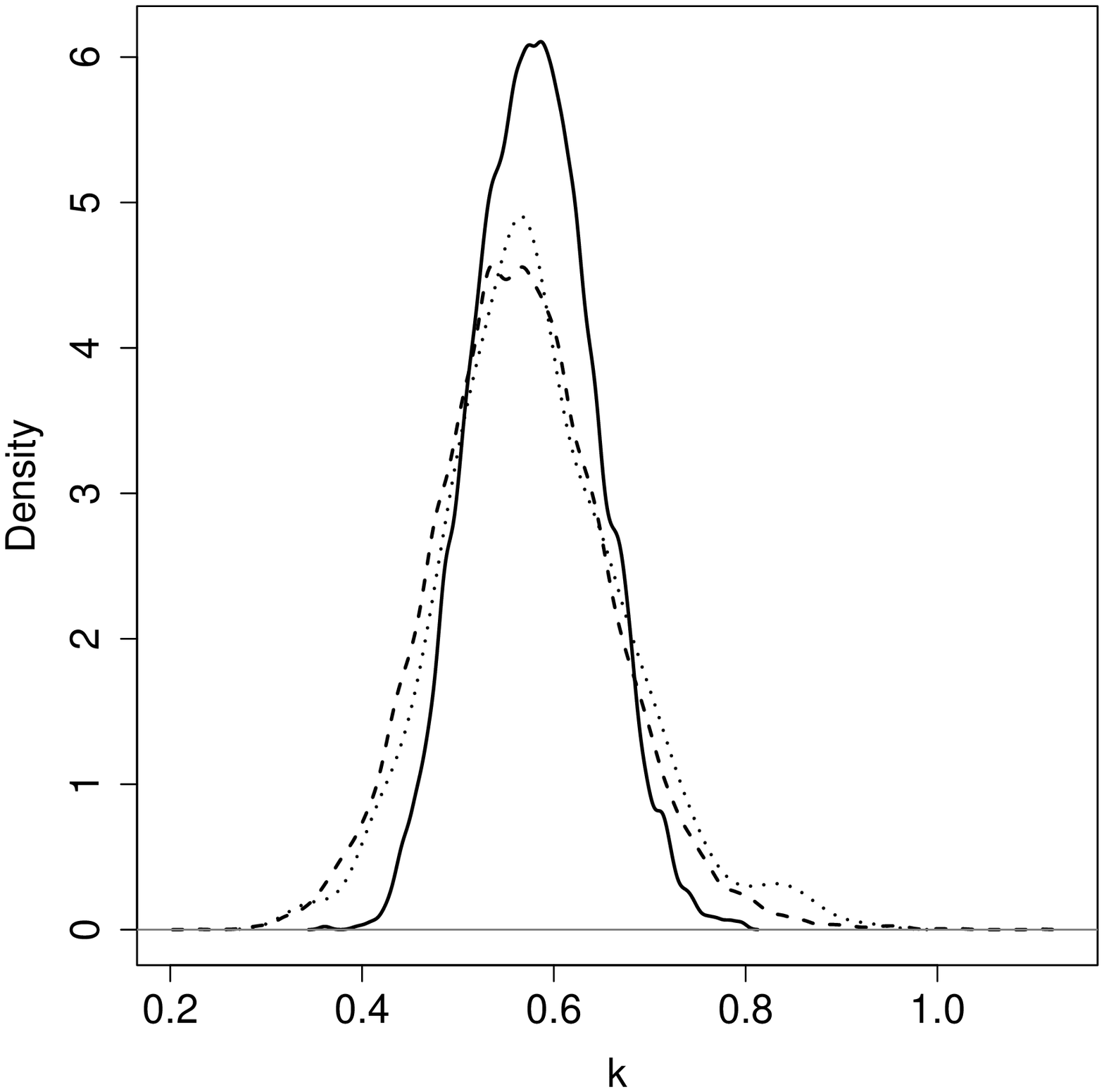}}
    \end{subfigure}
%\resizebox{4.25in}{4.25in}{\includegraphics{gk_comparison_david.eps}}
\caption{Estimated marginal posterior densities by proposed method (solid), synthetic likelihood (dashed) and regression ABC (dotted) for parameters of the $g$-and-$k$ model.}
\label{F2}
\end{center}
\end{figure}

\subsection{$g$-and-$k$ distribution}
Our third example concerns inference for the $g$-and-$k$ distribution \citep{haynes1997robustness}.  There is no closed form expression for the density function and the distribution is defined through its quantile function parametrised by four parameters $\theta=(A,B,g,k)$, \citet{allingham2009bayesian} and \citet{peters+s06}). 

\begin{align}
~&Q(p;A,B,g,k) =A+B \left[ 1+c\times\frac{1-\exp\left\{-gz(p)\right\}}{1+\exp\left\{-gz(p)\right\}} \right] \left\{ 1+{z(p)}^2 \right\} ^kz(p),\nonumber
\end{align} 
where $z(p)$ is the $p$th standard normal quantile and conventionally $c$ is fixed at $0.8$, which results in the constraint $k>-0.5$.
Simulation from this model can be performed by 
transforming uniform random variables on $[0,1]$ by the quantile function.  This feature, and the fact that 
there is no closed form expression for the density function, make likelihood-free inference methods attractive.
Components of the parameter vector $\theta=(A,B,g,k)$ are respectively related to location, scale, skewness and kurtosis of the distribution. In the  ABC context, this distribution was first considered in \citet{allingham2009bayesian}, 
with an analysis of the related $g$-and-$h$ distribution given earlier in \citet{peters+s06}.

A data set of size $n=1000$ was simulated from the distribution with $(A,B,g,k)=(3,1,2,0.5)$.  A uniform prior $U(0,10)^4$ for $\theta$ was assumed.  We approximate the proposed empirical likelihood and the synthetic
likelihood using $m=40$ data sets each of length $n$ for each value of $\theta$.  The mean and the three quartiles were used as summary statistics.  
Compared to the octile based summaries used in \citet{drovandi2011likelihood}, these summaries lead to a slightly better estimate for the parameter $k$.
%Some summary statistics used in \citet{drovandi2011likelihood} based on octiles were also considered, but 
%resulted in slightly inferior performance for estimation of the kurtosis parameter $k$.  
Posterior samples were drawn using a random walk Metropolis algorithm with normal proposal and diagonal
proposal covariance matrix, with the variances chosen based on a pilot run.
Posterior summaries are based on $100,000$ sampling
iterations after $100,000$ iterations burn in.  
 
The results are presented in Figure~\ref{F2}.  Estimated marginal posterior densities obtained from the synthetic likelihood and proposed empirical likelihood are shown as dashed and solid lines respectively. Also shown is 
a ``gold standard''  answer based on rejection ABC with a small tolerance and linear
regression adjustment \citep{beaumont+zb02}.  For the ABC approach, to improve computational efficiency, we restricted the prior for $\theta$ from $U(0,10)^4$ to $U(2,4)\times U(0,2)\times U(0,4)\times U(0,1)$.
This restricted prior is broad enough to contain the support of the posterior based on the original prior.  The ABC estimated marginal posterior densities (dotted) shown in Figure~\ref{F2} were based on $5,000,000$ samples, choosing the tolerance so that $2000$ samples are kept.
The summary statistics used here are asymptotically normal and $n$ is large, so the synthetic likelihood is expected to work well in this example, which it does. Our proposed method gives comparable results to synthetic likelihood and
the ``gold standard'' ABC analysis, although there does seem to be some slight underestimation of posterior uncertainty in the empirical likelihood method. %similar to the normal location example.

%%%% TABLE OF THE COVERAGE OF GK:

%\begin{table}
%\def~{\hphantom{0}}
%\tbl{The coverage in percentage of $95\%$ credible intervals for $A, B, g, k$.}{%
%\begin{tabular}{lcccc}
%\hline
%{} & A & B & g & k \\
%Empirical likelihood & $92.3$ & $90.1$ & $84.6 $& $ 89$ \\
%Synthetic likelihood  & $100$ & $98.9$ & $100 $& $ 100$ \\
%\hline
%\end{tabular}
%}
%\label{Tabgk1}
%\end{table}

\begin{figure}[t]
  \begin{center}
    \begin{subfigure}{.45\columnwidth}
      \resizebox{2.6in}{2.6in}{\includegraphics{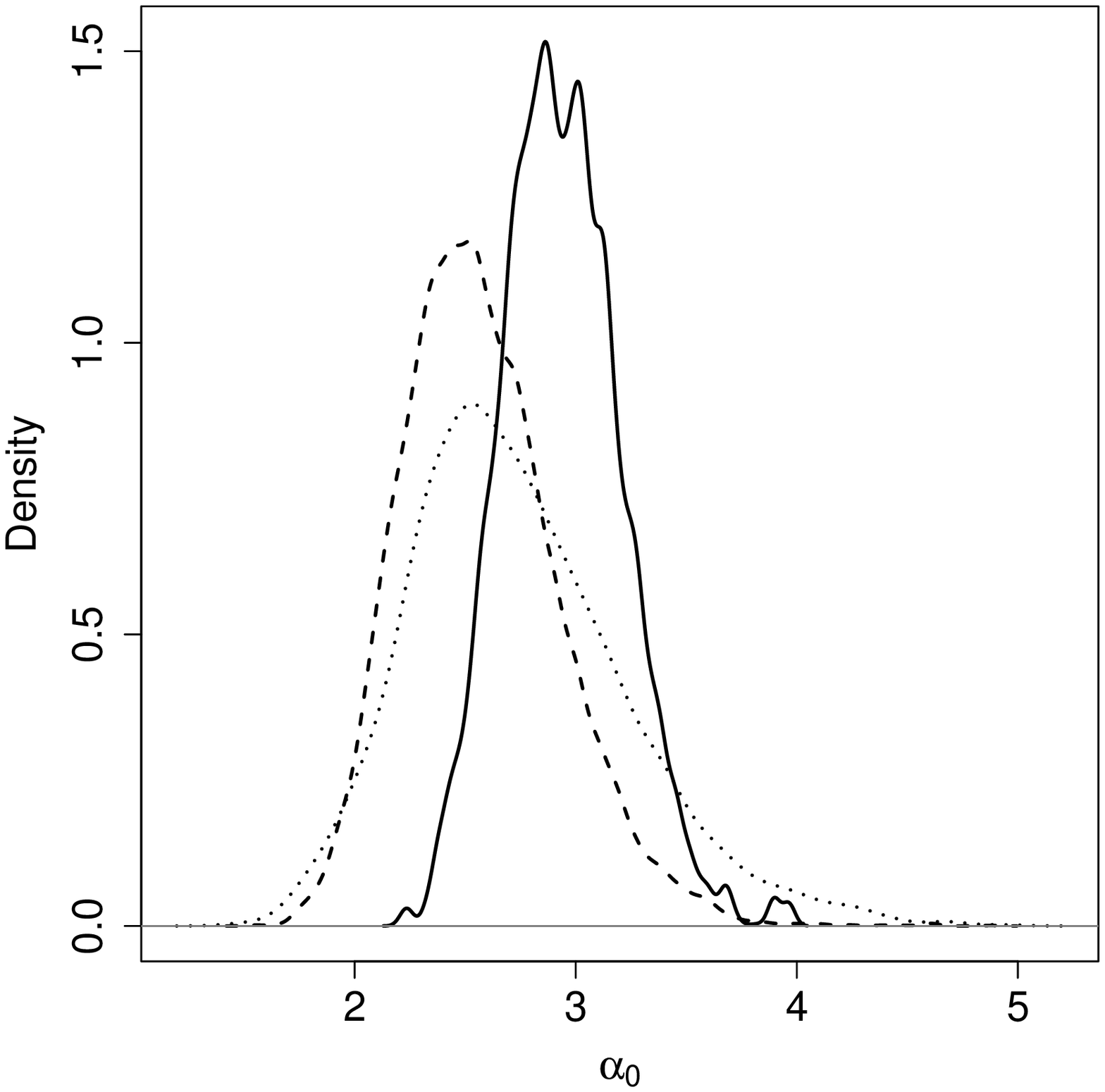}}
      \end{subfigure}\hfill\begin{subfigure}{.45\columnwidth}
    \resizebox{2.6in}{2.6in}{\includegraphics{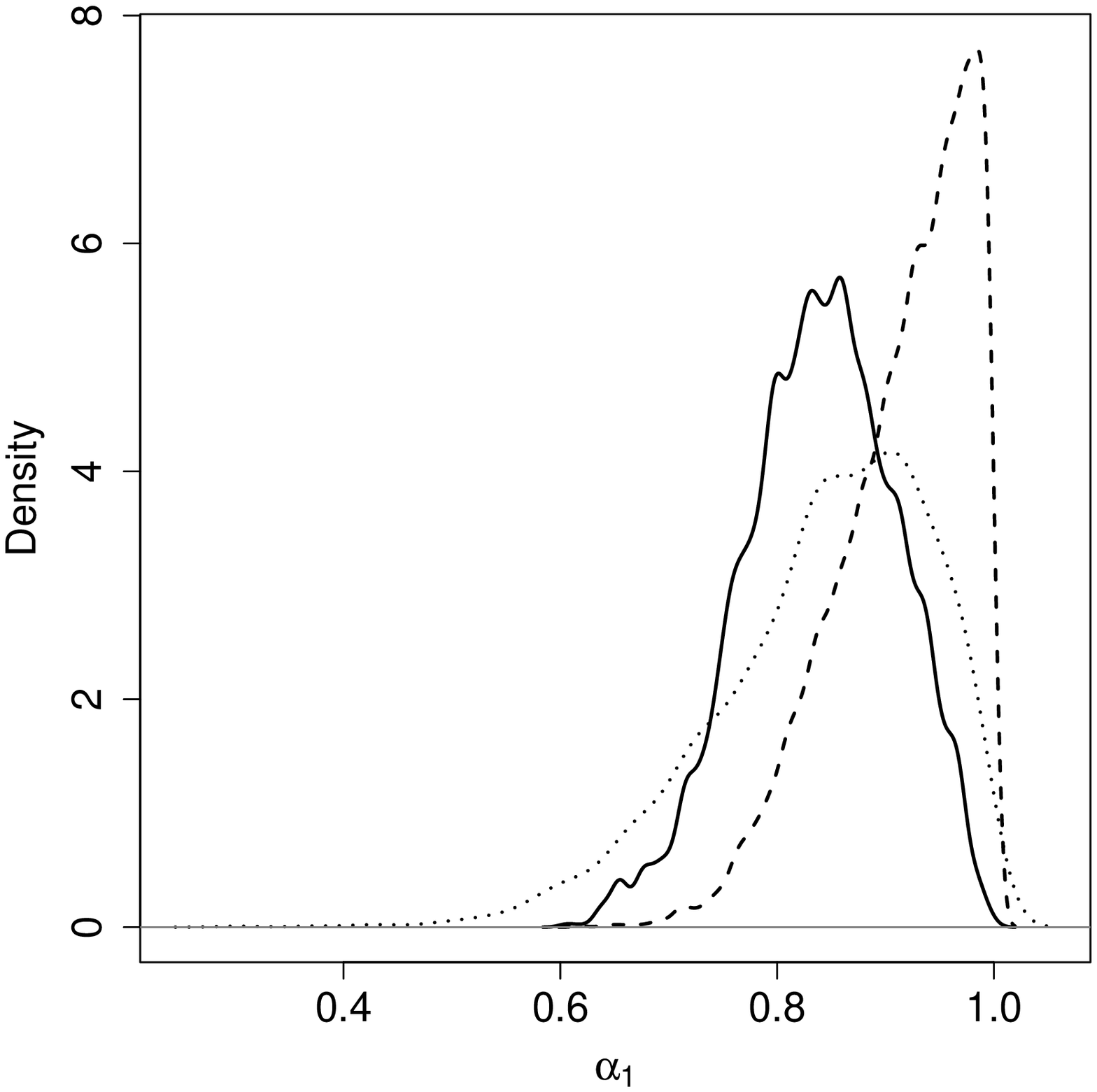}}
    \end{subfigure}\\
    \begin{subfigure}{.45\columnwidth}
      \resizebox{2.6in}{2.6in}{\includegraphics{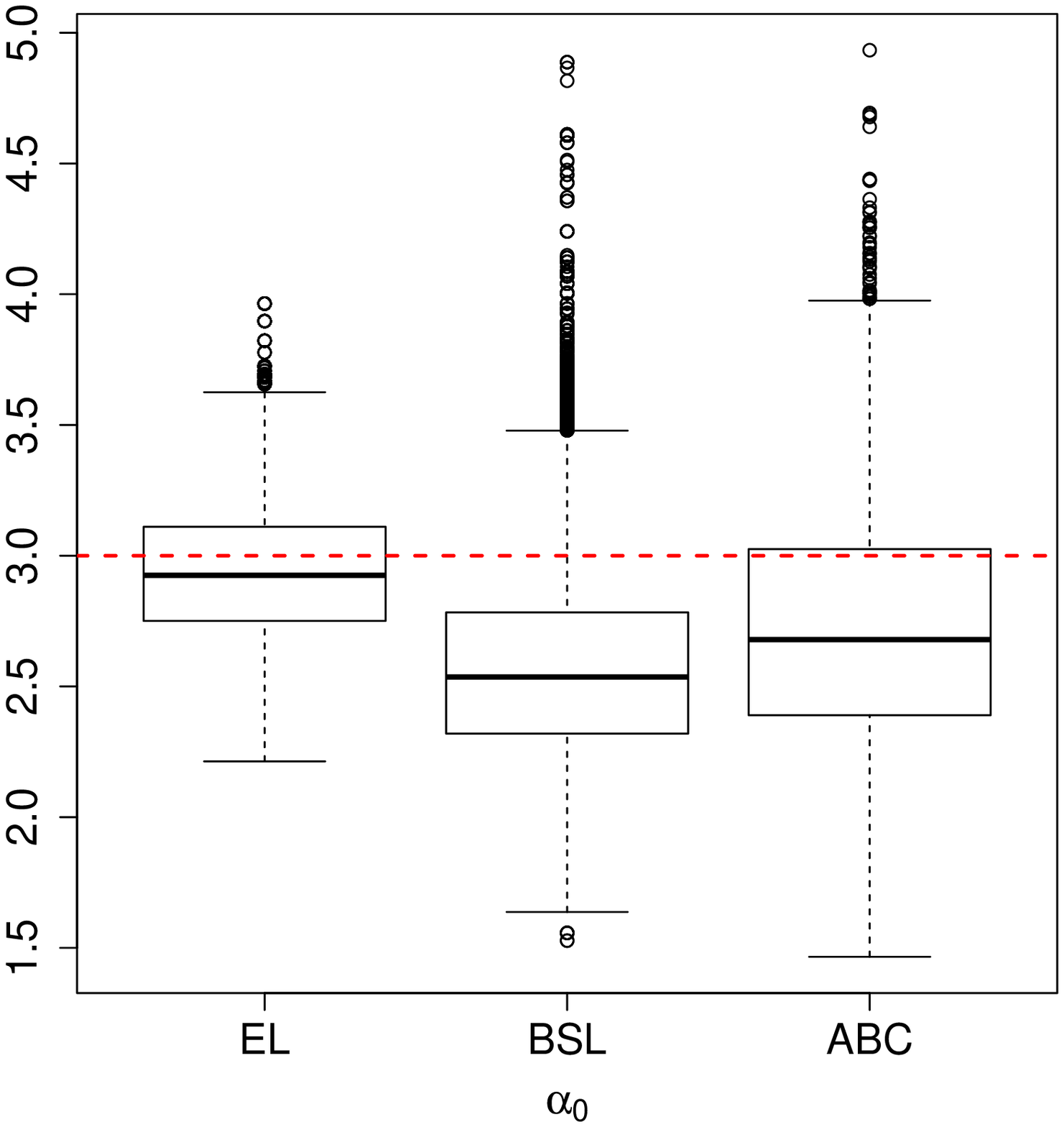}}
      \end{subfigure}\hfill\begin{subfigure}{.45\columnwidth}
    \resizebox{2.6in}{2.6in}{\includegraphics{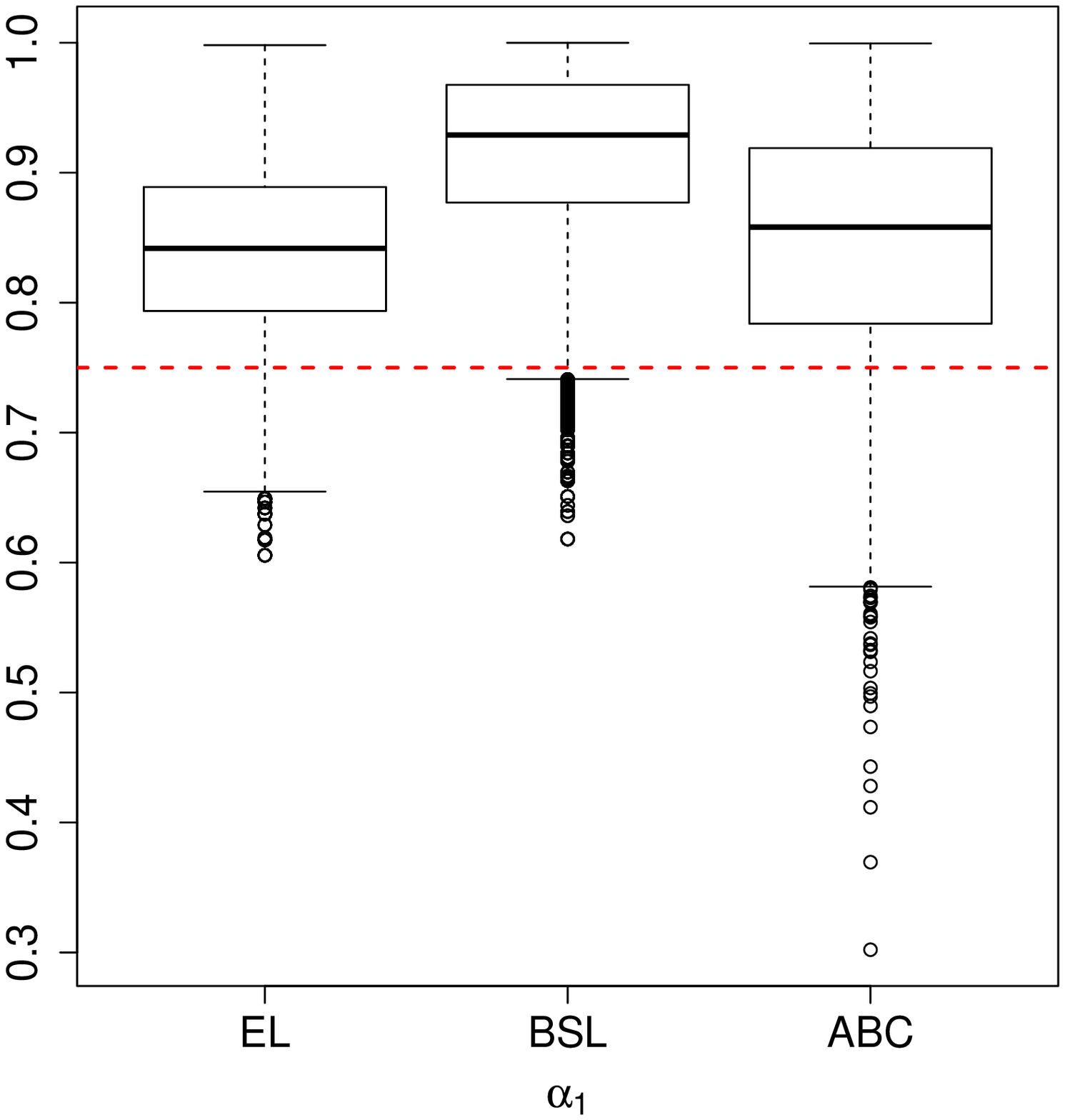}}
    \end{subfigure}
% The arguments in the next line are {height}{optional width}{used only by OUP for typesetting}[filename, in directory art]
%\resizebox{4.25in}{4.25in}{\includegraphics{arch_comparison_david.eps}}
% note that files may not be rotated
\caption{Estimated marginal posterior densities of parameters $\alpha_0$ and $\alpha_1$ in the ARCH(1) model.  The top row
shows kernel density estimates (empirical likelihood ABC (solid), synthetic likelihood (dashed), rejection ABC (dotted)), while the bottom row shows boxplots of posterior samples.  In the boxplots, the horizontal
dotted lines show the true parameter values.}
\label{F3}
\end{center}
\end{figure}

\subsection{An ARCH(1) model}\label{sec:arch}  
In contrast to the previous example, we now consider an example with summary statistics which are not close to normal, so that 
the assumptions behind the synthetic likelihood are not satisfied.  
We consider an autoregressive conditional heteroskedastic or ARCH(1) model, where for each $i=o,1,2,\ldots,m$, the components $X_{i1},X_{i2},\ldots,X_{in}$ are dependent.  This model was also considered in \citet{mengersen+pr13}. 
For each $i$, the time series ${X_{ij}}_{1\leq j \leq n}$ is generated by 
\begin{equation}\label{eq:arch_mod}
X_{ij}=\sigma_{ij}\epsilon_{ij},  \;\; {\sigma_{ij}}^2=\alpha_0+\alpha_1{X_{i(j-1)}}^2.
\end{equation}
where the $\epsilon_{ij}$ are i.i.d. $N(0,1)$ random variables. Here $\alpha_0,\alpha_1>0$ and stationarity requires
$\alpha_1<1$.
%To simulate $X_{ij}$, $j=1,\dots, n$ for each $i$, we first simulate $\epsilon_{ij}$, for $j=1,...,n$.  We set the initial standard deviation $\sigma_{i1}$ to $\surd\{\alpha_0/(1-\alpha_1)\}$, which is consistent
%with stationarity \citep[Section 21]{hamilton1994time}, and then the data can be generated following \eqref{eq:arch_mod}.
We assume a uniform prior over $(0,5)\times(0,1)$ for $(\alpha_0,\alpha_1)$. %%%%% is given a uniform prior over $(0,5)\times(0,1)$.

Our summary statistics include the three quartiles of the absolute values of the data.  Since the data is dependent we also use the following summary statistic.  Let, for a fixed $i$ and for each $j$, $Y_{ij}=X^2_{ij}-\sum^n_{j=1}X^2_{ij}/n$.% and suppose $\mathcal{L}$ is the backshift operator, i.e. $\mathcal{L}Y_{ij}=Y_{i(j-1)}$.
Then for each $i=1$, $2$, $\ldots$, $m$, we define,
\[
g_4(X_i)=\frac{1}{n}\sum^n_{j=2}\left(1_{\{(Y_{ij}\cdot Y_{i(j-1)})\ge0\}}-1_{\{(Y_{ij}\cdot Y_{i(j-1)})<0\}}\right).
\] 
That is, $g_4$ is the difference between the proportion of the concordant and that of the discordant pairs between series $Y_i$ with its lag-$1$ version. Empirical evidence suggests that $g_4$ performs better than the usual lag-$1$ autocovariance of the series $X^2_i$.  The quartiles of the absolute values of the data provide some information about the marginal distribution.
 
Our observed data were of size $n=1000$, with $(\alpha_0,\alpha_1)=(3,0.75)$ and we used $m=50$ replicates for each
likelihood approximation for both empirical and synthetic likelihoods in Bayesian computations.  Marginal posterior densities were estimated for the parameters based on $50,000$ sampling iterations with $50,000$ iterations burn in for both the synthetic likelihood and proposed empirical likelihood.
We compare these methods with the posterior obtained using rejection ABC with $1,000,000$ samples, a tolerance of $0.0025$ and linear regression adjustment.  
The estimated marginal densities in Figure \ref{F3} for the proposed method are quite close to the ABC gold standard.  However, the synthetic likelihood estimated marginal posterior densities are quite different to those obtained from ABC, 
especially for $\alpha_1$.  In this example 
the $g_4$ statistic is highly non-Gaussian, so the normality assumption made in the synthetic likelihood formulation is not satisfied.  

\subsection{Stereological data}  

Next we consider an example concerning the modelling of diameters of inclusions (microscopic particles introduced
in the steel production process) measured from planar cross-sections in a block of steel.  The size of the largest
inclusion in a block is thought to be important for steel strength.  
%The data considered here were first analysed by \citet{anderson2002largest}, and consist of measurements on inclusions
%from planar cross-sections.
We focus on an elliptical inclusion model due to \citet{bortot2007inference} here, which is an extension of the spherical model studied by \citet{anderson2002largest}.  Unlike the latter, the elliptcal model does not have tractable likelihood.

%\citet{anderson2002largest} considered a spherical model for the inclusions, which
%leads to a model with a tractable likelihood.  \citet{bortot2007inference} later extended this to an elliptical
%inclusion model which does not have tractable likelihood, and it is this model that we discuss.

%\citet{anderson2002largest}

It is assumed that the inclusion centres follow a homogeneous Poisson process with rate $\lambda$. 
%In the elliptical model,
For each inclusion, the three principal diameters of the ellipse 
are assumed independent of each other and of the process of inclusion centres.  
%Let $V$ be the largest inclusion diameter for a given inclusion.
Given $V$, the largest diameter for a given inclusion, the two other principal diameters are determined by multiplying $V$ with two independent uniform $U[0,1]$ random variables.
The diameter $V$, conditional on exceeding a threshold value $v_0$ ($5\mu m$ in \citet{bortot2007inference}) is assumed to follow a generalised Pareto distribution:
\begin{displaymath}
\operatorname{pr}(V\leq v|V>v_0)=1-\left\{1+\frac{\xi (v-v_0)}{\sigma}\right\} _{+}^{-\frac{1}{\xi}}.
\end{displaymath}
%Since the inclusion centres follow a homogeneous Poisson process, so do the inclusions 
%with $V>v_0$.
The parameters of the model are given by $\theta=(\lambda,\sigma,\xi)$.
We assume independent uniform priors %for $\lambda$, $\sigma$ and $\xi$
with ranges $(1,200)$, $(0,10)$ and $(-5,5)$ respectively. 
A detailed implementation of ABC for this example is discussed in \citet{erhardt+s15}.

%\begin{figure}[t]
%  \begin{center}
% The arguments in the next line are {height}{optional width}{used only by OUP for typesetting}[filename, in directory art]
%\resizebox{4.25in}{4.25in}{\includegraphics{Inclusion_el_syn_abc_112.eps}}
% note that files may not be rotated
%\caption{Estimated marginal posterior densities of $\lambda$, $\sigma$ and $\xi$ using empirical likelihood ABC (solid), rejection ABC (dotted) and synthetic likelihood (dashed).}
%\label{density_inclu}
%\end{center}
%\end{figure}

\begin{figure}[t]
  \begin{center}
  \begin{subfigure}{.45\columnwidth}
    \resizebox{2.5in}{2.5in}{\includegraphics{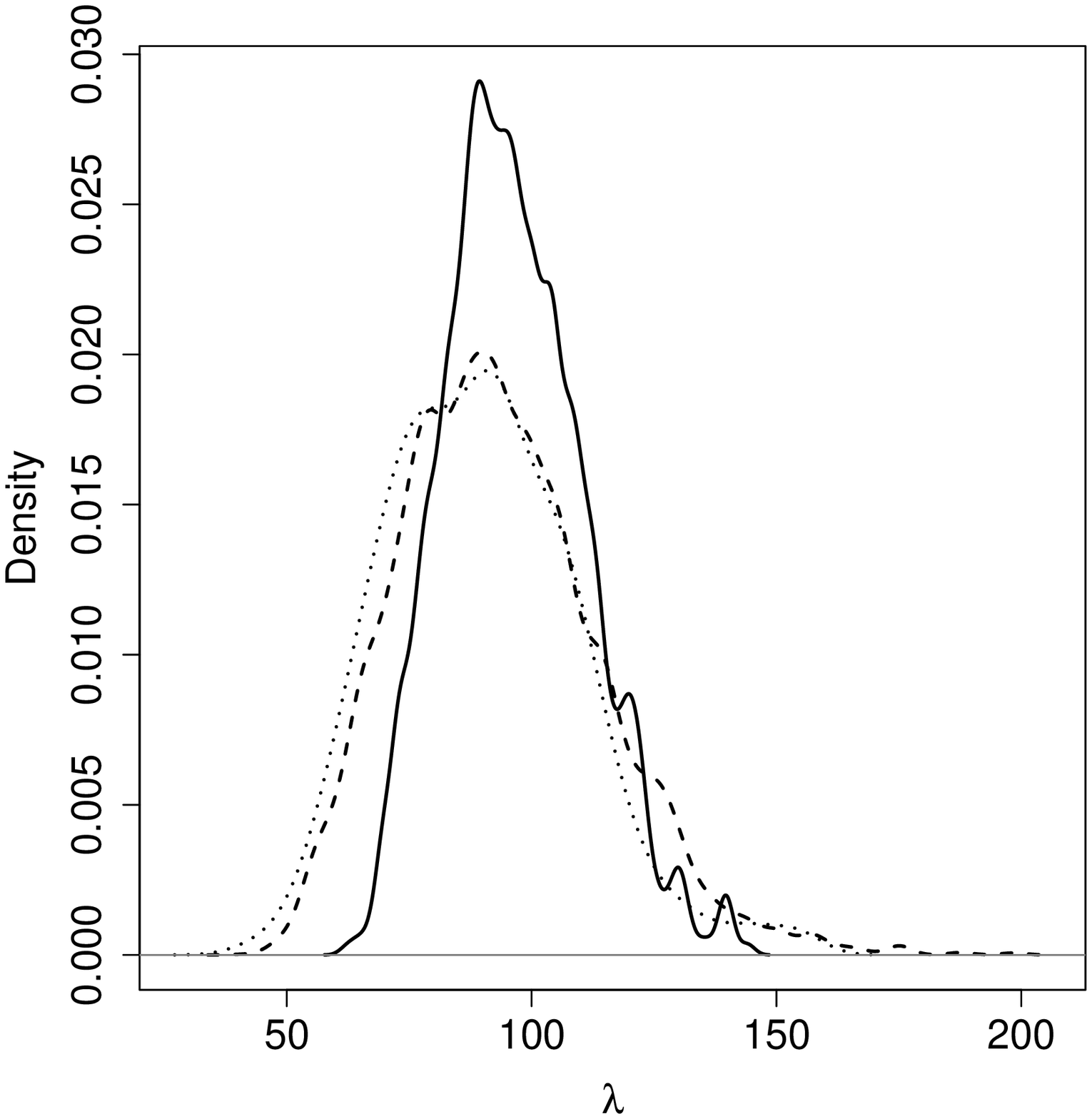}}
    \end{subfigure}\hfill\begin{subfigure}{.45\columnwidth}
  \resizebox{2.5in}{2.5in}{\includegraphics{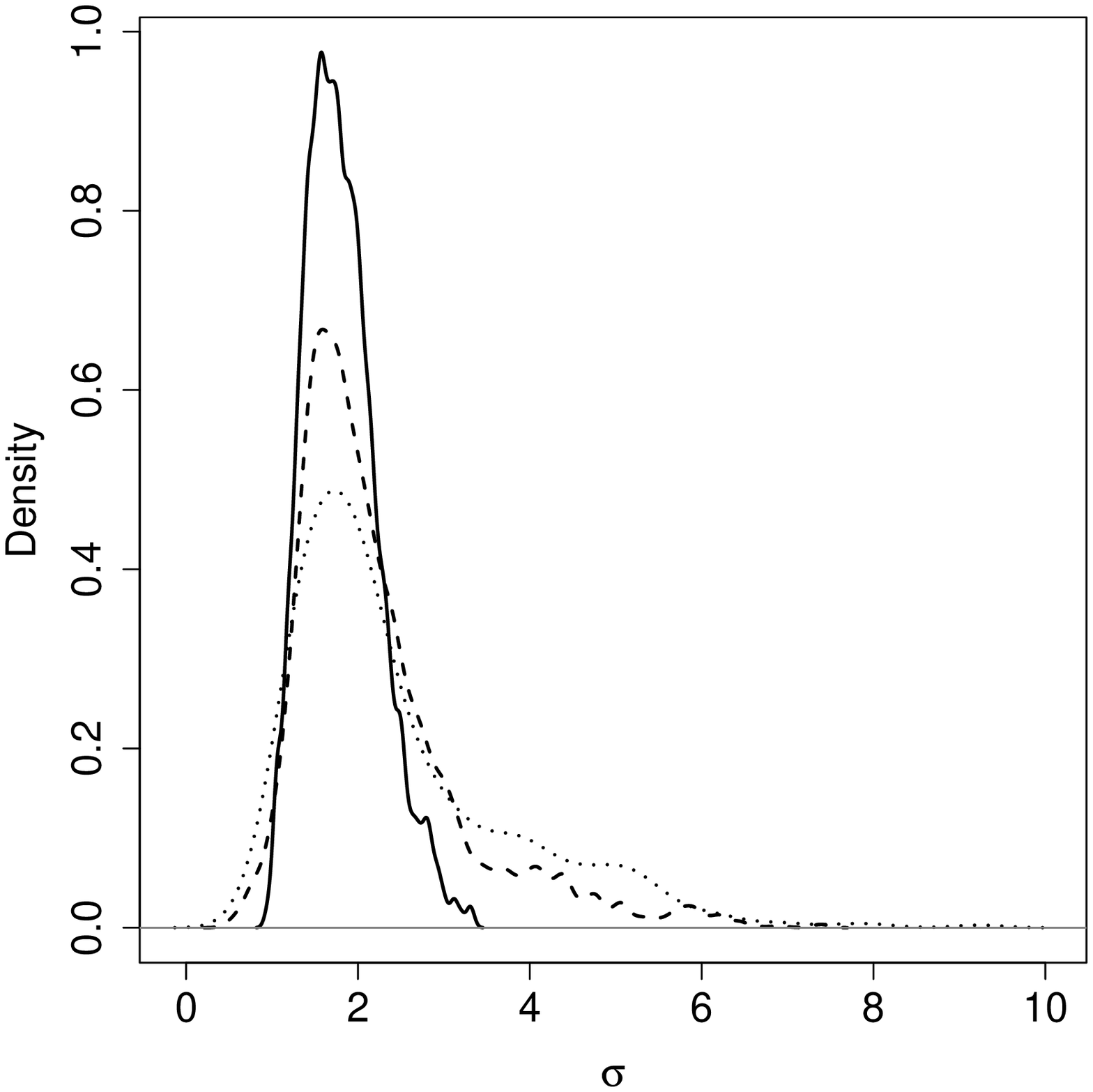}}
  \end{subfigure}\\
  \begin{subfigure}{.45\columnwidth}
    \resizebox{2.5in}{2.5in}{\includegraphics{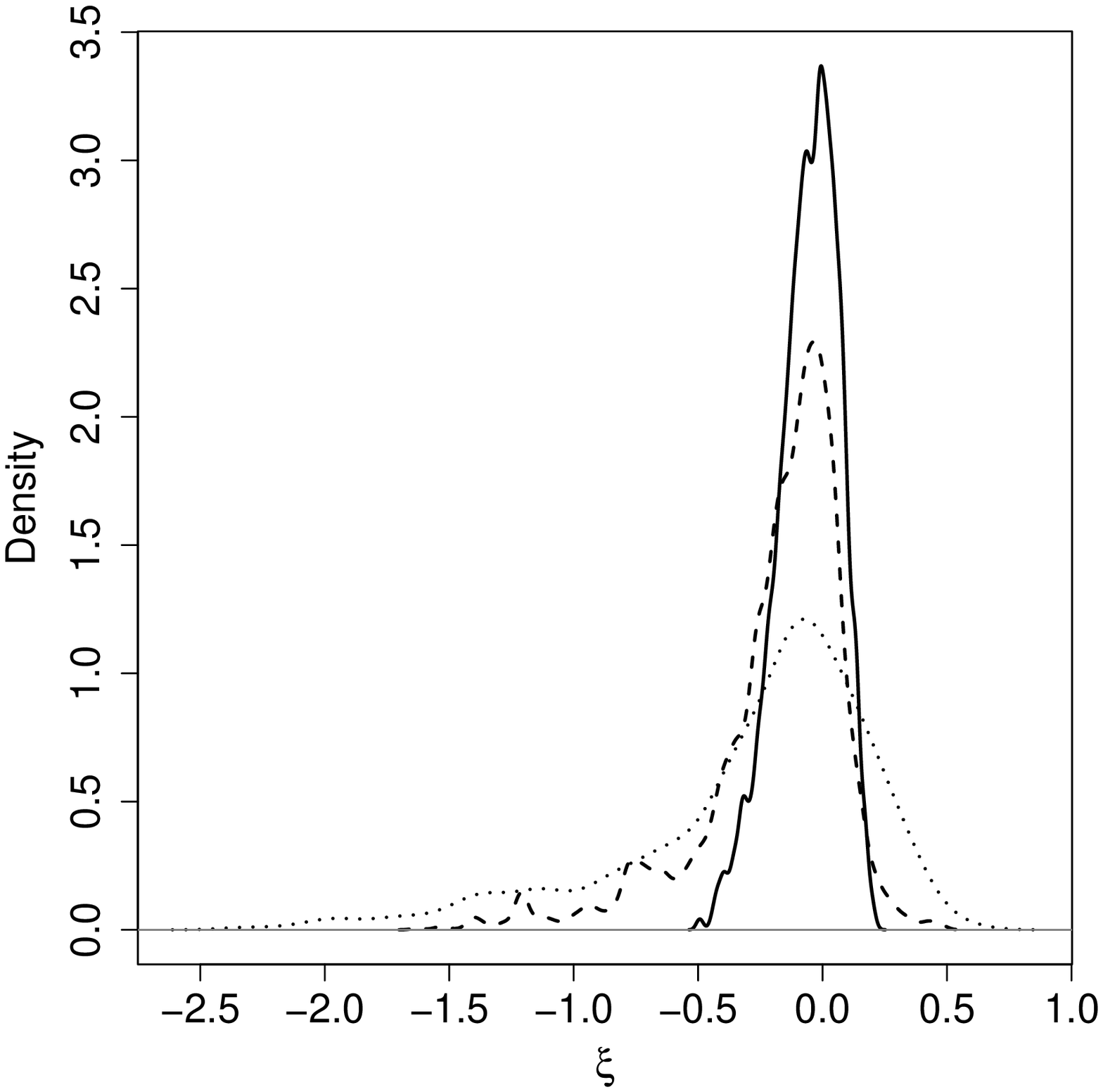}}
    \end{subfigure}
\caption{Estimated marginal posterior densities of $\lambda$, $\sigma$ and $\xi$ using empirical likelihood ABC (solid), rejection ABC (dotted) and synthetic likelihood (dashed).}
\label{density_inclu}
  \end{center}
  \end{figure}

The observed data has $112$ entries, measuring the largest principal diameters of elliptical cross-sections of inclusions for a planar slice. The number of inclusions $L$ in each dataset generated from the model is random.  %Writing $L$ for the number of inclusions,
The summary statistics used are $a)$ $(L-112)/100$, $b)$ the mean and $c)$ the median of the observed planar measurements, %$c)$ the median of the of the observed planar measurements,
and $d)$ the proportion of planar measurements less than or equal to six (approximately the median for the observed data). 
Even though %the number of observations
$L$ is itself random, the above estimating equations are unbiased under the truth.

Using the summary statistics described above, we compare the proposed empirical likelihood based method with the synthetic likelihood ($m=25$ for both) and a gold standard rejection ABC algorithm with small tolerance ($0.00005$) and linear regression adjustment. %For the 
%rejection ABC method we generated $10,000,000$ samples from the elliptic inclusion model and use a tolerance of $0.00005$ and linear regression adjustment.
%Both the proposed empirical likelihood and the synthetic likelihood methods use $m=25$ samples.  
%In total, $25,000$ samples were drawn from the empirical likelihood and synthetic likelihood posterior densities, following
%$25,000$ iterations burn in, 
%using the adaptive random walk Metropolis algorithm with normal proposal described in \citet{Pham2014note}. 
The resulting estimated marginal posterior densities for $\lambda, \sigma, \xi$  are shown in
Figure~\ref{density_inclu}. The results for the proposed empirical likelihood based method agree quite well with rejection ABC and synthetic likelihood. Similar to previous examples, however, there is a slight underestimation of posterior uncertainty in the empirical likelihood ABC method.  

The summary statistics in this example were judiciously chosen.  This dataset was also considered by \citet{Pham2014note}, who used $|L-112|/112$, the minimum, mean and maximum of the observed inclusions as summaries in their analysis.
%We realised that
For these summaries the observed values were too extreme for the values generated from the potentially mis-specified model for most values of $\theta$,  %That is, for most values of the parameter,
making the problem in \eqref{eq:w2} infeasible and the estimated empirical likelihood zero.
As a result, the MCMC scheme to sample from the resulting posterior mixed very slowly.
However, the performance of the proposed method was no worse than the synthetic likelihood for these summaries.  It is well-known that (see \citet{price+dln16}), for these summaries the synthetic likelihood covariance
matrix is often poorly estimated, resulting in gross over-estimation of the likelihood in the tail of the posterior, which leads to poor mixing in the MCMC algorithm.
%  In the marginal MCMC sampling scheme, this often leads to acceptance of those parameter values which should have been rejected.
It turns out that (see \citet{frazier+rr17}), the simple rejection ABC is more robust for such potentially mis-specified models.

\section{Discussion}

We have developed a new and easy-to-use empirical likelihood ABC method.  For implementation, all that is required
are some summary statistics, their observed values, and the ability to simulate from the model.  Properties of the approach
have been explored both empirically and theoretically.  The method enjoys posterior consistency under reasonable conditions, 
and shows good performance in simulated and real examples with appropriate summary statistic choices.

The proposed method is based on an interpretable empirical likelihood which is justified by a general variational approximation argument.  Unlike the conventional rejection ABC method, no tolerance or bandwidth needs to be specified.
  Furthermore, unlike the synthetic likelihood, the proposed method does not assume joint normaility of the summary statistics.  By using the variational approximation argument, we avoid any requirement of parameter dependent constraints to determine the empirical likelihood as well.  
  This directly contrasts with the previous empirical likelihood based ABC methods \citep{mengersen+pr13}.

In the proposed method, the empirical likelihood is approximated using data simulated from the underlying generative model.  Empirical evidence suggests that, like the synthetic likelihood \citep{price+dln16}, it is not sensitive to the number of generated replications.
  On the other hand, if the joint normality of the summary statistics is not satisfied (like in the ARCH(1) model above), the proposed approach is seen to work better than the synthetic likelihood.
  Since no distributional assumptions are made, the proposed approach can avoid the additional burden of searching for suitable marginal transformations to improve adherence to such assumptions.  As a result, it can be easily automated in practical applications.

Even though simple choices often work for our method, a judicious selection of summary statistics is required.  As we have demonstrated above, summaries which fit the model rather poorly, may result in failures of the empirical likelihood based ABC.
%  In many cases, it is no worse than the synthetic likelihood, which also fails and the conventional rejection ABC is the only way forward. 
However, for a poorly fitting model, such computational problems may arise for alternative methods as well.
It is important to diagnose poor model fit for the chosen summary statistics when this occurs (see \citet{frazier+rr17} for suggestions).
  Interestingly, synthetic likelihood can often down-weight unimportant summaries, which, as implemented, is not true for the proposed method. 
   Penalised empirical likelihood which can choose constraints has been recently considered. Such methods can be used in our proposed ABC as well. 

From the presented examples, it seems that the empirical likelihood slightly underestimates posterior uncertainty. Under-coverage of frequentist empirical likelihood confidence intervals is a well-known problem. This is most likely the Bayesian counterpart to that phenomenon.
The error would be small provided minimal and informative summary statistics are used.  Additionally, a wide variety of suggestions, similar to \citet{tsao2013,jing+tz17} etc. can be adapted in order to remedy this underestimation.  

%In the examples described here, we found that the empirical likelihood slightly underestimates posterior uncertainty, although
%only by a small amount if care is taken to use minimal and informative summary statistics, as in conventional ABC. 
%In the empirical likelihood literature, under-coverage of frequentist empirical likelihood confidence intervals is a well-known
%phenomenon, and what we observe seems to be the Bayesian counterpart to this.  There are a wide variety
%of existing suggestions about how to remedy the under-coverage problem - 
%see \citet{tsao2013,jing+tz17} for one recent suggestion
%and an up-to-date overview of the relevant literature.  Although the empirical likelihood used in this paper is not the conventional one, 
%we believe that some of the existing empirical likelihood modifications could be applied to our proposed empirical 
%likelihood ABC method.

%It would also be possible to investigate features of our empirical likelihood ABC method further
%theoretically.
Finally, similar to \citet{chernozhukov+h03}, it is likely that under suitable conditions, a Bernstein-von Mises theorem would hold for our posteriors, based on which asymptotic sandwich-type variance corrections might
also be considered.  We leave these investigations to future endeavours.  

\section*{Acknowledgement}

Sanjay Chaudhuri was supported by Singapore Ministry of Education Academic Research Fund Tier 1 grants R-155-000-194-114 and R-155-000-214-114.
Subhroshekhar Ghosh was suppoerted by Singapore Ministry of Education Academic Research Fund grants R-146-000-250-133 and R-146-000-312-114.
David Nott was supported by a Singapore Ministry of Education Academic Research
Fund Tier 1 grant (R-155-000-189-114).
Pham Kim Cuc was supported by the Singapore-Peking-Oxford Research Enterprise, COY-15-EWI-RCFSA/N197-1.

\appendix
\section*{Appendix}
%\section*{Proofs of the Asymptotic Properties}
\noindent{\it Proof of Theorem 1.}
 The proof proceeds by expanding the Kullback-Leibler divergence $D_{KL}\left(q(\theta,\gt)\mid\mid f(\theta,\gt\mid g_o)\right))$ when $q(\theta,g)=q^{\prime}(\theta)f_0(\gt\mid \theta)$.

For a $f\in\mathcal{F}$, suppose $f(g_o)$ is the marginal distribution of $g_o$.
It is well known that \citep{ormerodWand2010,faesOrmerodWand2011} the so called log evidence i.e. $\log f(g_o)$ can be expressed as:
\begin{equation}\label{eq:fund}
\log f(g_o)=D_{KL}\left(q(\theta,\gt)\mid\mid f(\theta,\gt\mid g_o)\right))+\int q(\theta,\gt)\log\left(\frac{f(\theta,\gt,g_o)}{q^{\prime}(\theta)f_0(\gt\mid \theta)}\right)d\gt~ d\theta.
\end{equation}

For the convenience of notation, for an $f\in\mathcal{F}$ we define:

\begin{align}
f^{\prime\prime}(\theta,g_o)&=\frac{\exp(\et_{\gt\mid\theta}\left[\log f(\theta,\gt,g_o)\right])}{\int\exp(\et_{\gt\mid t}\left[\log f(t,\gt,t^{\prime})\right])dtdt^{\prime}},~~f^{\prime\prime}(g_o)=\int f^{\prime\prime}(\theta,g_o)d\theta~\mbox{and}\nonumber\\
 f^{\prime\prime}(\theta\mid g_o)&=f^{\prime\prime}(\theta,g_o)/f^{\prime\prime}(g_o).\nonumber
\end{align}
 
By substituting the expression of $q(\theta,\gt)\in \mathcal{Q}^{\prime\prime}$ in \eqref{eq:fund} we get: 

\begin{align}
~D_{KL}\left(q(\theta,\gt)\mid\mid f(\theta,\gt\mid g_o)\right))=\log f(g_o)+\int q^{\prime}(\theta) f_0(\gt\mid \theta)\log f_0(\gt\mid \theta)d\gt&\nonumber\\
-\hfill\int q^{\prime}(\theta)\left\{\int\log f(\theta,\gt,g_o) f_0(\gt\mid \theta)d\gt-\log q^{\prime}(\theta)\right\}d\theta&\nonumber
\end{align}
\begin{align}
=&\log f^{\prime\prime}(g_o)-\int q^{\prime}(\theta)\log \left(\frac{\exp(\et_{\gt\mid\theta}\left[\log f(\theta,\gt,g_o)\right])}{q^{\prime}(\theta)}\right)d\theta-\int \htr_{\gt\mid\theta}(\theta)q^{\prime}(\theta)d\theta +\log \left(\frac{f(g_o)}{f^{\prime\prime}(g_o)}\right)&\nonumber\\
=&\log f^{\prime\prime}(g_o)-\int q^{\prime}(\theta)\left\{\log \left(\frac{f^{\prime\prime}(\theta,g_o)}{q^{\prime}(\theta)}\right)-\log\int \exp(\et_{\gt\mid t}\left[\log f(t,\gt,t^{\prime})\right])dt dt^{\prime}\right\}d\theta& \nonumber\\
&\hspace{.3\columnwidth}-\int \htr_{\gt\mid\theta}(\theta)q^{\prime}(\theta)d\theta+\log \left(\frac{f(g_o)}{f^{\prime\prime}(g_o)}\right)&\label{eq:varAprx}
\end{align} 

Similar to \eqref{eq:fund} one can show that:
\[
\log f^{\prime\prime}(g_o)=\int q^{\prime}(\theta)\log \left(\frac{f^{\prime\prime}(\theta,g_o)}{q^{\prime}(\theta)}\right)d\theta+D_{KL}\left(q^{\prime}(\theta)\mid\mid  f^{\prime\prime}(\theta\mid g_o)\right),
\]
where second addendum is the Kullback-Leibler divergence between the densities $q^{\prime}(\theta)$ and $f^{\prime\prime}(\theta\mid g_o)$.  Moreover, the third addendum in \eqref{eq:varAprx} depends on the hyper-parameters of $\pi(\theta)$ and thus independent of $\theta$.  Suppose we denote $C^{\prime}=\log\int \exp(\et_{\gt\mid t}\left[\log f(t,\gt,t^{\prime})\right])dt dt^{\prime}$. 

By substituting the above result in \eqref{eq:varAprx} and from \eqref{eq:fund} we get:
\begin{align}
~&D_{KL}\left(q(\theta,\gt)\mid\mid f(\theta,\gt\mid g_o)\right))=\log f(g_o)-\int q^{\prime}(\theta)f_0(\gt\mid \theta)\log\left(\frac{f(\theta,\gt,g_o)}{q^{\prime}(\theta)f_0(\gt\mid \theta)}\right)d\gt~ d\theta\nonumber\\
=&D_{KL}\left(q^{\prime}(\theta)\mid\mid  f^{\prime\prime}(\theta\mid g_o)\right)-\int \htr_{\gt\mid\theta}(\theta)q^{\prime}(\theta)d\theta-C^{\prime}+\log \left(\frac{f(g_o)}{f^{\prime\prime}(g_o)}\right)\label{eq:varAprx2}
\end{align}

Now by expanding the first two addenda in \eqref{eq:varAprx2} we get:
\begin{align}
~&D_{KL}\left(q^{\prime}(\theta)\mid\mid  f^{\prime\prime}(\theta\mid g_o)\right)-\int \htr_{\gt\mid\theta}(\theta)q^{\prime}(\theta)d\theta=\int q^{\prime}(\theta)\left\{\log\left(\frac{q^{\prime}(\theta)}{f^{\prime\prime}(\theta\mid g_o)}\right)-\htr_{\gt\mid\theta}(\theta)\right\}d\theta\nonumber\\
=&\int q^{\prime}(\theta)\left\{\log\left(\frac{q^{\prime}(\theta)}{f^{\prime\prime}(\theta\mid g_o)exp(\htr_{\gt\mid\theta}(\theta))}\right)\right\}d\theta\nonumber\\
=&\int q^{\prime}(\theta)\left\{\log\left(\frac{q^{\prime}(\theta)}{f^{\prime}(\theta\mid g_o)}\right)-\left(\log\int f^{\prime\prime}(t\mid g_o)\exp(\htr_{\gt\mid t}(t))dt\right) \right\}d\theta\label{eq:varAprx3}
\end{align}
The first addendum in \eqref{eq:varAprx3} is the Kullback-Leibler divergence between $q^{\prime}$ and $f^{\prime}(\theta\mid g_o)$. 
The second addendum is a function of $g_o$ and is independent of $\theta$.  By denoting it by $C(g_o)$ and collecting the terms from \eqref{eq:varAprx2} and \eqref{eq:varAprx3} we get:
\begin{equation}\label{eq:varAprx4}
D_{KL}\left(q(\theta,\gt)\mid\mid f(\theta,\gt\mid g_o)\right)=D_{KL}\left(q^{\prime}(\theta)\mid\mid f^{\prime}(\theta\mid g_o)\right)-C(g_o)-C^{\prime}+\log \left(\frac{f(g_o)}{f^{\prime\prime}(g_o)}\right).
\end{equation}

Note that, the R.H.S. of the equation \eqref{eq:varAprx4} is non-negative for all $q^{\prime}\in\mathcal{Q}_{\Theta}$. Furthermore, only the first addendum depends on $q^{\prime}$, which is also non-negative, with equality holding iff $q^{\prime}(\theta)=f^{\prime}(\theta\mid g_o)$.  
This implies the R.H.S. of \eqref{eq:varAprx4} attains its minimum at $q^{\prime}(\theta)=f^{\prime}(\theta\mid g_o)$.  So, it clearly follows that the variational approximation or the information projection of $f(\theta,\gt\mid g_o)$ is given by $f^{\prime}(\theta\mid g_o)f_0(\gt\mid \theta)$.\hfill $\square$

\bigskip

\noindent{\it Proof of Lemma \ref{lem:1}.} We show that for every $\epsilon>0$, there exists $n_0=n_0(\epsilon)$ such that for any $n\ge n_0$ for all $\theta\in\Theta_n$ the maximisation problem in \eqref{eq:w2} is feasible with probability larger than $1-\epsilon$.  

By assumption, for each $\theta$, random vectors $\xi^{(n)}_i(\theta)$ are i.i.d., put positive mass on each orthant and supremum of their lengths in each orthant diverge to infinity with $n$.  The random vectors $\left\{\xi^{(n)}_i(\theta)-\xi^{(n)}_o(\theta_o)\right\}$ will inherit the same properties.  
That is, there exists integer $n_0$, such that for each $n\ge n_0$, the convex hull of the vectors $\left\{\xi^{(n)}_i(\theta)-\xi^{(n)}_o(\theta_o)\right\}$, $i=1$, $\ldots$, $m(n)$, would contain the unit sphere with probability larger than $1-\epsilon/2$.  

%Let us define $h^{(n)}_i\left(\theta,\theta_0\right)=\left\{g\left(X^{(n)}_i(\theta)\right)-g\left(X^{(n)}_o(\theta_o)\right)\right\}$ and

We choose an $n\ge n_0$ and a $\theta\in\Theta_n$. For this choice of $\theta$:
\begin{align}
h^{(n)}_i(\theta,\theta_o)=&b_n\left\{\mathfrak{g}(\theta)-\mathfrak{g}(\theta_o)\right\}+\xi^{(n)}_i(\theta)-\xi^{(n)}_o(\theta_o)=c_n(\theta)+\xi^{(n)}_i(\theta)-\xi^{(n)}_o(\theta_o),\nonumber
\end{align}
where, $\mid\mid\mathfrak{g}(\theta)-\mathfrak{g}(\theta_o)\mid\mid\le b^{-1}_n$.  That is, $\mid\mid c_n(\theta)\mid\mid\le 1$.  
Now, since $-c_n(\theta)$ is in the convex hull of the vectors $\left\{\xi^{(n)}_i(\theta)-\xi^{(n)}_o(\theta_o)\right\}$, $i=1$, $\ldots$, $m(n)$, with probability larger than $1-\epsilon/2$, there exists weights $w\in\Delta_{m(n)-1}$ such that,

\[
-c_n(\theta)=\sum^{m(n)}_{i=1}w_i\left\{\xi^{(n)}_i(\theta)-\xi^{(n)}_o(\theta_o)\right\}.
\]
Now it follows that for the above choice of $w$ that
\[
\sum^{m(n)}_{i=1}w_ih^{(n)}_i(\theta,\theta_o)=c_n(\theta)+\sum^{m(n)}_{i=1}w_i\left\{\xi^{(n)}_i(\theta)-\xi^{(n)}_o(\theta_o)\right\}=0,
\] 
which shows that the problem in \eqref{eq:w2} is feasible.\hfill $\square$
%\end{proof}

\bigskip

\noindent{\it Proof of Lemma \ref{lem:2}.} Let $\epsilon$ be as in the statement.  By assumption (A1), for some $\delta>0$,  $\mid\mid\mathfrak{g}(\theta)-\mathfrak{g}(\theta_o)\mid\mid>\delta$ for all $\theta$ with $\mid\mid\theta-\theta_o\mid\mid >\epsilon$.

Consider $\eta>0$. We show that there exists $n_0=n_0(\eta)$ such that for any $n\ge n_0$, the constrained maximisation problem in \eqref{eq:w2} is not feasible for all $\mid\mid\theta-\theta_o\mid\mid >\epsilon$, with probability larger than $1-\eta$. %That is $\mathcal{W}_{\theta}=\emptyset$.

%Suppose that the problem in \eqref{eq:w2} is feasible for all values of $n$.  From the definition of $\mathcal{W}_{\theta}$ we get:

Let if possible $w\in\Delta_{m(n)-1}$ be a feasible solution.  Hence we get:

\begin{align}
0=&\sum^{m(n)}_{i=1}w_ih^{(n)}_i(\theta,\theta_o)=\sum^{m(n)}_{i=1}w_i\left\{g^{(n)}\left(X_i(\theta)\right)-g^{(n)}\left(X_o(\theta_o)\right)\right\}\nonumber\\
=&\left\{\mathfrak{g}^{(n)}(\theta)-\mathfrak{g}^{(n)}(\theta_o)\right\}+\left\{\sum^{m(n)}_{i=1}w_i\xi^{(n)}_i(\theta)\right\}-\xi^{(n)}_o(\theta_o),\nonumber
\end{align}
so that 
\begin{equation*}
-b_n\left\{\mathfrak{g}(\theta)-\mathfrak{g}(\theta_o)+o(1)\right\}=\sum^{m(n)}_{i=1}w_i\xi^{(n)}_i(\theta)-\xi^{(n)}_o(\theta_o).
\end{equation*}

By dividing both sides by $b_n$ we get:
\begin{equation}\label{eq:cons}
-\left\{\mathfrak{g}(\theta)-\mathfrak{g}(\theta_o)\right\}=\sum^{m(n)}_{i=1}w_i\left\{\frac{\xi^{(n)}_i(\theta)}{b_n}-\frac{\xi^{(n)}_o(\theta_o)}{b_n}\right\}-o(1).
\end{equation}
Now, $\mid\mid\xi^{(n)}_o(\theta_o)\mid\mid/b_n\le \sup_{i\in\{o,1,2\ldots,m(n)\}}\mid\mid\xi^{(n)}_o(\theta_o)\mid\mid/b_n$ and
\begin{align}
\left|\left|\sum^{m(n)}_{i=1}w_i\frac{\xi^{(n)}_i(\theta)}{b_n}\right|\right|\le\sum^{m(n)}_{i=1}w_i\frac{\mid\mid\xi^{(n)}_i(\theta)\mid\mid}{b_n}\le\sup_{i\in\{o,1,2\ldots,m(n)\}}\frac{\mid\mid\xi^{(n)}_i(\theta)\mid\mid}{b_n}.\nonumber
\end{align}
That is, by assumption (A3), there exists $n_0(\eta)$ such that for any $n\ge n_0$, the RHS of \eqref{eq:cons} is less than $\delta$ for all $\theta\in B(\theta_o,\epsilon)$, with probability larger than $1-\eta$.  However, $\mid\mid\mathfrak{g}(\theta)-\mathfrak{g}(\theta_o)\mid\mid>\delta$. %  That is for large $n$, the convex hull of the set of vectors $\left\{(\xi^{(n)}_i(\theta)b_-\xi^{(n)}_o(\theta_o))b^{-1}_n\right\}$, $i=1$, $2$, $\ldots$, $m(n)$ 
%cannot contain the vector $-(\mathfrak{g}(\theta)-\mathfrak{g}(\theta_o))$.  That is, for large values of $n$ there is no $w\in\Delta_{m(n)-1}$ such that equation \eqref{eq:cons} could be satisfied and 
We arrive at a contradiction.  Thus the problem is infeasible for every $\theta\in B(\theta_o,\epsilon)^C$ with probability larger than $1-\eta$.\hfill $\square$

\medskip

\noindent{\it Proof of Theorem \ref{thm:1}.}  Let $s(\theta)$ be a continuous, bounded function. We choose an $\epsilon>0$. Then by Lemma \ref{lem:2}, there exists $n(\epsilon)$, such that for any $n>n(\epsilon)$ and $\theta\in B\left(\theta_o,\epsilon\right)^C$, %such that $\mid\mid\theta-\theta_o\mid\mid>\epsilon$, 
$l_n(\theta)=0$ and by definition \eqref{eq:mpost} the posterior $\hat{\Pi}_n\left(\theta\mid g(X_o(\theta_o))\right)=0$.  That is for any $n>n(\epsilon)$, 
\begin{align}
~&\int_{\Theta}s(\theta)\hat{\Pi}_n\left(\theta\mid g(X_o(\theta_o))\right)d\theta=\int_{B\left(\theta_o,\epsilon\right)}s(\theta)\hat{\Pi}_n\left(\theta\mid g(X_o(\theta_o))\right)d\theta\nonumber\\
=&\int_{B\left(\theta_o,\epsilon\right)}\left\{s(\theta)-s(\theta_o)\right\}\hat{\Pi}_n\left(\theta\mid g(X_o(\theta_o))\right)d\theta +s(\theta_o)\int_{B\left(\theta_o,\epsilon\right)}\hat{\Pi}_n\left(\theta\mid g(X_o(\theta_o))\right)d\theta.\nonumber
\end{align}

Since the function $s(\theta)$ is bounded and continuous at $\theta_o$, the first term is negligible.  Furthermore, $\int_{B\left(\theta_o,\epsilon\right)}\hat{\Pi}_n\left(\theta\mid g(X_o(\theta_o))\right)d\theta=1$.  This implies the integral converges to $s(\theta_o)$.  This shows, the posterior converges weakly to $\delta_{\theta_o}$.  \hfill $\square$

\bibliographystyle{chicago}

%\bibliography{bib_abcel_new}

\begin{thebibliography}{}

\bibitem[\protect\citeauthoryear{{Akaike}}{{Akaike}}{1974}]{akaike74}
{Akaike}, H. (1974).
\newblock A new look at the statistical model identification.
\newblock {\em IEEE Transactions on Automatic Control\/}~{\em 19\/}(6),
  716--723.

\bibitem[\protect\citeauthoryear{Allingham, King, and Mengersen}{Allingham
  et~al.}{2009}]{allingham2009bayesian}
Allingham, D., R.~A.~R. King, and K.~L. Mengersen (2009).
\newblock Bayesian estimation of quantile distributions.
\newblock {\em Statistics and Computing\/}~{\em 19\/}(2), 189--201.

\bibitem[\protect\citeauthoryear{An, Nott, and Drovandi}{An
  et~al.}{2020}]{anNottDrovandi2020}
An, Z., D.~Nott, and C.~Drovandi (2020).
\newblock Robust {B}ayesian synthetic likelihood via a semi-parametric
  approach.
\newblock {\em Stat Comput\/}~{\em 30}, 543--557.

\bibitem[\protect\citeauthoryear{Anderson and Coles}{Anderson and
  Coles}{2002}]{anderson2002largest}
Anderson, C.~W. and S.~G. Coles (2002).
\newblock The largest inclusions in a piece of steel.
\newblock {\em Extremes\/}~{\em 5\/}(3), 237--252.

\bibitem[\protect\citeauthoryear{Andrieu and Roberts}{Andrieu and
  Roberts}{2009}]{andrieu+r09}
Andrieu, C. and G.~O. Roberts (2009).
\newblock The pseudo-marginal approach for efficient {M}onte {C}arlo
  computations.
\newblock {\em The Annals of Statistics\/}~{\em 37\/}(2), 697--725.

\bibitem[\protect\citeauthoryear{Beaumont}{Beaumont}{2003}]{beaumont03}
Beaumont, M.~A. (2003).
\newblock Estimation of population growth or decline in genetically monitored
  populations.
\newblock {\em Genetics\/}~{\em 164\/}(3), 1139--1160.

\bibitem[\protect\citeauthoryear{Beaumont, Robert, Marin, and Corunet}{Beaumont
  et~al.}{2009}]{beaumont+rmc09}
Beaumont, M.~A., C.~P. Robert, J.-M. Marin, and J.~M. Corunet (2009).
\newblock Adaptivity for {ABC} algorithms: {The ABC-PMC} scheme.
\newblock {\em Biometrika\/}~{\em 96}, 983--990.

\bibitem[\protect\citeauthoryear{Beaumont, Zhang, and Balding}{Beaumont
  et~al.}{2002}]{beaumont+zb02}
Beaumont, M.~A., W.~Zhang, and D.~J. Balding (2002).
\newblock Approximate {Bayesian} computation in population genetics.
\newblock {\em Genetics\/}~{\em 162}, 2025--2035.

\bibitem[\protect\citeauthoryear{Berrett, Samworth, and Yuan}{Berrett
  et~al.}{2019}]{berrettSamworthMing2019}
Berrett, T.~B., R.~J. Samworth, and M.~Yuan (2019).
\newblock Efficient multivariate entropy estimation via $k$-nearest neighbour
  distances.
\newblock {\em Ann. Statist.\/}~{\em 47\/}(1), 288--318.

\bibitem[\protect\citeauthoryear{Blum, Nunes, Prangle, and Sisson}{Blum
  et~al.}{2013}]{blum+nps13}
Blum, M. G.~B., M.~A. Nunes, D.~Prangle, and S.~A. Sisson (2013).
\newblock A comparative review of dimension reduction methods in approximate
  {B}ayesian computation.
\newblock {\em Statistical Science\/}~{\em 28}, 189--208.

\bibitem[\protect\citeauthoryear{Bortot, Coles, and Sisson}{Bortot
  et~al.}{2007}]{bortot2007inference}
Bortot, P., S.~Coles, and S.~Sisson (2007).
\newblock Inference for stereological extremes.
\newblock {\em Journal of the American Statistical Association\/}~{\em
  102\/}(477), 84--92.

\bibitem[\protect\citeauthoryear{Brown and Chen}{Brown and
  Chen}{1998}]{brown_chen_1998}
Brown, B.~M. and S.~X. Chen (1998).
\newblock Combined and least squares empirical likelihood.
\newblock {\em Ann. Inst. Statist. Math\/}~(4), 697--714.

\bibitem[\protect\citeauthoryear{Chaudhuri and Ghosh}{Chaudhuri and
  Ghosh}{2011}]{chaudhuri+g11}
Chaudhuri, S. and M.~Ghosh (2011).
\newblock Empirical likelihood for small area estimation.
\newblock {\em Biometrika\/}~{\em 98}, 473--480.

\bibitem[\protect\citeauthoryear{Chaudhuri, Mondal, and Yin}{Chaudhuri
  et~al.}{2017}]{chaudhuri+my17}
Chaudhuri, S., D.~Mondal, and T.~Yin (2017).
\newblock {Hamiltonian Monte Carlo sampling in Bayesian empirical likelihood}.
\newblock {\em Journal of the Royal Statistical Society, Series B\/}~{\em 79},
  293--320.

\bibitem[\protect\citeauthoryear{Chernozhukov and Hong}{Chernozhukov and
  Hong}{2003}]{chernozhukov+h03}
Chernozhukov, V. and H.~Hong (2003).
\newblock {An MCMC approach to classical estimation}.
\newblock {\em Journal of Econometrics\/}~{\em 115\/}(2), 293--346.

\bibitem[\protect\citeauthoryear{Cover and Thomas}{Cover and
  Thomas}{2012}]{coverThomasBook}
Cover, T. and J.~Thomas (2012).
\newblock {\em Elements of Information Theory}.
\newblock Wiley.

\bibitem[\protect\citeauthoryear{Doucet, Godsill, and Robert}{Doucet
  et~al.}{2002}]{doucet+gr02}
Doucet, A., S.~Godsill, and C.~Robert (2002).
\newblock {Marginal maximum a posteriori estimation using Markov chain Monte
  Carlo}.
\newblock {\em Statistics and Computing\/}~{\em 12}, 77--84.

\bibitem[\protect\citeauthoryear{Doucet, Pitt, Deligiannidis, and Kohn}{Doucet
  et~al.}{2015}]{doucet+pdk15}
Doucet, A., M.~K. Pitt, G.~Deligiannidis, and R.~Kohn (2015).
\newblock {Efficient implementation of Markov chain Monte Carlo when using an
  unbiased likelihood estimator}.
\newblock {\em Biometrika\/}~{\em 102\/}(2), 295--313.

\bibitem[\protect\citeauthoryear{Drovandi and Pettitt}{Drovandi and
  Pettitt}{2011}]{drovandi2011likelihood}
Drovandi, C.~C. and A.~N. Pettitt (2011).
\newblock Likelihood-free {B}ayesian estimation of multivariate quantile
  distributions.
\newblock {\em Computational Statistics \& Data Analysis\/}~{\em 55\/}(9),
  2541--2556.

\bibitem[\protect\citeauthoryear{Drovandi, Pettitt, and Lee}{Drovandi
  et~al.}{2015}]{drovandi+pl15}
Drovandi, C.~C., A.~N. Pettitt, and A.~Lee (2015).
\newblock Bayesian indirect inference using a parametric auxiliary model.
\newblock {\em Statistical Science.\/}~{\em 30\/}(1), 72--95.

\bibitem[\protect\citeauthoryear{Dutta, Corander, Kaski, and Gutmann}{Dutta
  et~al.}{2016}]{Dutta2016}
Dutta, R., J.~Corander, S.~Kaski, and M.~U. Gutmann (2016).
\newblock Likelihood-free inference by penalised logistic regression.
\newblock arXiv:1611.10242.

\bibitem[\protect\citeauthoryear{Erhardt and Sisson}{Erhardt and
  Sisson}{2015}]{erhardt+s15}
Erhardt, R. and S.~A. Sisson (2015).
\newblock Modelling extremes using approximate {B}ayesian computation.
\newblock In D.~K. Dey and J.~Yan (Eds.), {\em {Extreme Value Modelling and
  Risk Analysis: Methods and Applications}}, pp.\  281--306. Chapman and
  Hall/CRC Press.

\bibitem[\protect\citeauthoryear{Faes, Ormerod, and Wand}{Faes
  et~al.}{2011}]{faesOrmerodWand2011}
Faes, C., J.~T. Ormerod, and M.~P. Wand (2011).
\newblock Variational {Bayesian} inference for parametric and nonparametric
  regression with missing data.
\newblock {\em Journal of the American Statistical Association\/}~{\em
  106\/}(495), 959--971.

\bibitem[\protect\citeauthoryear{Fasiolo, Wood, Hartig, and Bravington}{Fasiolo
  et~al.}{2016}]{Fasiolo2016}
Fasiolo, M., S.~N. Wood, F.~Hartig, and M.~V. Bravington (2016).
\newblock An extended empirical saddlepoint approximation for intractable
  likelihoods.
\newblock arXiv:1601.01849.

\bibitem[\protect\citeauthoryear{Fearnhead and Prangle}{Fearnhead and
  Prangle}{2012}]{fearnhead+p12}
Fearnhead, P. and D.~Prangle (2012).
\newblock Constructing summary statistics for approximate {B}ayesian
  computation: {S}emi-automatic approximate {Bayesian} computation (with
  discussion).
\newblock {\em Journal of the Royal Statistical Society, Series B\/}~{\em 74},
  419--474.

\bibitem[\protect\citeauthoryear{Fellows and Handcock}{Fellows and
  Handcock}{2017}]{fellows_handcock_2017}
Fellows, I. and M.~Handcock (2017, 20--22 Apr).
\newblock {Removing Phase Transitions from Gibbs Measures}.
\newblock In A.~Singh and J.~Zhu (Eds.), {\em Proceedings of the 20th
  International Conference on Artificial Intelligence and Statistics},
  Volume~54 of {\em Proceedings of Machine Learning Research}, Fort Lauderdale,
  FL, USA, pp.\  289--297. PMLR.

\bibitem[\protect\citeauthoryear{Frazier and Drovandi}{Frazier and
  Drovandi}{2020}]{frazier2020robust}
Frazier, D.~T. and C.~Drovandi (2020).
\newblock Robust approximate {B}ayesian inference with synthetic likelihood.
\newblock arXiv:1904.04551.

\bibitem[\protect\citeauthoryear{Frazier, Martin, Robert, and Rousseau}{Frazier
  et~al.}{2018}]{frazier+mrr18}
Frazier, D.~T., G.~M. Martin, C.~P. Robert, and J.~Rousseau (2018).
\newblock Asymptotic properties of approximate {B}ayesian computation.
\newblock {\em Biometrika\/}~{\em 105\/}(3), 593--607.

\bibitem[\protect\citeauthoryear{Frazier, Robert, and Rousseau}{Frazier
  et~al.}{2017}]{frazier+rr17}
Frazier, D.~T., C.~P. Robert, and J.~Rousseau (2017).
\newblock Model misspecification in {ABC}: {C}onsequences and diagnostics.
\newblock arXiv:1708.01974.

\bibitem[\protect\citeauthoryear{Ghosh and Chaudhuri}{Ghosh and
  Chaudhuri}{2019}]{ghosh2019empirical}
Ghosh, S. and S.~Chaudhuri (2019).
\newblock Empirical likelihood under mis-specification: Degeneracies and random
  critical points.
\newblock arxiv:1910.01396.

\bibitem[\protect\citeauthoryear{Gouri\'eroux and Monfort}{Gouri\'eroux and
  Monfort}{1996}]{gourieroux1996}
Gouri\'eroux, C. and A.~Monfort (1996).
\newblock {\em Simulation-based Econometric Methods}.
\newblock Oxford, United Kingdom: Oxford University Press.

\bibitem[\protect\citeauthoryear{Hall and Morton}{Hall and
  Morton}{1993}]{hallMorton1993}
Hall, P. and S.~Morton (1993).
\newblock On the estimation of entropy.
\newblock {\em Annals of Institute of Statistical Mathematics\/}~{\em 45},
  69--88.

\bibitem[\protect\citeauthoryear{Haynes, MacGillivray, and Mengersen}{Haynes
  et~al.}{1997}]{haynes1997robustness}
Haynes, M.~A., H.~L. MacGillivray, and K.~L. Mengersen (1997).
\newblock Robustness of ranking and selection rules using generalised g-and- k
  distributions.
\newblock {\em Journal of Statistical Planning and Inference\/}~{\em 65\/}(1),
  45--66.

\bibitem[\protect\citeauthoryear{Jing, Tsao, and Zhou}{Jing
  et~al.}{2017}]{jing+tz17}
Jing, B.-Y., M.~Tsao, and W.~Zhou (2017).
\newblock Transforming the empirical likelihood towards better accuracy.
\newblock {\em Canadian Journal of Statistics\/}~{\em 45\/}(3), 340--352.

\bibitem[\protect\citeauthoryear{{Kozachenko} and {Leonenko}}{{Kozachenko} and
  {Leonenko}}{1987}]{kozLeo87}
{Kozachenko}, L.~F. and N.~N. {Leonenko} (1987).
\newblock Sample estimate of the entropy of a~random vector.
\newblock {\em Probl. Peredachi Inf.\/}, 9--16.

\bibitem[\protect\citeauthoryear{Lazar}{Lazar}{2003}]{lazar03}
Lazar, N.~A. (2003).
\newblock Bayesian empirical likelihood.
\newblock {\em Biometrika\/}~{\em 90}, 319--326.

\bibitem[\protect\citeauthoryear{Lele, Dennis, and Lutscher}{Lele
  et~al.}{2007}]{lele+dl07}
Lele, S.~R., B.~Dennis, and F.~Lutscher (2007).
\newblock {Data cloning: easy maximum likelihood estimation for complex
  ecological models using Bayesian Markov chain Monte Carlo methods}.
\newblock {\em Ecology Letters\/}~{\em 10}, 551--563.

\bibitem[\protect\citeauthoryear{Li and Fearnhead}{Li and
  Fearnhead}{2018a}]{li+f18b}
Li, W. and P.~Fearnhead (2018a).
\newblock {Convergence of regression-adjusted approximate Bayesian
  computation}.
\newblock {\em Biometrika\/}~{\em 105\/}(2), 301--318.

\bibitem[\protect\citeauthoryear{Li and Fearnhead}{Li and
  Fearnhead}{2018b}]{li+f18a}
Li, W. and P.~Fearnhead (2018b).
\newblock {On the asymptotic efficiency of approximate Bayesian computation
  estimators}.
\newblock {\em Biometrika\/}~{\em 105\/}(2), 285--299.

\bibitem[\protect\citeauthoryear{Marin, Pudlo, Robert, and Ryder}{Marin
  et~al.}{2011}]{marin+prr11}
Marin, J.-M., P.~Pudlo, C.~P. Robert, and R.~Ryder (2011).
\newblock Approximate {B}ayesian computational methods.
\newblock {\em Statistics and Computing\/}~{\em 21}, 289--291.

\bibitem[\protect\citeauthoryear{Marjoram, Molitor, Plagnol, and
  Tavar{\'e}}{Marjoram et~al.}{2003}]{marjoram+mpt03}
Marjoram, P., J.~Molitor, V.~Plagnol, and S.~Tavar{\'e} (2003).
\newblock Markov chain {Monte Carlo} without likelihoods.
\newblock {\em Proceedings of the National Academy of Sciences of the
  USA\/}~{\em 100}, 15324--15328.

\bibitem[\protect\citeauthoryear{Mengersen, Pudlo, and Robert}{Mengersen
  et~al.}{2013}]{mengersen+pr13}
Mengersen, K.~L., P.~Pudlo, and C.~P. Robert (2013).
\newblock Bayesian computation via empirical likelihood.
\newblock {\em Proceedings of the National Academy of Sciences\/}~{\em
  110\/}(4), 1321--1326.

\bibitem[\protect\citeauthoryear{Monahan and Boos}{Monahan and
  Boos}{1992}]{monahan+b92}
Monahan, J.~F. and D.~D. Boos (1992).
\newblock Proper likelihoods for {B}ayesian analysis.
\newblock {\em Biometrika\/}~{\em 79}, 271--278.

\bibitem[\protect\citeauthoryear{Ormerod and Wand}{Ormerod and
  Wand}{2010}]{ormerodWand2010}
Ormerod, J.~T. and M.~P. Wand (2010).
\newblock Explaining variational approximation.
\newblock {\em The American Statistics\/}~{\em 64\/}(2), 140--153.

\bibitem[\protect\citeauthoryear{Owen}{Owen}{2001}]{owen01}
Owen, A.~B. (2001).
\newblock {\em {Empirical Likelihood}}.
\newblock London: Chapman and Hall.

\bibitem[\protect\citeauthoryear{{Paninski} and {Yajima}}{{Paninski} and
  {Yajima}}{2008}]{paninskiYazima2008}
{Paninski}, L. and M.~{Yajima} (2008).
\newblock Undersmoothed kernel entropy estimators.
\newblock {\em IEEE Transactions on Information Theory\/}~{\em 54\/}(9),
  4384--4388.

\bibitem[\protect\citeauthoryear{Peters and Sisson}{Peters and
  Sisson}{2006}]{peters+s06}
Peters, G. and S.~Sisson (2006).
\newblock Bayesian inference, {M}onte {C}arlo sampling and operational risk.
\newblock {\em Journal of Operational Risk\/}~{\em 1\/}(3), 27--50.

\bibitem[\protect\citeauthoryear{Pham, Nott, and Chaudhuri}{Pham
  et~al.}{2014}]{Pham2014note}
Pham, K.~C., D.~J. Nott, and S.~Chaudhuri (2014).
\newblock A note on approximating {ABC-MCMC} using flexible classifiers.
\newblock {\em Stat\/}~{\em 3\/}(1), 218--227.

\bibitem[\protect\citeauthoryear{Price, Drovandi, Lee, and Nott}{Price
  et~al.}{2018}]{price+dln16}
Price, L.~F., C.~C. Drovandi, A.~C. Lee, and D.~J. Nott (2018).
\newblock Bayesian synthetic likelihood.
\newblock {\em Journal of Computational and Graphical Statistics\/}~{\em
  27\/}(1), 1--11.

\bibitem[\protect\citeauthoryear{Priddle and Drovandi}{Priddle and
  Drovandi}{2020}]{priddleDrovandi2020}
Priddle, J.~W. and C.~Drovandi (2020).
\newblock Transformations in semi-parametric {Bayesian} synthetic likelihood.
\newblock arxiv:2007.01485.

\bibitem[\protect\citeauthoryear{Robins, Pattison, Kalish, and Lusher}{Robins
  et~al.}{2007}]{robins_pattison_kalish_lusher_2007}
Robins, G., P.~Pattison, Y.~Kalish, and D.~Lusher (2007).
\newblock An introduction to exponential random graph (p*) models for social
  networks.
\newblock {\em Social Networks\/}~{\em 29\/}(2), 173 -- 191.
\newblock Special Section: Advances in Exponential Random Graph (p*) Models.

\bibitem[\protect\citeauthoryear{Sisson, Fan, and Tanaka}{Sisson
  et~al.}{2007}]{sisson+ft07}
Sisson, S.~A., Y.~Fan, and M.~M. Tanaka (2007).
\newblock Sequential {Monte Carlo} without likelihoods.
\newblock {\em Proceedings of the National Academy of Sciences of the
  USA\/}~{\em 104}, 1760--1765. Errata (2009), {\it 106}, 16889.

\bibitem[\protect\citeauthoryear{Snijders, Pattison, Robins, and
  Handcock}{Snijders et~al.}{2006}]{snijders_pattison_robins_handcock_2006}
Snijders, T. A.~B., P.~E. Pattison, G.~L. Robins, and M.~S. Handcock (2006).
\newblock New specifications for exponential random graph models.
\newblock {\em Sociological Methodology\/}~{\em 36\/}(1), 99--153.

\bibitem[\protect\citeauthoryear{Tavar\'e, Balding, Griffiths, and
  Donnelly}{Tavar\'e et~al.}{1997}]{tavare+bgd97}
Tavar\'e, S., D.~J. Balding, R.~C. Griffiths, and P.~Donnelly (1997).
\newblock Inferring coalescence times from {DNA} sequence data.
\newblock {\em Genetics\/}~{\em 145}, 505--518.

\bibitem[\protect\citeauthoryear{Tsao and Wu}{Tsao and Wu}{2013}]{tsao2013}
Tsao, M. and F.~Wu (2013, 08).
\newblock Empirical likelihood on the full parameter space.
\newblock {\em Ann. Statist.\/}~{\em 41\/}(4), 2176--2196.

\bibitem[\protect\citeauthoryear{Tsybakov and van~der Meulen}{Tsybakov and
  van~der Meulen}{1996}]{tsybakovMeulen1996}
Tsybakov, A.~B. and E.~C. van~der Meulen (1996).
\newblock Root-n consistent estimators of entropy for densities with unbounded
  support.
\newblock {\em Scandinavian Journal of Statistics\/}~{\em 23\/}(1), 75--83.

\bibitem[\protect\citeauthoryear{Wood}{Wood}{2010}]{wood10}
Wood, S.~N. (2010).
\newblock Statistical inference for noisy nonlinear ecological dynamic systems.
\newblock {\em Nature\/}~{\em 466\/}(7310), 1102--1104.

\bibitem[\protect\citeauthoryear{Zhou and Yang}{Zhou and Yang}{2016}]{zhou+y16}
Zhou, M. and Y.~Yang (2016).
\newblock {\em emplik: Empirical Likelihood Ratio for Censored/Truncated Data}.
\newblock R package version 1.0-3.

\end{thebibliography}

\end{document}